\documentclass[preprint,authoryear,10pt]{elsarticle}
\usepackage{natbib}
\biboptions{authoryear}
\usepackage{geometry}
\usepackage[colorlinks,linkcolor=blue]{hyperref}
\usepackage{amsmath}
\usepackage[utf8]{inputenc}
\usepackage[T1]{fontenc}
\usepackage{url}
\usepackage{booktabs}
\usepackage{amsfonts}
\usepackage{nicefrac}
\usepackage{microtype}
\usepackage{lipsum}
\usepackage{graphicx}
\usepackage{tikz}
\usepackage[labelsep=period, figurename=Fig., tablename=Table]{caption}
\usepackage{subfigure}
\usepackage{subcaption}
\usepackage{float}  
\usepackage{hyperref, wasysym}
\usepackage{multicol}
\usepackage{booktabs}
\usepackage{makecell}
\usepackage{array}

\usepackage{doi}
\usepackage{amssymb}
\usepackage{enumerate}
\usepackage{algorithm, algpseudocode}

\usepackage{mathrsfs}
\usepackage{threeparttable}
\usepackage{appendix}
\usepackage{multirow}
\usepackage{xcolor}
\makeatletter
\renewcommand{\paragraph}{%
    \@startsection{paragraph}{4}{\z@}%
    {3.25ex \@plus1ex \@minus.2ex}%
    {-1em}%
    {\normalfont\normalsize}%
}
\makeatother

\newcommand{\subsubsubsection}[1]{
    \paragraph{#1}
    \par
    \indent
}
\begin{document}

\begin{frontmatter}

\title{Online Prediction-Assisted Safe Reinforcement Learning for Electric Vehicle Charging Station Recommendation in Dynamically Coupled Transportation-Power Systems}

\author[mymainaddress]{Qionghua Liao}
\ead{qionghua.liao@connect.polyu.hk}

\author[mymainaddress]{Guilong Li}
\ead{guilong97.li@connect.polyu.hk}

\author[mymainaddress]{Jiajie Yu\corref{mycorrespondingauthor}}
\ead{jiajie.yu@polyu.edu.hk}

\author[mysecondaryaddress]{Ziyuan Gu}
\ead{ziyuangu@seu.edu.cn}

\author[mymainaddress]{Wei Ma\corref{mycorrespondingauthor}}
\cortext[mycorrespondingauthor]{Corresponding author}
\ead{wei.w.ma@polyu.edu.hk}

\address[mymainaddress]{Department of Civil and Environmental Engineering, The Hong Kong Polytechnic University\\Hung Hom, Kowloon, Hong Kong SAR, China}
\address[mysecondaryaddress]{Jiangsu Key Laboratory of Urban ITS, Jiangsu Province Collaborative Innovation Center of Modern Urban Traffic Technologies\\School of Transportation, Southeast University, Nanjing 210096, China}

\begin{abstract}
With the proliferation of electric vehicles (EVs), the transportation network and power grid become increasingly interdependent and coupled via charging stations. The concomitant growth in charging demand has posed challenges for both networks, highlighting the importance of charging coordination. However, existing literature largely overlooks the interactions between power grid security and traffic efficiency, where the deterioration of grid security also leads to a decrease in traffic efficiency. In view of this, we study the en-route charging station (CS) recommendation problem for EVs in dynamically coupled transportation-power systems. The system-level objective is to maximize the overall traffic efficiency while enhancing the safety of the power grid. This problem is for the first time formulated as a constrained Markov decision process (CMDP), and an online prediction-assisted safe reinforcement learning (OP-SRL) method is proposed to learn the optimal and secure policy. To be specific, we mainly address two challenges. First, the constrained optimization problem is converted into an equivalent unconstrained optimization problem by applying the Lagrangian method, and then the Proximal Policy Optimization (PPO) method is extended to incorporate the constraint in the sequential decision process through the inclusions of cost critic and Lagrangian multiplier. Second, to account for the uncertain long-time delay between performing charging station recommendation and commencing charging, we put forward an online sequence-to-sequence (Seq2Seq) predictor for state augmentation, offering foresightful information to guide the agent in making forward-thinking decisions. Finally, we conduct comprehensive experimental studies based on the Nguyen-Dupuis network and a large-scale real-world road network, coupled with IEEE 33-bus and IEEE 69-bus distribution systems, respectively. Results demonstrate that the proposed method outperforms baselines in terms of road network efficiency, power grid safety, and EV user satisfaction. The case study on the real-world network also illustrates the applicability in the practical context.
\end{abstract}

\begin{keyword}
Electric Vehicle (EV) \sep Charging Station Recommendation \sep Coupled Transportation-Power Systems \sep Constrained Markov Decision Process \sep Safe Reinforcement Learning
\end{keyword}

\end{frontmatter}

\section{Introduction}
Transportation electrification through the adoption of electric vehicles (EVs) is regarded as a promising step for environmentally sustainable urban development, which has garnered increasing interest worldwide in recent years {\citep{das2020multi, Stockkamp.2021, dai2023data, li2024reinforcement}}. As reported in the Global EV Outlook by the International Energy Agency \citep{iea2023}, global EV exhibits an exponential growth trend in sales and is expected to reach almost 250 million in 2030. Concurrently, the number of publicly accessible chargers is expected to rise from 2.7 million in 2022 to about 13 million in 2030 across the world. With the growing penetration of EVs and the increasing stock of public chargers, the operations of urban transportation network (UTN) and power distribution network (PDN) are becoming increasingly interconnected via charging stations (CSs) \citep{sun2021spatial, ding2022optimal, aghajan2023optimal}.

Despite the support of government policies and advancements in battery technology, the large-scale development of EVs, especially private vehicles, is still facing a series of obstacles under the context of coupled transportation-power systems: 1) EVs face multiple barriers such as perceived range anxiety, prolonged charging duration, and uncertain waiting time at public charging stations \citep{guo2018battery, dastpak2024dynamic}; 2) the massive adoption of EVs poses additional risks and challenges to the closely coupled transportation and power grid systems due to their inherent demands for both mobility and recharging \citep{cui2021optimal, sun2021spatial}. Specifically, the routing behavior of EVs towards/from designated CSs constitutes a crucial component of traffic flow and exerts a significant impact on traffic congestion. The duration of queuing and charging at CSs, meanwhile, also affects the overall travel efficiency. Furthermore, the charging process of EVs introduces extra loads to the power grid, forming part of the power flow and posing pressure to the secure and reliable operation of PDN, such as incurring an increase in voltage drops at PDN nodes (also termed electricity buses and used interchangeably hereafter) \citep{nour2020review, gao2024charging} and even leading to blackouts \citep{khalid2021comprehensive}. Therefore, it is imperative to examine the EVs charging station recommendation strategy considering the dynamically coupled transportation-power systems. However, most existing studies focus only on the performance of a single EV \citep[e.g.,][]{jin2022shortest,huang2023electric,jiang2024electric} or multiple EVs \citep[e.g.,][]{xing2022graph,lin2022toward,bachiri2023multi} from the UTN's perspective (e.g., time consumption and/or charging cost), while failing to consider the system-level traffic efficiency and overlooking the safety of the PDN.

In this paper, we focus on the task of public charging station recommendation for en-route EVs from an interdisciplinary perspective. The system-level objectives pertaining to various stakeholders are taken into account, i.e., maximizing the traffic efficiency of UTN while ensuring the safety of PDN. Hence, the task can be regarded as a constrained optimization problem. Specifically, the overall traffic efficiency is characterized using the total travel time of all vehicles in the road network, including EVs with charging needs and other vehicles without charging requirements. A vehicle's trip covers from its origin to its destination, consisting of the driving time on the road, possible queuing time at CSs, and potential charging time at CSs. Given that nodal voltage magnitude is commonly considered one of the most critical security indicators for PDN, where large deviations from nominal voltage indicate the degraded performance of the system \citep{dixit2019integration}. Thus voltage violation (i.e., the difference between the operation voltage and the nominal voltage) is utilized to signify the security of the power grid, which should be kept as small as possible.

It is worth noting that the consideration of voltage violation has a positive impact on both ensuring grid safety and boosting traffic efficiency in the context of coupled systems. When the security of the power grid is under threat, the control mechanism for voltage stability will be triggered \citep{deconinck2015combining}, which might cause a decrease in charging power and, subsequently, a decline in transportation efficiency. The voltage controller is considered the first step towards the integration of EVs into the power grid instead of merely an auxiliary facility \citep{clement2011impact}. For example, voltage droop control \citep{deconinck2015combining} is a conventional method to reduce the severity of voltage violation, the basic idea of which is to linearly reduce the output power when the nodal operation voltage falls below the reference value until the minimum power is reached. In this way, when the security of UTN is jeopardized by voltage drops, the charging power of EVs will also be affected, which in turn causes the increase of queuing and charging times of EVs at charging stations. Eventually, the overall traffic efficiency will be impaired (as illustrated on the right side of Figure \ref{fig:coupled_sys}). However, the inclusion of voltage-responsive charge controllers increases the complexity of the interaction behavior between the components of the coupled systems, thus making it more challenging to solve the CS recommendation problem.

The charging station recommendation problem considered here entails real-time control for the coupled systems of UTN and PDN, which is inherently a sequential decision-making task. In recent years, reinforcement learning (RL), especially deep RL (DRL), has attracted considerable attention as model-free methods for tackling complex tasks in real-time, including charging station recommendation for an EV/EVs \citep[e.g.,][]{zhang2020effective,lee2020deep,xing2022graph,jin2022shortest,xu2022real,li2023multi,jiang2024electric} and other traffic control tasks \citep[e.g.,][]{chen2021graph,dong2021space,chow2021adaptive,AHAMED2021227,liu2022deep,su2023hierarchical,xie2023two,WANG2023104309,liu2024distributed,ma2024providing}. {In RL-related studies, the targeted problem is often modeled as a Markov Decision Process (MDP), aiming at maximizing the long-term cumulative rewards regarding the objective. {However, the task confronting us is a sequential decision-making problem where both the objective and the constraint need to be taken into account. Given that regular RL methods typically address objective-oriented tasks, they are inadequate for the constrained optimization problem.} Safe reinforcement learning (SRL) \citep{garcia2015comprehensive} is regarded as a sub-field within RL to learn the optimal policy while respecting safety constraints during the learning process. The SRL method uses the constrained MDP (CMDP) model \citep{altman2021constrained} by extending regular MDP to account for the constraint, making it essentially suitable for addressing constrained optimization problems.} Yet, even though SRL has gained growing interest in control tasks involving constraint, it is still in the early development stage with relatively limited application \citep{gu2022review}. In particular, there is hardly any research on applying SRL methods to the CS recommendation task via a CMDP model.

This paper proposes an SRL-based method to deal with the constrained CS recommendation problem and formulate it as a CMDP model for the first time. When applying the SRL method to this constrained charging station recommendation problem, we need to tackle two key challenges: 1) The first challenge is how to incorporate the constraint in the sequential decision-making process, since regular RL methods only focus on long-term objectives. To address this problem, we implement the Lagrangian method \citep{chow2018risk} to convert the constrained optimization problem of the CMDP into an equivalent unconstrained optimization problem; and then the proximal policy optimization (PPO) method \citep{schulman2017proximal} is extended to incorporate constraint in the learning process through the inclusions of cost critic and Lagrangian multiplier, to learn the desired charging station recommendation policy. 2) The second challenge is the uncertain long-time delay between performing the en-route charging station recommendation and commencing charging at CS. This challenge is due to the complicated and ever-changing system state under the interaction of the two networks, i.e., UTN and PDN. The delay may lead to misleading strategy results and cause instability in training when the agent only focuses on the current state. In this regard, we put forward an online sequence-to-sequence (Seq2Seq) predictor \citep{keneshloo2019deep} to augment the state space of the PPO agent with predicted charging demands at CSs. The Seq2Seq model is based on the recurrent neural network (RNN) and can provide estimated future state sequences given historical information, thus offering insightful information to guide the agent in making more forward-thinking decisions. 

Accordingly, an online prediction-assisted safe reinforcement learning (OP-SRL) method is developed to find the effective CS recommendation policy for en-route EVs. The policy is gradually updated via experiences from repeated interactions between the SRL agent (i.e., the EV coordination center) and the environment in the context of dynamically coupled systems. Upon receiving a charging request, the control center will promptly make a decision regarding CS selection for the EV user based on the current system status and specific condition of the EV. 

To validate the effectiveness of the proposed OP-SRL method for CS recommendation and examine the additional value of incorporating the Lagrangian multiplier and state augmentation, we conduct comprehensive numerical studies based on the Nguyen-Dupuis network and a large-scale real-world transportation network coupled with IEEE 33-bus and IEEE 69-bus distribution systems, respectively. Three performance metrics are designed for performance evaluation with regard to road network efficiency, power grid safety, and EV user satisfaction. Based on the Nguyen-Dupuis network, the effectiveness of the Lagrangian multiplier and state augmentation in the proposed method is unveiled and quantified through a comparison with six baselines. Besides, sensitivity analysis for several parameters (i.e., the variation of EV penetration, time resolution of the charge controller, and decoder length in the predictor) are implemented to show the robustness and flexibility of the proposed method. Furthermore, numerical experiments on a large-scale real-world road network are performed to showcase the applicability of the proposed method in the practical context.

To summarize, this paper makes the following innovative contributions:

\begin{itemize}
\item {The dynamically coupled systems of the transportation network and the power distribution network are introduced to better capture the effects of the CS recommendation strategy on different stakeholders, considering that the deterioration of grid safety (i.e., voltage violation) also leads to a decrease in traffic efficiency.}
\item {For the first time, to the best of our knowledge, the constrained CS recommendation problem is formalized as a CMDP model, where the system-level objective and constraint (i.e., overall traffic efficiency and grid security) are accounted for in the decision-making process.}
\item {A novel online prediction-assisted safe reinforcement learning (OP-SRL) method is proposed to solve the CMDP by extending the PPO method, where 1) the Lagrangian method is introduced for the constraint-free conversion of the CMDP; and 2) an online Seq2Seq predictor is put forward for state augmentation.}
\item {Extensive computational experiments are conducted on the Nguyen-Dupuis network and a large-scale real-world road network coupled with IEEE 33-bus and IEEE 69-bus distribution systems. The efficacy and practical applicability of the proposed method are substantiated in terms of road network efficiency, power grid safety, and EV user satisfaction.}
\end{itemize}

The remainder of the paper is organized as follows. In Section \ref{sec:lr}, we review the related literature. In Section \ref{Problem statement and model formulation}, the en-route charging station recommendation problem, the coupled transportation-power systems, and the formulated CMDP model are developed and delineated. We further present the proposed OP-SRL method in Section \ref{sec:OP-SRL} and show the numerical studies and results in Section \ref{sec:cases}. Finally, we draw our conclusions and provide final thoughts in Section \ref{sec:conc}.

\section{Literature review} \label{sec:lr}

Over the past decade, extensive studies have investigated the charging station recommendation problem of the EV, also known as charging navigation or charging guidance. In this section, we review the relevant literature from three perspectives (i.e., coupled systems, policy objectives, and solution methods) and then pinpoint the major research gaps.

\subsection{Coupled systems}

In the scenarios of charging station recommendation, the majority of studies have only considered a single UTN system and focused on the interaction process among the traffic network, single EV or EVs, and CSs \citep{liu2017electric, zhang2019congestion, qian2019deep, lee2020deep, zhang2020effective, lin2022toward, jin2022shortest, xing2022graph, basso2022dynamic, bachiri2023multi, huang2023electric, jiang2024electric}, while overlooking the development of strategies for coupled UTN and PDN systems. 

Some research has taken into account coupled systems of UTN and PDN, with emphases on different aspects. One category of studies tended to integrate the traffic assignment model with the optimal power flow (OPF) model to achieve traffic flow and power flow that meet the constraints of system operation, where the selection of CSs is dictated by the resulting traffic flow distribution \citep[e.g.,][]{geng2019smart, zhang2020power, ding2022optimal}. For example, \citet{ding2022optimal} developed two market competition models, the Nash game model and the Stackelberg game model, to decide the discharge pricing strategy; and then the routing and discharging behaviors of shared EVs were derived. These studies mainly focused on the distribution of traffic flow in the context of coupled systems, failing to capture individual characteristics and randomness of charging requests.

In another category of studies, \citet{shi2020distributed} and \citet{li2023multi} focused on the interaction of the UTN and the PDN through hourly locational marginal price (LMP) and aimed at optimizing charging costs along with time consumption, where LMP is achieved by solving the optimal power flow (OPF) problem. Different from these two studies which considered the electricity pricing for coupling systems, \citet{xu2022real} explored the effect of EVs charging load on the operation stability of the grid in terms of voltage deviation. However, to the best of our knowledge, there is a lack of research exploring the effect of charge controllers in the problem of charging station recommendation, even though voltage controllers are regarded as potentially compulsory due to the need to ensure the reliable operation of power grid \citep{clement2011impact}, especially in light of the rising penetration rate of EVs and the expansion of CSs construction.

\subsection{Strategy objectives}

Various objectives have been considered in the literature for the CS recommendation problem. Most of the existing research has examined various types of time consumption at the individual level or aggregate level. From the individual EV's perspective, \citet{shi2020distributed} and \citet{liu2017electric} accounted for the time consumption from the generation of charging request to charging completion, which includes driving time in the road network, waiting time, and charging time at CSs. \citet{jin2022shortest} and \citet{jiang2024electric} considered driving time to the target CS and waiting time at the CS as the time cost for consideration, while disregarding the variations of charging time. In \citet{huang2023electric}, the travel time of an individual EV from origin to destination along with charging time was taken into account, while overlooking the queuing process at the CS. At the aggregate level, similarly, some studies took into account the elapsed time of EV fleet from decision-making to charging completion \citep[e.g.,][]{lin2022toward, xing2022graph, li2023multi, bachiri2023multi}. In this category, \citet{qian2019deep}, \citet{xu2022real} and \citet{zhang2019congestion} assumed a constant charging time, while the charging dynamics are omitted. For the entire OD trip, the travel time that entails charging behavior was accounted for in \citet{lee2020deep}. In addition to elapsed time, some of the existing research has concentrated on charging cost at CSs \citep[e.g.,][]{li2023multi, shi2020distributed, jiang2024electric}, battery energy consumption while driving \citep[e.g.,][]{jiang2024electric, qian2019deep, basso2022dynamic}, and driving distance \citep[e.g.,][]{zhang2020effective, jin2022shortest}. Nevertheless, previous studies mainly focused on individual or multiple EVs, with a scarcity of research examining the impact of CS recommendation strategy on the overall traffic flow. 

According to the number of objectives under consideration for the optimal strategy, existing research can be categorized into single-objective \citep[e.g.,][]{huang2023electric, bachiri2023multi, basso2022dynamic} and multi-objective \citep[e.g.,][]{li2023multi, xu2022real, shi2020distributed}. Typically, \citet{xu2022real} considered the minimization of driving time and waiting time of EVs, service balance of CSs, traffic congestion, and voltage deviation. However, most studies that deal with multiple objectives adopted the weighted sum approach to convert them into a single objective, which has limitations such as difficulty in adjusting weights and poor robustness. Besides, some research has been dedicated to constrained optimization problems for charging station recommendation. In \citet{xu2022real}, a multi-objective optimization problem with a lower voltage bound constraint was considered. When handling the constraint, a fixed penalty was added to the objective function to deal with the case of constraint violation. \citet{basso2022dynamic} proposed a chance-constrained optimization problem to minimize the expected energy consumption while considering the failure probability of route completion. However, they focused on routing planning for a single EV to service customer requests, with charging planning as a subsidiary product. 

\subsection{Solution methods}

Charging station recommendation involves dynamically processing charging requests from EVs, which necessitates the use of real-time control methods for resolution. In the existing literature, \citet{shi2020distributed} developed a multi-agent system utilizing the distributed biased min-consensus algorithm to solve the charging navigation problem. \citet{huang2023electric} developed a Mixed-integer nonlinear programming (MINLP) model for the charging navigation of a single EV. Due to the ability to deal with uncertainty without prior knowledge and strike a balance between short-term rewards and long-term returns \citep{qiu2023reinforcement}, the model-free RL methods have gained notable success in real-time traffic control tasks in recent years, including vehicle trajectory control \citep[e.g.,][]{chen2021graph, dong2021space, liu2024distributed}, perimeter control \citep[e.g.,][]{su2023hierarchical, hu2024guided, LI2024103016}, signal control \citep[e.g.,][]{chow2021adaptive, su2023hierarchical}, dynamic pricing \citep[e.g.,][]{LEI2023102848}, control strategies in metro systems \citep[e.g.,][]{WANG2023244,YING202236,YING2020210} and railway systems \citep[e.g.,][]{SEMROV2016250,LI2022230}, etc. In the meantime, the majority of studies on CS recommendation for an EV or EV fleet tends to employ RL \citep[e.g.,][]{zhang2019congestion, basso2022dynamic} or DRL \citep[e.g.,][]{jin2022shortest, xu2022real, bachiri2023multi} to model the sequential decision-making problem as a MDP. Since RL is capable of expressing personalized attributes of different EVs, such as origin point \citep[e.g.,][]{qian2019deep}, initial state of charge (SoC) \citep[e.g.,][]{qian2019deep, xu2022real}, and destination point \citep[e.g.,][]{lee2020deep}, they are well suited to make online decisions for heterogeneous EVs.

When addressing the CS recommendation problem via RL methods, existing research mainly concentrates on unconstrained optimization problems \citep[e.g.,][]{li2023multi, jiang2024electric, xing2022graph, jin2022shortest}, which can be addressed via regular DRL methods. For instance, \citet{xing2022graph} integrated graph convolutional network (GCN) with a modified Rainbow algorithm to develop the optimal charging navigation strategy, where two objectives regarding travel time and charging cost of EV fleet are considered in a weighted average manner. \citet{jiang2024electric} proposed a hierarchical RL framework based on two DQNs to solve the charging destination and route problems for a single EV, aiming at minimizing the charging cost and travel cost. In \citet{bachiri2023multi}, a multi-agent DDPG method is proposed for optimal EV charging station recommendations, focusing on a single objective of time consumption. Although RL methods have been demonstrated to be appropriate for addressing the problem under study, a significant practical issue still confronts us, i.e., the safety of the agent in a constrained context. In \citet{xu2022real}, the DQN($\lambda$) with graph attention networks (GATs) was applied to find the optimal recommendation strategy, where a weighted method was utilized to deal with multiple objectives and the penalty for violating constraints with constant penalty factors. While regular RL methods can address the constrained optimization problem via incorporating a fixed penalty factor into the reward function, it requires tedious manual tuning of the penalty coefficient, and a fixed factor is ill-suited to accommodate varying levels of constraint violation.

To address the sequential decision-making problem in a constrained context, the SRL has received growing attention recently, which, yet, is still in the early stages with relatively limited applications \citep{gu2022review}. At present, SRL methods are primarily applied in some scenarios such as robot control \citep{garcia2020teaching} and autonomous driving \citep{gu2022constrained}, while the application in the scenario of charging station recommendation for EVs is scarce. In \citet{basso2022dynamic}, they developed an SRL algorithm to solve the chance-constrained optimization problem by designing two rule-based safety layers to manage the risk of failure, with the focus on routing planning for a single EV to service customer requests. Yet, there is still a lack of research investigating the direct handling of constraints in this context.

\subsection{Summary}

Table \ref{tab:review} summarizes the related works on CS recommendation and highlights the consideration of coupled systems, objective level in terms of vehicles involved, processing of constraint in the method, and design of the method. {Upon reviewing the relevant literature on charging station recommendation, we identify and summarize four main research gaps: 1) There is a scarcity of studies that have considered the synergistic effect of CS recommendation strategy on the coupled transportation-power systems, especially in the context of dynamically integrated scenarios. 2) The positive effect of considering voltage violation on grid safety and traffic efficiency has been greatly underestimated. 3) The literature has overlooked the optimization and evaluation of management strategies from a system-level perspective, as most studies focus on an individual EV or multiple EVs within the road network, rather than the entire traffic flow. 4) The prevailing research tends to simplify the developed multi-objective or constrained optimization problem through the weighted sum method with fixed penalty factors.}

To address these gaps, this paper examines the constrained charging station recommendation problem in the context of dynamically coupled transportation-power systems, taking into account the congestion effect of the UTN, time-varying charging loads of power flow, and the dynamic interaction among different stakeholders. We aim at maximizing the overall traffic efficiency of UTN while meeting the safety requirement of PDN in terms of voltage violation. Notably, the effect of the deterioration of power grid safety (i.e., voltage violation) on traffic efficiency due to the potentially compulsory charge controllers is also considered in this paper, which complicates the CS recommendation problem to be addressed. To this end, the problem is formulated as a CMDP model and the novel OP-SRL method is proposed to solve it. The details will be described in Section \ref{Problem statement and model formulation} and Section \ref{sec:OP-SRL}.

\begin{table}[h]
  \small
  \centering
  \caption{Summary of some recent studies related to charging station recommendation}
  \label{tab:review}
\begin{threeparttable}
  \begin{tabular}{m{2cm}<{\centering}m{2.5cm}<{\centering}m{3cm}<{\centering}m{3cm}<{\centering}m{4cm}<{\centering}}
    \toprule
    \\[-20pt]
    \centering Reference & \makecell[c]{Coupled systems} & \makecell[c]{Objective level\tnote{1}} & \makecell[c]{Constraint processing\tnote{2}} & \makecell[c]{Method\tnote{3}}\\
    \\[-20pt]
    \midrule
    \citet{li2023multi} &\checkmark & EV fleet & -- & MARL\\
      \citet{xu2022real} & \checkmark & EV fleet & Weighted method & DRL\\
      \citet{shi2020distributed} & \checkmark & Individual EV & -- & DBMC\\
      \citet{jiang2024electric} & $\times$ & Individual EV & -- & DRL\\
      \citet{huang2023electric} & $\times$ & Individual EV & Optimization solver & MINLP\\
      \citet{xing2022graph} & $\times$ & EV fleet & -- & DRL\\
      \citet{jin2022shortest} & $\times$ & Individual EV & -- & DRL\\
      \citet{qian2019deep} & $\times$ & EV fleet & -- & DRL \\
      \citet{liu2017electric} & $\times$ & Individual EV & -- & DP\\
      \citet{bachiri2023multi} & $\times$ & EV fleet & -- & MARL\\
      \citet{basso2022dynamic} & $\times$ & Individual EV & Two rule-based safety layers & SRL\\
      \citet{lin2022toward} & $\times$ & EV fleet & -- & Multi-phase MDP\\
      \citet{zhang2020effective} & $\times$ & EV fleet & -- & DRL\\
      \citet{lee2020deep} & $\times$ & EV fleet & -- & DRL\\
      \citet{zhang2019congestion} & $\times$ & EV fleet & -- & RL\\
      \textbf{This paper} & \checkmark & System-level & Lagrangian method & SRL\\
    \bottomrule
  \end{tabular}
\begin{tablenotes}
\footnotesize
\item[1] This field indicates which level of objective in terms of vehicles involved is considered in the system. At the system level, studies aimed at investigating the impact of CS recommendation strategy on all vehicles in the road network (including EVs and non-EVs); whereas at the individual level, studies focused exclusively on a single EV, and the consideration at the EV fleet-level is limited to the impact on multiple EVs (without other types of vehicles).
\item[2] Refer specifically to the constraints in CS recommendation modeling.
\item[3] Multi-agent RL (MARL); Distributed biased min-consensus (DBMC); Dynamic programming (DP); Mixed-integer nonlinear programming (MINLP).
\end{tablenotes}
\end{threeparttable}
\end{table}

\section{Problem statement and model formulation} \label{Problem statement and model formulation}
In Section \ref{Problem statement}, we formally define the charging station recommendation problem and introduce the dynamically coupled transportation-power systems in Section \ref{Coupled transportation-power systems}. In Section \ref{CMDP model}, we mathematically formulate the proposed problem in terms of a CMDP.

\subsection{Problem setup} \label{Problem statement}
This paper studies the public charging station recommendation problem for en-route EVs with specific destinations. The goal is to obtain the system-level optimal strategy for CS guidance, with the objective of maximizing overall traffic efficiency while {enhancing the security of the PDN operation in the context of dynamically coupled transportation-power systems.} More specifically, maximizing traffic efficiency is equivalent to minimizing the total travel time of all vehicles within a certain period, instead of only one EV or several EVs as the research reviewed in Table \ref{tab:review}. The security constraint of the power grid is defined as the total voltage deviation from the nominal value at the grid nodes, owing to the fact that nodal voltage is regarded as one of the most crucial safety and service metrics for the power grid \citep{ghasemi2014multi, dixit2019integration}. As mentioned above, the consideration of PDN security has a positive impact on both ensuring grid safety and boosting traffic efficiency (see Section \ref{Coupled transportation-power systems} for details). 

In this study, we consider the EV coordination center responsible for the comprehensive management and processing of sequential charging requests from EVs. Upon receiving a new charging request from an EV, the coordination center will promptly recommend a CS to the EV user based on the current system status and the specific condition of the EV, such as its location and battery level. Assume that EVs always follow the guidance \citep{lin2022toward, xu2022real}; for example, with the support of autonomous electric vehicle technology. After determining the target CS, the EV users will plan routes considering real-time traffic conditions. Once the EV reaches the CS, it will charge directly if there are available charging spots; otherwise, it must wait in line until it becomes the lead car in the queue, and meanwhile, a charging spot becomes available. Finally, when the charging EV achieves the desired battery level, it will depart from the CS and travel to the destination based on route planning.

\subsection{Coupled transportation-power systems} \label{Coupled transportation-power systems} 

In this subsection, we develop a general framework of the dynamically coupled transportation-power system to capture the dynamic and intimate interactions between distinct stakeholders. {Differing from the widespread use of macroscopic models for aggregated traffic flow modeling in existing literature \citep[e.g.,][]{wang2024coordinating}, we craft the coupled systems from a microscopic perspective, allowing for the implementation and evaluation of detailed individual-level management strategies, such as the charging station recommendation strategy considered in this study.}

\begin{figure}[h!]
\center
\includegraphics[width=1\textwidth]{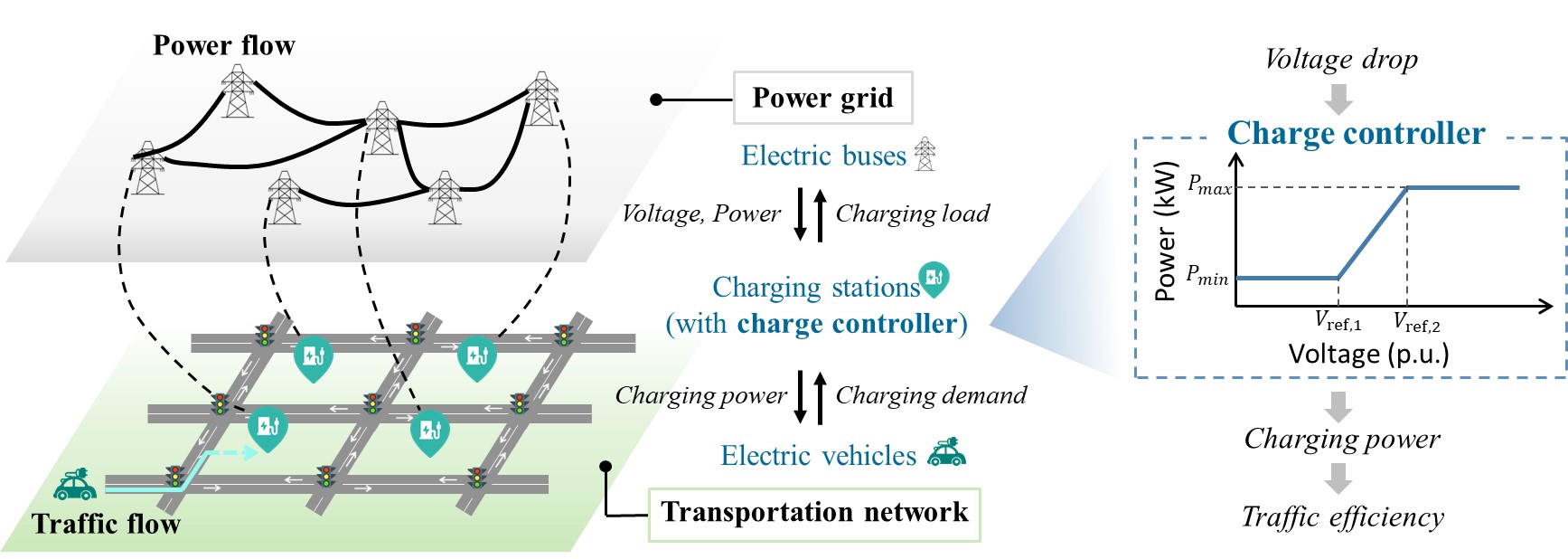}
\caption{Illustration of the synergy in the dynamically coupled transportation-power systems. (1) The left diagram illustrates the coupling relationships between UTN and PDN through EVs, CSs, and buses, as well as the information flow among various components in the coupled systems. (2) The diagram on the right explains the working mechanism of the charge controller and its impact on traffic efficiency.}
\label{fig:coupled_sys}
\end{figure}

As depicted in the left part of Figure \ref{fig:coupled_sys}, the coupled systems consist of two sub-systems, transportation network (i.e., UTN) and power grid (i.e., PDN), which are intricately integrated through EVs and CSs. Specifically, the connection between the transportation network and CSs is established via EV routing behavior to/from CSs. {When arriving at the CS, the EV needs to draw power from the grid to fulfill charging demand and thus introduces additional charging loads to the PDN. In other words, each CS aggregates the charging power load of the connected EVs and provides it to the PDN through electricity buses (i.e., the interconnection nodes connected to loads or other components of the power system). The PDN then distributes the charging power to electrical users, including CSs, to meet their electricity demands. As such, the CSs and the PDN are interconnected through EV charging behaviors.} 

As a result of dynamic interactions, EV routing behavior contributes to the formation of traffic flow and imposes an effect on traffic congestion. On the other hand, the charging process of EVs constitutes the power flow and is highly correlated with voltage instability \citep{dharmakeerthi2014impact}. {For example, if there is no coordinated charging guidance, a large number of EVs may charge simultaneously at a single CS and potentially overload the PDN, especially under high EV penetration scenarios. Such uncoordinated EV charging behaviors can lead to unexpected voltage drops and a decrease in power quality \citep{wu2020online}, suggesting a potential threat to power grid security.}

We highlight that EVs routing process and charging behavior both contribute to the traffic efficiency. Given the adverse effect of EV charging load on the power grid, various control mechanisms have been developed and employed in the existing literature to reduce the voltage deviation caused by charging behavior \citep{shareef2016review}. For example, voltage-responsive charge control methods have been identified as viable solutions to address the voltage deviation issues \citep{deconinck2015combining}. As depicted in the right part of Figure \ref{fig:coupled_sys}, when the voltage of the grid system falls below the reference value (i.e., $V_{\text{ref},2}$), the charging power at CSs will decrease to prevent further voltage violation, resulting in longer waiting time and charging time. In this way, when the security of the grid is threatened in terms of voltage drop, the charging power of EVs is also affected, thereby further impacting the overall transportation efficiency, especially at high EV penetration.

\begin{figure}[h!]
\center
\includegraphics[width=0.85\textwidth]{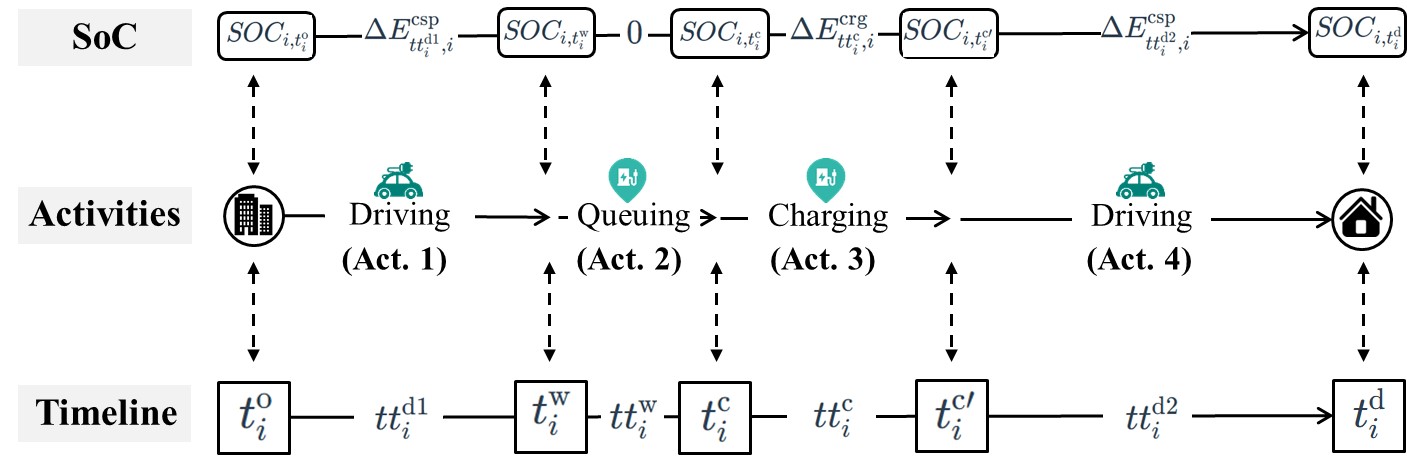}
\caption{The correspondence diagram of trip activities, timeline, and state of charge (SoC) change for exemplary EV $i$ ($i \in \mathcal{I}^\mathrm{EV}$) with charging request. (1) The activities chain in the middle illustrates the sequence of events of EV $i$ from the starting point to the trip destination, including the driving process from origin to target CS, the queuing process (if present), the charging process, and the driving process to the destination. (2) The timeline chain at the bottom shows the start time (inside the boxes) and duration (between two boxes) of each activity using variable representations. (3) The SoC chain at the top represents the SoC change at each key time point (in the boxes) and the variations in battery energy (between two boxes).}
\label{fig:act_time_soc}
\end{figure}

The overall traffic efficiency is quantified through the total travel time of all vehicles in the UTN. Let $\mathcal{I}^\mathrm{EV}$ be the set of EVs that require charging, and $\mathcal{I}^\mathrm{CV}$ be the set of all other vehicles. Then the set of all vehicles involved in the UTN is defined as $\mathcal{I}^\mathrm{ALL} = \mathcal{I}^\mathrm{EV} \cup \mathcal{I}^\mathrm{CV}$. As shown in Figure \ref{fig:act_time_soc}, the trip of a typical EV $i \in \mathcal{I}^\mathrm{EV}$ with charging request is composed of the driving process, queuing process, and charging process. While the trips for $i \in \mathcal{I}^\mathrm{CV}$ only entail driving behavior. As such, the total travel time for all vehicles, $TTT$, can be obtained via Equation \ref{eq:TTT-1}, based on the travel time for each vehicle in $\mathcal{I}^\mathrm{EV}$ and $\mathcal{I}^\mathrm{CV}$, $tt_i$, calculated using Equation \ref{eq:tt-1}. 
\begin{align}
\label{eq:TTT-1}
TTT&\,=\,\sum_{i\in\mathcal{I}^{\mathrm{ALL}}}tt_{i},\\
\label{eq:tt-1}
tt_{i}&\,=\,
\begin{cases}
tt_i^\text{d} + tt_i^\text{w} + tt_i^\text{c}, \, \text{for }i \in \mathcal{I}^\mathrm{EV}, \\
tt_i^\text{d}, \qquad\qquad\quad \text{for }i \in \mathcal{I}^\mathrm{CV}.
\end{cases}
\end{align}
where $tt_i^\text{d}$, $tt_i^\text{w}$, and $tt_i^\text{c}$ represent the driving time, waiting time, and charging time of vehicle $i$ during the trip from the starting point to destination; As shown in Figure \ref{fig:act_time_soc}, $tt_i^\text{d}$ for $i \in \mathcal{I}^\mathrm{EV}$ is the sum of two parts, i.e., the driving time from starting point to target CS $tt_i^\text{d1}$ and the driving time from CS to destination after completing the charging supplement $tt_i^\text{d2}$, which can be expressed as $tt_i^\text{d} = tt_i^\text{d1} + tt_i^\text{d2} (i \in \mathcal{I}^\mathrm{EV})$. Denote the start time of vehicle $i$'s trip as $t_i^\text{o}$. Then the starting times of queuing process $t_i^\text{w}$ and charging process $t_i^\text{c}$, the time when charging completion $t_i^{\text{c}\prime}$, as well as the end time of trip $t_i^\text{d}$ for $i \in \mathcal{I}^\mathrm{EV}$ can be derived based on $tt_i^\text{d1}$, $tt_i^\text{w}$, $tt_i^\text{c}$, and $tt_i^\text{d2}$.

The safety performance of the PDN is accounted for by the total voltage violation at the buses over the operating time period. Let $\mathcal{U}^\mathrm{PDN}$ be the set of all buses in the PDN. The time period for evaluation is divided into multiple time steps, the set of which is represented as $\mathcal{T}^\text{as}$. Then the total voltage violation averaged over all buses, $CVV$, can be depicted in Equation \ref{eq:CVV-1} based on the operating voltage magnitudes in each time segment.
\begin{align}
\label{eq:CVV-1}
CVV &\,=\, \frac{1}{\lvert \mathcal{U}^\mathrm{PDN} \rvert} \sum_{t \in \mathcal{T}^\text{as}} \sum_{b \in \mathcal{U}^\mathrm{PDN}} v_{b,t}^\text{dev}, \\
\label{eq:CVV-2}
v_{b,t}^\text{dev} &\,=\, |v_{b,t}-v_\text{ref}|,
\end{align}
where $\lvert \mathcal{U}^\mathrm{PDN} \rvert$ indicates the number of buses; $v_{b,t}^\text{dev}$ denotes the voltage violation at bus $b \in \mathcal{U}^\mathrm{PDN}$ in time segment $t$, which can be computed by Equation \ref{eq:CVV-2}; $v_{b,t}$ is the operating voltage magnitude at bus $b$ during time $t$; $v_\text{ref}$ denotes the reference value of voltage magnitude, which is typically set to 1.0 per unit (p.u.).

To estimate the time duration of each trip activity and the operating voltage of PDN under the influence of the CS recommendation strategy, the dynamically coupled systems are modeled in terms of the involved key components including UTN, EVs, PDN, and CSs, as depicted in the coupled systems diagram in Figure \ref{fig:OP_SRL}.

(1) UTN: Dynamic network loading model

Dynamic network loading (DNL) models endogenously encapsulate the traffic flow propagation and congestion effect in the transportation network from a microscopic or macroscopic perspective, given road network and traffic demand \citep{ma2020estimating}. {Since we focus on each charging request in the overall traffic flow as well as the CS guidance for each EV, the microscopic DNL model is utilized for traffic flow propagation. Generally, microscopic-level DNL can be represented by individual trip-based simulation models, e.g., SUMO \citep{behrisch2011sumo}. To be specific, the microscopic DNL model carries out traffic flow propagation according to individual travel demand featuring the departure time, origin, and destination. With the generation and evolution of traffic demand, the congestion effect in each link can then be obtained from the DNL model, such as travel time and speed.} Denote the UTN used for modeling as $\mathcal{N}et^\mathrm{UTN}$ with the set of road links $\mathcal{E}$. For a specific time segment $t$, the DNL model $\Lambda_t(\cdot)$ can be represented as Equation \ref{eq:UTN-1}.
\begin{equation}
\left\{tt_{i}^\mathrm{d} 
\right\}_{i \in \mathcal{I}^\mathrm{ALL}}, \left\{k_{e, t} \right\}_{e \in \mathcal{E}}=\Lambda_t\left(\mathcal{I}^\mathrm{EV}, \mathcal{I}^\mathrm{CV}, \mathcal{N}et^\mathrm{UTN} \right),
\label{eq:UTN-1}
\end{equation}
where $tt_{i}^\mathrm{d}$ denotes the driving time of vehicle $i$; and $k_{e,t}$ represents the traffic density of road link $e$ at time $t$, which serves as the performance index of traffic congestion level.

(2) EVs: Route-planning method and battery energy variation model

For route planning, all vehicles in $\mathcal{I}^\mathrm{EV} \cup \mathcal{I}^\mathrm{CV}$ are involved since we focus on the overall trip time spent, {including the trip from the starting point to the destination for vehicles in $\mathcal{I}^\mathrm{CV}$, the trip from the starting point to target CS for vehicles in $\mathcal{I}^\mathrm{EV}$ after CS recommendation, and the trip from the CS to destination for vehicles in $\mathcal{I}^\mathrm{EV}$ upon completion of charging process.} Assume that all vehicles involved follow the shortest path based on the Dijkstra algorithm \citep[e.g.,][]{xing2022graph, zhang2020effective}. {The shortest path means the path with the minimum travel time which is derived from the DNL model based on current traffic condition. Subsequently, the resulting trajectories from route planning serve as the inputs for the DNL model in Equation \ref{eq:UTN-1}. In this way, the DNL model and the route-planning method are performed alternately.}

Therefore, the DNL model and the selected route determine the arrival time of EVs at target CSs according to the traffic state, which potentially affects the temporal distribution of EV charging demands and further influences the charging load variation in the power grid, finally impacting the operating state of the power grid.

The battery energy variation model is used to describe the changes in the EV's SoC throughout the trip. The SoC of vehicle $i \in \mathcal{I}^\mathrm{EV}$ at time $t$ is defined as the percentage of the EV's battery capacity that is currently available, as expressed in Equation \ref{eq:EVs-1}. {The variation process of SoC, as represented in Equation \ref{eq:EVs-2}, consists of the energy consumption process during driving (equation's second part) and the charging process at CSs (the third part).}

\begin{equation}
SOC_{i,t} := \frac{E_{i,t}^\mathrm{btr}}{E_i^\mathrm{cap}} \times 100\%,
\label{eq:EVs-1}
\end{equation}
\begin{equation}
SOC_{i,t} = SOC_{i,t_i^\text{o}} - \sum_{\tau_\mathrm{d}} \frac{{\Delta E_{\tau_\mathrm{d}, i}^\mathrm{csp}} }{E_i^\mathrm{cap}} \times 100\% + \sum_{\tau_\mathrm{c}} \frac{{\Delta E_{\tau_\mathrm{c}, i}^\mathrm{crg}} }{E_i^\mathrm{cap}} \times 100\%,
\label{eq:EVs-2}
\end{equation}
where $SOC_{i,t}$ is the SoC of vehicle $i$ ($i \in \mathcal{I}^\mathrm{EV}$) at time $t$; $SOC_{i,t_i^\text{o}}$ is the initial SoC of vehicle $i$ at the starting point at the departure time $t_i^\text{o}$; $E_{i,t}^\mathrm{btr}$ indicates the battery energy of vehicle $i$ at time $t$; ${E_i^\mathrm{cap}}$ is the battery capacity of vehicle $i$; $\Delta E_{\tau_\mathrm{d}, i}^\mathrm{csp}$ is the energy consumption of vehicle $i$ during the driving time slot $\tau_{\mathrm{d}}$ ($\tau_{\mathrm{d}} \subseteq [t_i^\text{o}, t]$), which is approximated via Equation \ref{eq:EVs-3}; $\Delta E_{\tau_\mathrm{c}, i}^\mathrm{crg}$ is the charging energy of vehicle $i$ during the charging time slot $\tau_{\mathrm{c}}$ 
 ($\tau_{\mathrm{c}} \subseteq [t_i^\text{o}, t]$), as expressed in Equation \ref{eq:EVs-4}. 
\begin{align}
\label{eq:EVs-3}
\Delta E_{\tau_\mathrm{d}, i}^\mathrm{csp} &\,=\, \rho \Delta d_{\tau_\mathrm{c}}, \\
\label{eq:EVs-4}
\Delta E_{\tau_\mathrm{c}, i}^\mathrm{crg} &\,=\, \eta p_{m(i),\tau_\mathrm{c}} \Delta \tau_\mathrm{c},
\end{align}
where $\Delta d_{\tau_\mathrm{c}}$ denotes the driving distance during time slot $\tau_\mathrm{c}$, which can be obtained from the DNL model in Equation \ref{eq:UTN-1}; $\rho$ is the power consumption per kilometer; in Equation \ref{eq:EVs-4}, $\Delta \tau_\mathrm{c}$ is the time duration in time slot $\tau_\mathrm{c}$; $\eta$ corresponds to the battery charging efficiency coefficient; $m(i)$ represents the target CS for vehicle $i$ when $i \in \mathcal{I}^\mathrm{EV}$; $p_{m(i),\tau_\mathrm{c}}$ denotes the charging power at the target CS $m(i)$ at time slot $\tau_\mathrm{c}$, the calculation process of which will be presented in the following models for CSs.

On the other hand, the charging time $tt_i^\text{c}$ can be derived via the calculation process as shown in Algorithm \ref{alg:calc_t_c}, given the expected SoC of the EV $SOC_{i}^\text{ep}$. This process is iterative until $SOC_{i}^\text{ep}$ is reached.
\begin{algorithm}
\caption{Calculation process of charging time ($tt_i^\text{c}$) for the EV}
\label{alg:calc_t_c}
\begin{algorithmic}[1]
\State Initialize $t \gets t_i^\text{c}$, $SOC_{i,t} \gets SOC_{i,t_i^\text{c}}$, and $tt_i^\text{c} \gets 0$
\While{$SOC_{i,t} < SOC_{i}^\text{ep}$}
    \State $tt_i^\text{c} \gets tt_i^\text{c} + \Delta \tau_\mathrm{c}$
    \State $t \gets t + \Delta \tau_\mathrm{c}$
    \State $\tau_\mathrm{c} \gets \tau_\mathrm{c} + 1$
    \State $SOC_{i,t} \gets SOC_{i,t} + \frac{\eta p_{m(i),\tau_\mathrm{c}} \Delta \tau_\mathrm{c}}{E_i^\mathrm{cap}}$
\EndWhile
\end{algorithmic}
\end{algorithm}

(3) PDN: Power flow analysis

As mentioned above, the charging demand of EVs at CSs brings extra charging loads to the power grid, which may jeopardize the stable and safe operation of the PDN, such as large voltage deviation. To investigate the impact of EVs charging behavior on the PDN, the power flow (PF) analysis, or power flow study, is adopted to determine the operating condition of the PDN. 

Given the distribution network configuration, generation, and load (including both EV charging loads and other types of loads), the PF analysis is performed by solving a set of balance equations of active and reactive power for each bus in the PDN. As a result, the voltage magnitude for each bus can be obtained for evaluation \citep{stevenson1994power}. Denote $\mathcal{U}^\mathrm{PDN}_\mathrm{cs}$, $\mathcal{U}^\mathrm{PDN}_\mathrm{ot}$ as the sets of buses connected with and without CSs, respectively. That is, $\mathcal{U}^\mathrm{PDN} = \mathcal{U}^\mathrm{PDN}_\mathrm{cs} \cup \mathcal{U}^\mathrm{PDN}_\mathrm{ot}$. For a bus $b_\mathrm{ot} \in \mathcal{U}^\mathrm{PDN}_\mathrm{ot}$, the active power at time $t$ refers to the basic load (e.g, residential electricity consumption), i.e., $P_{b_\mathrm{ot},t}^\mathrm{basic}$. {While regarding the load buses that are connected with CSs, in addition to the basic load, charging loads from EVs also contribute significantly and dynamically to the active power, which are subsequently considered in the PF equations.} For a bus $b_\mathrm{cs} \in \mathcal{U}^\mathrm{PDN}_\mathrm{cs}$, the active power at time $t$, $P_{b_\mathrm{cs},t}$, is calculated as the sum of the basic load $P_{b_\mathrm{cs},t}^\mathrm{basic}$ and charging load from EVs, as expressed in Equation \ref{eq:PDN-1}.
\begin{align}
\label{eq:PDN-1}
P_{b_\mathrm{cs},t} &\,=\, P_{b_\mathrm{cs},t}^\mathrm{basic} + p_{m(b_\mathrm{cs}),t} \sum_{i\in \mathcal{I}^\mathrm{EV}} Index(i),\\
\label{eq:PDN-2}
Index(i) &\,=\,
\begin{cases}
1, \text{if } t_i^\text{c} \leq t < t_i^{\text{c}\prime}\\
0, \text{others,}
\end{cases}
\end{align}
where ${m(b_\mathrm{cs})}$ is the index of CS connected to bus $b_\mathrm{cs}$; $p_{m(b_\mathrm{cs}),t}$ denotes the charging power at CS ${m(b_\mathrm{cs})}$ and at time $t$; $Index(i)$ is an indicator function that represents whether or not vehicle $i \in \mathcal{I}^\mathrm{EV}$ is charging at CS $m(b_\mathrm{cs})$ at time $t$, calculated by Equation \ref{eq:PDN-2}.

In the power system, each bus is associated with four quantities, i.e., active power $P$, reactive power $Q$, voltage magnitude $v$, and voltage phase $\theta$. According to the type of bus, only two variables are specified and the remaining two are obtained via the PF analysis \citep{low2014convex}. {For example, for load buses related to electricity loads (e.g., $b_\mathrm{cs} \in \mathcal{U}^\mathrm{PDN}_\mathrm{cs}$), only $P$ and $Q$ are specified while $v$ and $\theta$ are the quantities to be determined.} 
Both $P$ and $Q$ are determined by basic loads and EV charging loads. For EVs, we assume EVs only consume active power (i.e., the corresponding $Q$ value is considered to be 0) \citep{liu2017decentralized}. At a specific time $t$, the active power is derived in Equation \ref{eq:PDN-1} according to current basic loads and EV charging loads, based on which the PF analysis is calculated, as shown in Equations \ref{eq:PDN-3}. Consequently, varying spatiotemporal distributions of EV charging demands across different CSs generated by CS recommendation strategies (as inputs in the PF analysis) determine different operating states of the power grid (as outputs in the PF analysis, including the voltage magnitude). Typically, the Newton-Raphson method \citep{tinney1967power} can be used to solve the non-linear power flow equations.
\begin{equation} 
\begin{cases}
P_{b,t}=\sum_{b^{\prime}}v_{b,t}v_{b^{\prime}, t}[G_{b{b^{\prime}}}\cos(\theta_b-\theta_{b^{\prime}})+B_{b{b^{\prime}}}\sin(\theta_b-\theta_{b^{\prime}})]\\
Q_{b,t}=\sum_{b^{\prime}}v_{b,t}v_{b^{\prime}, t}[G_{b{b^{\prime}}}\sin(\theta_b-\theta_{b^{\prime}})+B_{b{b^{\prime}}}\cos(\theta_b-\theta_{b^{\prime}})],
\end{cases}
\label{eq:PDN-3}
\end{equation}
for each $b \in \mathcal{U}^\mathrm{PDN}$. In Equations \ref{eq:PDN-3}, $P_{b,t}$, $Q_{b,t}$ represent the active power and reactive power of bus $b$ at time $t$, respectively; $b^{\prime}$ belongs to the set of buses directly connected to bus $b$ via a transmission line; $v_{b,t}$, $v_{b^{\prime},t}$ are the voltage magnitude at bus $b$ and bus $b^{\prime}$ at time $t$; $\theta_b$, $\theta_{b^{\prime}}$ are the voltage angle at bus $b$ and bus $b^{\prime}$; $G_{b{b^{\prime}}}$, $B_{b{b^{\prime}}}$ represent the conductance and susceptance between bus $b$ and bus $b^{\prime}$.

(4) CSs: Queuing model and charge controller

The queuing model determines the waiting time for a vehicle to charge when there are not enough charging spots at CSs. It is assumed that CSs provide service on a First-In, First-Out (FIFO) basis. Then the waiting time ($tt_i^\text{w}$ for $i \in \mathcal{I}^\mathrm{EV}$) can be estimated based on the ordered queue and the charging duration of the vehicles that are being charged.

Besides, the voltage-based droop controller \citep{ireshika2021voltage} is employed at CSs for charge control, assuming a rectified linear relationship between the voltage magnitude and the charging power output. Note that we focus on the impact of the charge controller on the coupled systems, and thus more sophisticated non-linear controllers are beyond the scope of our investigation. Let $\mathcal{M}^\mathrm{cs}$ be the set of available CSs in the road network. Also, we define the time interval of the controller (e.g., 10 minutes) as the duration in which the charging power remains constant. The time period during the control interval is divided into $N_{t_\text{c}}$ time intervals for the controller, and the index of time interval is denoted as $t_\text{c}$. Besides, we denote $V_{\text{ref,1}}$ and $V_{\text{ref,2}}$ as the lower and upper reference values of voltage magnitude to determine the control range of voltage deviation, as shown in the right part of Figure \ref{fig:coupled_sys}. Specifically, the input voltage considered by the controller is the average voltage on all buses of the PDN, $\overline{v}_{t_\text{c}}$, based on the PF calculation results for the maximum load during the time interval $t_\text{c}$, as given in Equation \ref{eq:CSs-2}. For each $m \in \mathcal{M}^\mathrm{cs}$ and each time interval $t_\text{c}$, the relationship between the charging power $p_{m,t_\text{c}}$ and average nodal voltage $\overline{v}_{t_\text{c}}$ in the voltage-based charge controller can be expressed in Equation \ref{eq:CSs-1}. If $\overline{v}_{t_\text{c}}$ lies between $V_{\text{ref,1}}$ and $V_{\text{ref,2}}$, the charging power and voltage demonstrate a negative linear correlation between the minimum power $P_\text{min}$ and the maximum power $P_\text{max}$. Otherwise, the charging power is limited at the boundary values. As a result, the existence of the charging controller will increase the interdependence of the two systems and contribute to a more complex evolution of the system states, thereby increasing the difficulty in handling the CS recommendation problem.
\begin{align}
\overline{v}_{t_\text{c}} &\,=\, \frac{1}{\lvert \mathcal{U}^\mathrm{PDN} \rvert} \sum_{b \in \mathcal{U}^\mathrm{PDN}}v_{b,t_\text{c}},
\label{eq:CSs-2}\\
p_{m,t_\text{c}} &\,=\,
\begin{cases}
P_\text{min},\,\quad \qquad\qquad\qquad\quad \text{if }\overline{v}_{t_\text{c}} \leq V_{\text{ref,1}} \\
\alpha (\overline{v}_{t_\text{c}} - V_{\text{ref,1}}) + P_\text{min} ,\quad \text{if }V_{\text{ref,1}}<\overline{v}_{t_\text{c}}<V_{\text{ref,2}}\\
P_\text{max},\,\quad \qquad\qquad\qquad\quad \text{if }\overline{v}_{t_\text{c}} \geq V_{\text{ref,2}}, 
\end{cases}
\label{eq:CSs-1}
\end{align}
for $m \in \mathcal{M}^\mathrm{cs}$ and $t_\text{c} \in \{1, 2, ..., N_{t_\text{c}}\}$. Where $P_\text{max}$ denotes the maximum charging power at CSs; $P_\text{min}=\delta P_\text{max}$ is the minimum charging power at CSs, with $\delta$ as the minimum proportion of charging power; and $\alpha = \frac{P_\text{max}-P_\text{min}}{V_{\text{ref,2}}-V_{\text{ref,2}}}$ defines the control rate.

\subsection{CMDP model} \label{CMDP model}
Solving the proposed problem of charging station recommendation entails making sequential decisions regarding which charging station should be recommended in response to each charging request of EVs in the road network. We consider the aforementioned coupled systems as the environment and designate the EV coordination center as the intelligent agent that interacts with the environment to learn the optimal policy. At each decision step, the agent performs an action based on the perceived current environmental state, specifically selecting a charging station, and subsequently receives the instant reward and cost from the environment as feedback for action evaluation. Meanwhile, the environment transitions to the next state. In the long term, the goal of the agent is to maximize overall traffic efficiency while enhancing the security of the PDN operation related to voltage deviation.

{Given that conventional MDP models focus on optimizing objectives and struggle with constraints, CMDP, as an extension of MDP to accommodate constraints, is employed to model the constrained optimization problem for the charging station recommendation task. The CMDP \citep{altman2021constrained} model offers a mathematical framework for modeling the sequential decision-making process of a constrained agent, defined by the seven-element tuple,}
\begin{equation}
M=<\mathcal{S},\mathcal{A},\mathcal{P},r,c,b,\gamma>,
\label{eq:mdp-1}
\end{equation}
where $\mathcal{S}$ is a finite state space; $\mathcal{A}$ is the action space; $\mathcal{P}:\mathcal{S}\times \mathcal{A}\to \mathcal{S}$ is the state transition function, which is given by the coupled system; $r:\mathcal{S}\times \mathcal{A}\to \mathbb{R}$ is the reward function to guide the maximization of the agent's objective; $c:\mathcal{S}\times \mathcal{A}\to \mathbb{R}$ is the cost function, instructing the agent to identify feasible solutions; $b \in \mathbb{R}$ is the desired maximum constraint offset. And $\gamma \in (0,1)$ is the discount factor to balance the importance of immediate and long-term rewards and costs. 

(1) Action space and state space

The action of the agent indicates the CS choice for the specific EV at the current decision step. Denote the set of action steps or decision steps as $\mathcal{T}^\text{as}$. For a decision step $t \in \mathcal{T}^\text{as}$, let $i_t$ ($i_t \in \mathcal{I}^\mathrm{EV}$) be the specific EV that initiates a charging request at time step $t$, abbreviated as $i$ in the following. Then the action at the $t$-th step is represented as $a_t$ ($a_t \in \mathcal{A}$). For multiple charging requests concurrently initiated, a sequential processing approach is employed to make decisions one by one.

The environmental state $s_t \in \mathcal{S}$ in the coupled systems is comprised of relevant information from different components, including EV state $s_t^{ev}$, road network traffic state $s_t^{tf}$, and charging station state $s_t^{cs}$, that is,
\begin{equation}
s_t = \{s_t^{ev}, s_t^{tf}, s_t^{cs}\},
\label{eq:st-1}
\end{equation}

As expressed in Equation \ref{eq:st-2}, the EV's state consists of its current position ($O_i$), destination location ($D_i$), and current SoC ($SOC_{i,t}$), which are uploaded to the coordination center by the EV user when initiating a charging request. The traffic state contains the congestion level of each link ($\left\{k_{e, t} \right\}_{e \in \mathcal{E}}$) as shown in Equation \ref{eq:st-3}, defined by the traffic density, which can be estimated via traffic detectors, such as loop detectors and automatic vehicle identification technology \citep{larionov2017uhf}. At each CS $m \in \mathcal{M}^\mathrm{cs}$, the related state includes the numbers of EVs queuing and charging ($n_{m,t}^{q}$ and $n_{m,t}^{c}$ respectively), the SoC distributions of EVs in the queue ($\mu_{m,t}^{q}$ for mean value and $\delta_{m,t}^{q}$ for standard deviation) and charging ($\mu_{m,t}^{c}$ for mean value and $\delta_{m,t}^{c}$ for standard deviation), the distribution of waiting times already spent by queuing EVs ($\mu_{m,t}^{w}$ for mean value and $\delta_{m,t}^{w}$ for standard deviation), together with the number of EVs that have chosen this CS but are still en route ($n_{m,t}^{pt}$). The $s_t^{cs}$ can be defined via Equation \ref{eq:st-4}. These occupancy data and queue information within the CSs can be collected through various sensors, cameras, or mobile applications \citep{dastpak2024dynamic}. In this case, we utilize distribution parameters to describe the state of battery energy and waiting time, rather than specific individual values, to reduce the state dimensionality as well as enhance the generalization ability of the agent. 
\begin{align}
s_t^{ev} &\,=\, \{O_i, D_i, SOC_{i,t}\},
\label{eq:st-2}\\
s_t^{tf} &\,=\, \left\{k_{e, t} \right\}_{e \in \mathcal{E}},
\label{eq:st-3}\\
s_t^{cs} &\,=\, \{n_{m,t}^{q}, n_{m,t}^{c}, \mu_{m,t}^{q}, \delta_{m,t}^{q}, \mu_{m,t}^{c}, \delta_{m,t}^{c}, \mu_{m,t}^{w}, \delta_{m,t}^{w}, n_{m,t}^{pt}\}_{m \in \mathcal{M}},
\label{eq:st-4}
\end{align}

(2) Reward function

The reward function, $r(s_t,a_t)$, is utilized to evaluate the performance of the agent's behavior. Through immediate reward feedback, the agent can be guided toward the desired goal by improving its strategy. The objective for the CS recommendation problem is to maximize the overall traffic efficiency, which is represented by minimizing the total travel time of all vehicles involved in an episode $TTT$, as defined in Equation \ref{eq:TTT-1}.

When computing the value of $TTT$, Equation \ref{eq:TTT-1} accumulates the time duration from the vehicle dimension, whereas an equivalent calculation method is to sum up the number of all vehicles in the UTN from the time dimension, as depicted in Equation \ref{eq:reward-new-eq}. 
\begin{equation}
{TTT} = \sum_{i\in\mathcal{I}^{\mathrm{ALL}}}tt_{i}
\iff 
{TTT} = \sum_{p \in \mathcal{P}} n_{p} ,
\label{eq:reward-new-eq}
\end{equation}
where $\mathcal{P}$ is the set of time points involved in an episode, ranging from the departure time of the first vehicle at its origin point to the arrival time of the last vehicle at its destination. $n_p$ is the number of vehicles in the road network at time point $p$, which can be calculated via Equations \ref{eq:reward-new-eq-add1} and \ref{eq:reward-new-eq-add2}. The $Index_\text{T}$ in Equation \ref{eq:reward-new-eq-add2} is an indicator function of whether or not the vehicle $i^\prime$ is loaded in the road network at time point $p$.
\begin{align}
n_{p} &\,=\, \sum_{i^\prime \in \mathcal{I}^\mathrm{ALL}}Index_\text{T}(i^\prime, p)
\label{eq:reward-new-eq-add1},\\
Index_\text{T}(i^\prime, p) &\,=\,
\begin{cases}
1, \text{if } t_{i^\prime}^\text{o} \leq p < t_{i^\prime}^{\text{d}},\\
0, \text{others.}
\end{cases}
\label{eq:reward-new-eq-add2}
\end{align}

By further partitioning the $TTT$ expressed in Equation \ref{eq:reward-new-eq} according to decision steps, another alternative formulation of $TTT$ can be defined based on the number of vehicles associated with each decision step, as indicated in Equation \ref{eq:reward-new-eq2}.
\begin{equation}
\begin{aligned}
TTT &= \sum_{p \in \mathcal{P}} n_{p} \\
&= \sum_{t \in \mathcal{T}^\text{as}} \sum_{p_t \in \mathcal{P}_t} n_{p_t},
\label{eq:reward-new-eq2}
\end{aligned}
\end{equation}
where $\mathcal{P}_t$ is the set of time points involved in decision step $t$. Then we can obtain the number of vehicles associated with each decision step ($TTT_t$), depicted as Equation \ref{eq:reward-new-eq2-add1}.
\begin{equation}
TTT_t = \sum_{p_t \in \mathcal{P}_t} n_{p_t},
\label{eq:reward-new-eq2-add1}
\end{equation}

Based on the extracted $TTT_t$ ($t \in \mathcal{T}^\text{as}$) corresponding to each decision step, the reward function can be derived. Since a larger ${TTT}_{t}$ implies a smaller reward value, the sign of $TTT_t$ is changed to obtain the opposite value. However, if negative ${TTT}_{t}$ value is directly used as the reward function, it will cause an issue of imbalanced rewards, where ${TTT}_{t}$ value of the last action step is significantly greater than that of other steps. This is because, the last decision step encompasses the period from the initiation moment of the last charging request to the moment when all vehicles in $\mathcal{V}$ reach their destinations (e.g., 30 or 50 minutes), while the duration for non-final steps only covers a short interval between two consecutive charging request initiation moments (e.g., 6 seconds, 12 seconds). To avoid misleading the agent due to imbalanced reward representation, we need to reshape the reward function in a manner that guides the agent to learn the desired strategy and also prevent the problem of unbalanced rewards.

Consider that a well-suited reward function should reflect the sensitivity of the objective function to various strategies, implying that the designed reward function should capture the fluctuations in ${TTT}_{t}$ values in our context. Since ${TTT}_{t}$ is determined by the number of vehicles in the road network at each moment, a more congested traffic condition with a larger vehicle count corresponds to a greater ${TTT}_{t}$ value. For non-final steps, it can be considered that the traffic condition remains relatively stable within a short period. Hence, the average number of vehicles in the network can be used as the indicator of non-final reward. But for the last step, the average vehicle quantity fails to reflect the changes in the final ${TTT}_{t}$ value due to the extended and highly fluctuating time span. For instance, a small average vehicle count but with a larger interval duration could lead to a higher last ${TTT}_{t}$ value. To ensure the comparability of rewards associated with the final step between two episodes, we utilize the ${TTT}_{t}$ scaled by a fixed coefficient $w_2$ as the final reward. Furthermore, the parameter $w_2$ should be fine-tuned during the experiment to ensure that the reward for the final decision step does not differ significantly from the rewards of other steps, thus effectively guiding the learning of the agent. Finally, the derived reward function, $r(s_t, a_t)$, is illustrated as Equations \ref{eq:reward-6} and \ref{eq:reward-7}.
\begin{equation}
r(s_t, a_t):={w_1} (R_{max} - {R}_t),
\label{eq:reward-6}
\end{equation}
\begin{equation}
{R}_t=
\begin{cases}
\frac{1}{\lvert \mathcal{P}_t \rvert} {TTT}_t, \,\,\, \text{for non-final steps},\\
{w_2} {TTT}_t, \quad \text{for the final step}.
\end{cases}
\label{eq:reward-7}
\end{equation}
where $R_t$ is the designed indicator that can approximate the changes in ${TTT}_t$ values, defined via Equation \ref{eq:reward-7}; $R_{max}$ and $w_1$ are adjustable hyper-parameters used to scale the reward; $w_2$ ($0 < w_2 < 1$) is another tunable hyper-parameter for scaling the ${TTT}_t$ value in the last decision step; $\lvert \mathcal{P}_t \rvert$ is the number of time points in the set $\mathcal{P}_t$.

(3) Cost function

The cost function $c(s,a)$ is related to the safety constraint of the agent. In this context, the constraint of the proposed problem refers to the total deviation of operating voltage from the nominal voltage at the load buses of the power grid system, i.e., $CVV$, as expressed in Equation \ref{eq:CVV-1}. {Despite the fact that voltage deviation cannot be completely avoided due to the complex and fluctuating nature of the power grid \citep{ireshika2021voltage}, lower voltage offsets is beneficial to achieve better performance of the power system. In other words, the constraint involved in this problem is a soft constraint.} Therefore, the long-term constraint of the agent can be expressed as minimizing the $CVV$. 

Specifically, the $CVV$ is calculated based on the operating voltage magnitudes at each bus can be obtained from power flow analysis, given the dynamic power grid load including EVs charging load. Instead of carrying out computationally expensive real-time computations, the power flow analysis in each decision step focuses on the scenarios associated with the peak load as typical. Therefore, the cost for each decision step can be defined as the average voltage deviation corresponding to the time point of maximum load. Denote the cost as $c(s_t,a_t)$ and let $p_t^*$ ($p_t^* \in \mathcal{P}_t$) represent the peak load moment at the $t$-th step, then the cost function can be formulated as the following equation.
\begin{equation}
c(s_t,a_t) := \frac{1}{\lvert \mathcal{U}^\mathrm{PDN} \rvert} \sum_{b \in \mathcal{U}^\mathrm{PDN}} v_{b,p_t^*}^\text{dev},
\label{eq:cost-3}
\end{equation}
where $v_{b,p_t^*}^\text{dev}$ is the voltage violation at bus $b$ based on the charging load at time point $p_t^*$. Accordingly, the $CVV$ defined in Equation \ref{eq:CVV-1} can be explicitly reformulated based on the peak load moment in each action step, as depicted in Equation \ref{eq:CVV-final}.
\begin{equation}
CVV = \frac{1}{\lvert \mathcal{U}^\mathrm{PDN} \rvert} \sum_{t \in \mathcal{T}^\text{as}} \sum_{b \in \mathcal{U}^\mathrm{PDN}} v_{b,p_t^*}^\text{dev}.
\label{eq:CVV-final}
\end{equation}

(4) Policy and model formulation

The policy to choose an action is defined by the conditional probability $\pi(a\mid s)$, which maps a state $s$ ($s \in \mathcal{S}$) to an action $a$ ($a \in \mathcal{A}$). The state-value function in terms of reward or cost, i.e., $V^r_\pi(s)$, $V^c_\pi(s)$, can then be expressed as Equation \ref{eq:mdp-1-value}. In the CMDP context, the objective is to maximize the expected reward function $J_r(\pi)$ while ensuring that the expected cost function $J_c(\pi)$ remains below the constraint threshold, which leads to the generation of a decision policy. $J_r(\pi)$, $J_c(\pi)$ are defined as the expectation of state-value functions of reward and cost, respectively, as depicted in Equation \ref{eq:mdp-1-expt}. 
\begin{align}
V^{[\cdot]}_\pi(s) &\,=\, \mathbb{E}_\pi[\sum_{t=0}^\infty\gamma^{t}[\cdot]_{t+1})|s_0=s],
\label{eq:mdp-1-value}\\
J_{[\cdot]}(\pi) &\,=\, \mathbb{E}_{s}[V^{[\cdot]}_\pi(s)],
\label{eq:mdp-1-expt}
\end{align}
where $[\cdot]$ can represent $r$ or $c$, indicating values calculated based on reward or cost, respectively.

Let $\Pi$ denote the set of all stationary policies, and then the proposed problem can be mathematically formulated as follows:
\begin{equation}
\begin{aligned}
\text{\textbf{P1}: } \max_{\pi\in\Pi}\,&J_r(\pi)\\
\text{\ s.t.\ } &J_c(\pi) \leq b,
\end{aligned}
\label{eq:mdp-3}
\end{equation}
Here, $b$ ($b \geq 0$) denotes the constraint offset. Given that the grid safety constraint regarding $CVV$ is a soft constraint, a more favorable way for the agent to learn secure strategies is to set $b$ to 0. Despite the fact that loads always exist and result in voltage violations, making $J_c(\pi)=0$ hard to attain. The purpose of setting $b$ to 0 here is to encourage the agent to explore secure strategies.

\section{Online prediction-assisted SRL method} \label{sec:OP-SRL}

\begin{figure}[h!]
\center
\includegraphics[width=1\textwidth]{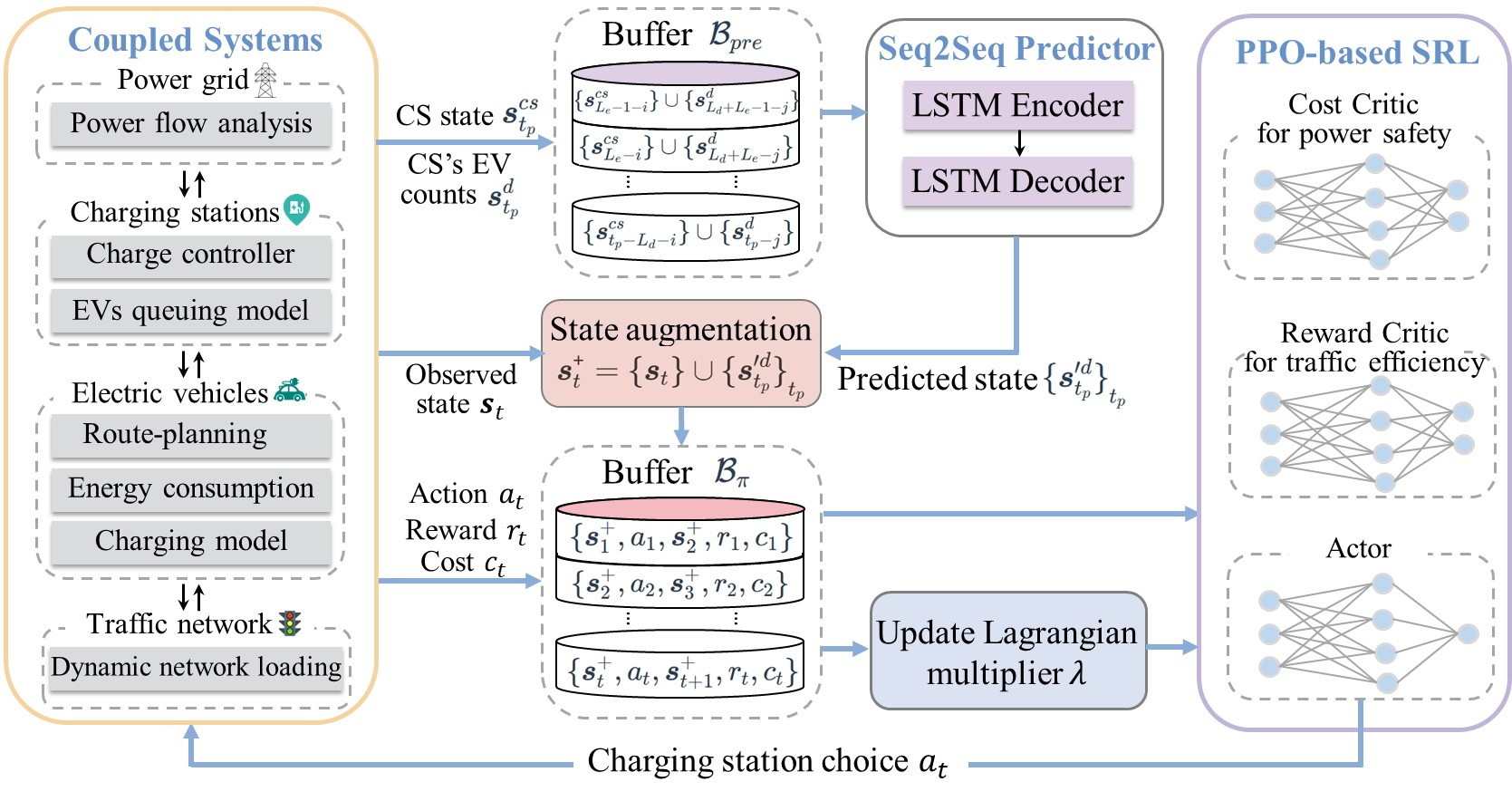}
\caption{A schematic overview of the proposed Online Prediction-Assisted SRL (OP-SRL) method.}
\label{fig:OP_SRL}
\end{figure}

In this section, we develop an online prediction-assisted safe reinforcement learning (OP-SRL) method to solve the proposed charging station recommendation problem, where two key challenges are considered and addressed. Firstly, {the issue confronted by us is how to enhance the policy safety considering the long-term constraint in the sequential decision-making process.} To address this problem, we implement the Lagrangian method to convert the constrained optimization problem into an equivalent unconstrained optimization problem, and then the proximal policy optimization (PPO) method is extended to incorporate constraint in the learning process through the inclusions of cost critic and Lagrangian multiplier. Secondly, we need to tackle the challenge of uncertain long-time delay between performing en-route charging station recommendation and commencing charging at CS, caused by the complicated and ever-changing system state under the interaction of the two networks, i.e., UTN and PDN. This delay may lead to misleading strategy results and cause instability in training. In this regard, we put forward an online sequence-to-sequence (Seq2Seq) predictor based on encoder-decoder architecture for state augmentation, thus offering insightful information to guide the agent in making more forward-thinking decisions. Finally, the Lagrangian-based PPO method with Seq2Seq predictor is trained to find the optimal policy; wherein the predictor and the policy are trained in a synchronized manner to achieve efficient and real-time control, by utilizing online predicted results. Figure \ref{fig:OP_SRL} depicts the schematic overview of the proposed OP-SRL method. The implementation details of each module and interaction process will be elaborated in this section.

\subsection{Lagrangian-based SRL for solving CMDP} 

Based on the Lagrangian method \citep{chow2018risk}, the constrained optimization model of CMDP in Equation \ref{eq:mdp-1} (\textbf{P1}) can be converted into an equivalent unconstrained form (\textbf{P2}), where the objective function combines a penalty term for constraint violation by introducing Lagrangian multiplier as the variable penalty coefficient.
\begin{equation}
\text{\textbf{P2}: }\max_{\theta}\min_{\lambda\geq0}\, J_r(\pi_\theta)-\lambda J_c(\pi_\theta),
\label{eq:lag-1}
\end{equation}
where $\theta$ denotes the parameters of policy $\pi$; $\lambda$ is the Lagrangian multiplier of the inequality constraint.

{To solve the maxmini problem, the two parameters involved in P2 are updated iteratively in an alternating manner. This involves a two-stage updating process:} 
\begin{itemize}
\item Stage 1: As shown in Equation \ref{eq:lag-add1}, $\lambda$ is updated using the gradient descent method with fixed policy, based on the collected trajectory data. The update rule is expressed as: $\lambda \gets \lambda + \alpha J_c(\pi_\theta)$. As the degree of constraint violation increases, the updating magnitude of $\lambda$ also increases, implying a greater penalty on the constraints, and vice versa.
\begin{equation}
\min_{\lambda\geq0}\,\mathcal{L}(\lambda)=J_r(\pi_\theta)-\lambda J_c(\pi_\theta).
\label{eq:lag-add1}
\end{equation}
\item Stage 2: As illustrated in Equation \ref{eq:lag-add2}, the Lagrangian multiplier is kept constant, and the parameters of policy $\pi$ are updated through the gradient ascent method. The objective of updating the policy is to optimize both cumulative rewards and constraint violation reduction simultaneously. This indicates that a policy with a high reward value but violates constraints may have a lower value in the objective function, making it sub-optimal, which is different from the situation of solely considering the objective of $J_r(\pi_\theta)$.
\begin{equation}
\max_{\theta}\,\mathcal{J}(\theta)=J_r(\pi_\theta)-\lambda J_c(\pi_\theta).
\label{eq:lag-add2}
\end{equation}
\end{itemize}

The RL-based methods can then be utilized to handle the second stage. We use the PPO method \citep{schulman2017proximal} to derive the optimal policy, considering its efficient and stable training performance. Generally, PPO consists of two neural networks as key components. One is an actor that functions as the agent, interacting with the environment by generating actions through the policy $\pi$. The other is a critic used to evaluate the performance of the actor based on rewards, which maps each state $s$ to a state value $V_{\pi}^{r}(s)$. To guide the RL agent accounting for both the objective represented by $V_{\pi}^{r}(s)$ and the constraint associated with $V_{\pi}^{c}(s)$, we need to extend the architecture of PPO and modify its objective based on the Lagrangian method. As illustrated in Figure \ref{fig:OP_SRL}, apart from using a reward-related critic for evaluation, another critic is introduced to assess the actor's performance by approximating $V_{\pi}^{c}(s)$. For the sake of differentiation, we refer to these two critics as the reward critic and the cost critic. 

In PPO, the update of stochastic policy is commonly to maximize the clipped surrogate objective function $J_\text{PPO}{ ( \pi_\theta ) }$, as expressed in Equation \ref{eq:ppo-1}. The key component is the estimator of advantage function $\hat{A}_{\pi}(s_t,a_t)$. The purpose of the advantage function is to measure the effectiveness of taking an action in a given state, which can be defined as the difference between the action value function and the state value function. The value of $\hat{A}_{\pi}(s_t,a_t)$ indicates the advantageous extent. Besides, the clip function is used to impose restrictions on policy updates, and the minimum method is utilized to obtain the lower bound of the objective to ensure safe updates.
\begin{equation}
\max J_\text{PPO}{ ( \pi_\theta ) }=\mathbb{E}[\min(\zeta(\theta)\hat{A}_{\pi}(s_t,a_t),\operatorname{clip}(\zeta(\theta),1-\varepsilon,1+\varepsilon)\hat{A}_{\pi}(s_t,a_t))],
\label{eq:ppo-1}
\end{equation}
where $\zeta(\theta)=\frac{\pi_\theta(a|s)}{\pi_{\theta_{old}}(a|s)}$ is the probability ratio derived from importance sampling; $\varepsilon > 0$ is a hyperparameter for clip fraction; $\operatorname{clip}(\cdot)$ represents a clipping operation that limits $\zeta(\theta)$ within the range of $1-\varepsilon$ and $1+\varepsilon$.

We use the generalized advantage estimator (GAE) \citep{schulman2015high} as $\hat{A}_{\pi}(s_t,a_t)$, which strikes a balance between bias and variance in estimating the action advantage. However, the original GAE only considers evaluating actions based on the state value of the reward. In our context, we need to evaluate actions taking into account both reward and cost, thus necessitating adjustments to the GAE function. Similar to the Lagrangian method provided in Equation \ref{eq:lag-add2}, the modified GAE, $\hat{A}_{\pi}(s_t,a_t)$, can be expressed as the weighted average of reward advantage and cost advantage using Equation \ref{eq:gae-1}. As such, the objective function of PPO is extended by incorporating both $V_{\pi}^r$ and $V_{\pi}^c$, with Lagrangian multiplier $\lambda$ as the penalty factor.
\begin{equation}
\hat{A}_{\pi}(s_t,a_t)=\hat{A}_t^r-\lambda \hat{A}_t^c,
\label{eq:gae-1}
\end{equation}
\begin{equation}
\hat{A}^{[\cdot]}_t=\sum_{k=0}^{T-t}(\gamma^{[\cdot]}\eta^{[\cdot]})^k\delta^{[\cdot]}_{t+k},
\label{eq:gae-2}
\end{equation}
where $\hat{A}_t^r$ and $\hat{A}_t^c$ denote the GAE functions based on reward and cost respectively, calculated as the exponentially weighted average over time steps $t$ to $T$ using Equation \ref{eq:gae-2}; In Equation \ref{eq:gae-2}, $[\cdot]$ serves as a placeholder for either $r$ or $c$; with ${V}_{\pi}^{[\cdot]}(s_t)$ as the state value function with relation to reward critic or cost critic, $\delta^{[\cdot]}_{t}=r_t+\gamma^{[\cdot]} {V}_{\pi}^{[\cdot]}(s_{t+1})-{V}_{\pi}^{[\cdot]}(s_t)$ represents the corresponding temporal-difference (TD) error, with $\gamma^{r}$ and $\gamma^{c}$ as the discount factors; $\eta^r$ is used to balance bias and variance in estimating the advantage on reward, and similarly, $\eta^c$ is the trade-off coefficient for cost-related advantage estimator.

By incorporating Equation \ref{eq:gae-1} into Equation \ref{eq:ppo-1}, the objective function of the Lagrangian-based PPO can be obtained, which serves as the surrogate objective of the second stage. Finally, the two-stage updating process for $\theta$ and $\lambda$ can be reformulated as Equations \ref{eq:ppo-2} (\textbf{P3}).
\begin{equation}
\text{\textbf{P3}: }
\begin{cases}
\min_{\lambda\geq0}\,\mathcal{L}(\lambda)=J_r(\pi_\theta)-\lambda J_c(\pi_\theta),\, \text{given }\theta,\\ 
\max_{\theta}\,J_\text{PPO}(\theta),\, \text{given }\lambda.
\end{cases}
\label{eq:ppo-2}
\end{equation}

The two critics are updated by minimizing the temporal-difference error between the target value and the predicted value, namely $\mathscr{L}_{cri}^{r}$ for reward critic and $\mathscr{L}_{cri}^{c}$ for cost critic, as expressed by Equations \ref{eq:critic-1}.
\begin{equation}
{\min}\, \mathscr{L}_{cri}^{[\cdot]}(\boldsymbol{W}_{cri}^{[\cdot]})=\|r_t+\gamma^{[\cdot]} {V}_{\pi}^{[\cdot]}(s_{t+1})-{V}_{\pi}^{[\cdot]}(s_t)\|_2^2,
\label{eq:critic-1}
\end{equation}
where $\boldsymbol{W}_{cri}^{r}$, $\boldsymbol{W}_{cri}^{c}$ denote the weight matrices of two critics.

\subsection{Seq2Seq predictor for state augmentation}

As mentioned earlier, the charging station recommendation strategy encounters a lag effect, where an uncertain delay exists between executing the en-route charging station recommendation and the charging commencement for the EV, due to the complex system evolution process under the interaction of the two systems. This delay may result in misleading strategy results and cause instability during the training process of the agent. To address this issue, a basic intuition is to mitigate the delay by estimating future system status as augmented states for the RL agent. This leads us to another challenge. In particular, given that the distribution of charging requests in the road network is characterized by randomness, and the travel time from EVs to different charging stations is also uncertain, it is inadequate for capturing effective future information with the estimate just from one time step.

To solve this challenge, we put forward a recurrent neural network (RNN)-based sequence-to-sequence (Seq2Seq) predictor that can forecast future state sequences using historical sequential information. The Seq2Seq model is built upon the encoder-decoder architecture \citep{keneshloo2019deep}. The encoder takes in a sequence of variable length as input and converts it into a fixed-length encoding state (i.e., context vector). Subsequently, the decoder utilizes the derived context vector to generate the target sequence, thereby enhancing the retention and utilization of the contextual information from the input sequence. Specifically, the Long Short-Term Memory (LSTM) \citep{hochreiter1997long} is employed as the encoder and decoder model due to its ability to capture long-term dependencies in sequential data. The charging station-related state defined in Section \ref{CMDP model} is set as the input, and the charging demand (i.e., the number of vehicles in the queue and vehicles being charged at each CS) averaged over a past time window is identified as the output of the Seq2Seq model for augmenting the agent's state. This is because, unlike the stochastic information contained in an individual time moment, the average charging demand can provide more consistent information, implicitly benefiting the guidance of the agent. For a time resolution $t_p$, the state related to CSs is represented by $\boldsymbol{s}_{t_p}^{cs}$. And the average charging demand $s_{m,t_p}^d$ at CS $m \in  \mathcal{M}^\mathrm{cs}$ can be obtained by applying Equation \ref{eq:s2s-add1}. Then the set of average charging demand in all CSs can be defined as $\boldsymbol{s}_{t_p}^d={\{s_{m,t_p}^d\}}_{m \in  \mathcal{M}^\mathrm{cs}}$.
\begin{equation}
s_{m,t_p}^d = \frac{1}{\lvert \mathcal{P}_{t_p}^\text{d} \rvert} \sum_{p \in \mathcal{P}_{t_p}^\text{d}} (n_{m,p}^q + n_{m,p}^c),\, \forall m \in \mathcal{M}^\mathrm{cs},
\label{eq:s2s-add1}
\end{equation}
where $\mathcal{P}_{t_p}^\text{d}$ is the collection of time points sampled within $t_p$; $\lvert \mathcal{P}_{t_p}^\text{d} \rvert$ denotes the sampling frequency within $t_p$; $n_{m,p}^q$ and $n_{m,p}^c$ denote the numbers of EVs queuing and charging at CS $m$ and time point $p$.

The calculation process of the LSTM-based Seq2Seq model is formulated via Equation \ref{eq:s2s-1} for the encoder and Equation \ref{eq:s2s-2} for the decoder. 
\begin{equation}
\textit{\textbf{h}}_{t_{p}}^e, \textit{\textbf{c}}_{t_{p}}^e = {LSTM}_{enc}(\boldsymbol{s}_{t_p}^{cs},\textit{\textbf{h}}_{t_{p}-1}^e,\textit{\textbf{c}}_{t_{p}-1}^e;\textit{\textbf{W}}_e)
,\, t_{p}=1,...,L_e
\label{eq:s2s-1}
\end{equation}
\begin{equation}
\left.\left
\{\begin{array}{l}
\textit{\textbf{o}}_{t_{p}}^d, \textit{\textbf{h}}_{t_{p}}^d, \textit{\textbf{c}}_{t_{p}}^d={LSTM}_{dec}(\boldsymbol{s}_{t_p}^d,\textit{\textbf{h}}_{t_{p}-1}^d,\textit{\textbf{c}}_{t_{p}-1}^d;\textit{\textbf{W}}_d)\\
 {\boldsymbol{s}_{t_{p}}^{\prime d}} = {Linear}(\textit{\textbf{o}}_{t_{p}}^d;\textit{\textbf{W}}_{l})\\
\end{array}
,\, t_{p}=1,...,L_d
\right.\right.
\label{eq:s2s-2}
\end{equation}
where mapping functions ${LSTM}_{enc}$, ${LSTM}_{dec}$ represent LSTM neural networks of encoder and decoder, respectively; $L_e$, $L_d$ denote the length of the input sequence in encoder and decoder; $\boldsymbol{s}_{t_p}^{cs}$, $\boldsymbol{s}_{t_p}^d$ are the inputs of encoder and decoder at time step $t_p$; $\textit{\textbf{h}}_{t_p}^e$, $\textit{\textbf{c}}_{t_p}^e$ denote the hidden state and cell state of the encoder, respectively; $\textit{\textbf{h}}_{t_p}^d$, $\textit{\textbf{c}}_{t_p}^d$ are defined similarly for the decoder; $\textit{\textbf{o}}_{t_p}^d$ is the output of the LSTM network in decoder; $\boldsymbol{s}_{t_p}^{\prime d}$ is the predicted charging demand at step $t_p$; $\textit{\textbf{W}}_{e}$, $\textit{\textbf{W}}_{d}$, and $\textit{\textbf{W}}_{l}$ are the weight matrices.

By utilizing the square error between predicted values ${\boldsymbol{s}_{t_{p}}^{\prime d}}$ and true values $\boldsymbol{s}_{t_{p}}^d$ as the loss function $\mathscr{L}_{p}$, the optimization objective of the predictor can be expressed as Equation \ref{eq:s2s-3}. Finally, the state of the proposed CMDP model at time step $t$ can be augmented as the concatenation of observed state and predicted results, i.e., $\boldsymbol{s}_t^{\texttt{+}}=\{\boldsymbol{s}_t\} \cup {\{\boldsymbol{s}_{t_{p}}^{\prime d}\}}_{t_{p}}$. Note that $t$ and $t_p$ represent the indexes of time steps for the agent and the predictor's time series, respectively, while referring to the same moment.

\begin{equation}
{\min}\, \mathscr{L}_{pre}(\boldsymbol{W}_{pre})=\|{\boldsymbol{s}}_{t_{p}}^d- {\boldsymbol{s}_{t_{p}}^{\prime d}}\|_2^2, 
\label{eq:s2s-3}
\end{equation}
where $\textit{\textbf{W}}_{pre}=\{\textit{\textbf{W}}_e,\textit{\textbf{W}}_d,\textit{\textbf{W}}_l\}$.

\subsection{Online training and execution for efficient control}
To learn the optimal charging station recommendation strategy, we employ Lagrangian-based PPO as the SRL method and additionally enhance the state using the inferred results of the Seq2Seq predictor. This leads us to the question of how to train the agent and the predictor. A straightforward approach is to train the predictor first and subsequently train the RL agent. However, since the policy keeps updating during the agent training process, the dataset utilized to train the predictor could be derived from strategies that differ significantly from the current policy of the agent, resulting in substantial errors in the predictor's inference results and misleading the learning process of the agent. Taking into account training efficiency and prediction accuracy, we propose an online approach for training and execution, where the predictor is trained during the agent's learning process and utilized online. As a result, the predictor can make use of the newly generated data for updates, thereby enhancing both training efficiency and stability.

The complete training process of OP-SRL is summarized in Algorithm \ref{alg:alg-1}. We first initialize the parameters of the SRL method and Seq2Seq predictor, along with two empty replay buffers for storing experience dataset, $\mathcal{B}_{\pi}$ for SRL and $\mathcal{B}_{pre}$ for training predictor. In each epoch for updating the policy, there could be multiple episodes of interaction between the agent and the environment. Depending on whether control actions are required, we divide the running process of an episode into two stages, i.e., the warm-up stage and the control stage. The warm-up stage aims to pre-collect an adequate amount of charging station information, which will serve as part of the inputs to the predictor in the control process. During the control process, the transition information with augmented states of each step $\{\boldsymbol{s}_{t}^{\texttt{+}}, a_t, \boldsymbol{s}_{t+1}^{\texttt{+}}, r_t, c_t\}$ is collected and stored in buffer $\mathcal{B}_{\pi}$, as shown in lines 9-15. Again, the time steps of the agent ($t$) and that of the predictor ($t_p$) are distinct, even though they correspond to the same moment. In addition, the time scales for updating the two buffers are also different. Buffer $\mathcal{B}_{\pi}$ is updated at a fast scale at each transition step, whereas buffer $\mathcal{B}_{pre}$ typically has a lower update frequency with longer sampling interval $w$, see lines 16-20. When updating parameters, the predictor updates its parameters every time it collects $n_2^{th}$ samples until convergence, as specified in lines 21 to 23. However, {the updating process of SRL-related parameters is implemented at each epoch upon collecting several episodes of transition information, sequentially updating the Lagrange multiplier, the actor network, and the critic network, as indicated in lines 28-31.}

\begin{algorithm}[h!]
\caption{Online Prediction-Assisted Safe Reinforcement Learning (OP-SRL) based on PPO}
\label{alg:alg-1}
\begin{algorithmic}[1]

\Require{Coupled systems of transportation network and power grid, traffic demand, and charging requests.}
\Ensure{Optimal charging station recommendation policy $\pi_\theta$.}
\State Initialize policy $\pi_{\theta}$, Lagrangian multipiler $\lambda$, parameter $\boldsymbol{W}_{cri}^{r}$ for reward critic, parameter $\boldsymbol{W}_{cri}^{c}$ for cost critic, parameter $\boldsymbol{W}_{pre}$ for Seq2Seq predictor, and two empty replay buffers $\mathcal{B}_{\pi} \leftarrow \emptyset$, $\mathcal{B}_{pre}\leftarrow\emptyset$.
\For{each epoch of policy learning $k=1,...,{K}$}
\For{each episode of policy learning $e=1,...,{E}$}
\State Reset the environment.
\Statex \qquad \quad {\textit{\textbf{Warm-up Process:}}}
\For{each time step $t_w=0,...,T_w-1$}
\State Obtain the charging station-related state $\boldsymbol{s}_{t_w}^{cs}$.
\EndFor
\Statex \qquad \quad {\textit{\textbf{Control Process:}}}
\For{each transition step $t=0,...,T-1$} \textcolor{blue}{\Comment{$t$ and $t_{p}$ refer to distinct scales of the same moment}}
\State Obtain the environmental state $\boldsymbol{s}_t=\{\boldsymbol{s}_t^{ev}, \boldsymbol{s}_t^{tf}, \boldsymbol{s}_t^{cs}\}$.
\State Predict ${\{\boldsymbol{s}_{t_{p}\texttt{+}j}^{\prime d}\}}_{j=1:L_d}$ based on $\{\boldsymbol{s}_{t_{p}\texttt{-}i}^{cs}\}_{i=0:L_e\texttt{-}1}$.
\State Obtain environmental augmented state $\boldsymbol{s}_t^{\texttt{+}}=\{\boldsymbol{s}_t\} \cup  {\{\boldsymbol{s}_{t_{p}\texttt{+}j}^{\prime d}\}}_{j=1:L_d}$.
\State Sample a control action from policy $a_t\sim {\pi_{\theta_{old}}}(a_t|\boldsymbol{s}_t^{\texttt{+}})$.
\State Similarly obtain the next-step augmented state $\boldsymbol{s}_{t+1}^{\texttt{+}}$.
\State Compute the reward $r_t$ and cost $c_t$ based on Equations \ref{eq:reward-6} and \ref{eq:cost-3}.
\Statex \qquad\qquad\, \textit{\textbf{Fast Scale:}}
\State Update buffer $\mathcal{B}_{\pi}\leftarrow\mathcal{B}_{\pi}\cup\{\boldsymbol{s}_{t}^{\texttt{+}}, a_t, \boldsymbol{s}_{t+1}^{\texttt{+}}, r_t, c_t\}$.
\Statex \qquad\qquad\, \textit{\textbf{Slow Scale:}}
\If{predictor does not converge and $t\bmod{w}=0$} \textcolor{blue}{\Comment{$w$: sampling interval for the predictor}}
\State Compute average charging demand $\boldsymbol{s}_{t_{p}}^{d}$ based on Equation \ref{eq:s2s-add1}.
\If{$t_{p} \geq L_d+L_e-1$}
\State Update buffer $\mathcal{B}_{pre}\leftarrow\mathcal{B}_{pre} \cup 
\{\boldsymbol{s}_{t_{p}\texttt{-}L_d\texttt{-}i}^{cs}\}_{0:L_e\texttt{-}1} \cup 
\{\boldsymbol{s}_{t_{p}\texttt{-}j}^{d}\}_{0:L_d\texttt{-}1}$.
\EndIf
\If{$\left| \mathcal{B}_{pre} \right| \geq n^{th}_1$ and $\left| \mathcal{B}_{pre} \right| \bmod{n^{th}_2} = 0$}  \textcolor{blue}{\Comment{$n^{th}_1$: minimum training sample size}}
\State Update $\boldsymbol{W}_{pre}$ w.r.t. the objective function \ref{eq:s2s-3}.
\textcolor{blue}{\Comment{$n^{th}_2$: sample size in a training interval}}
\EndIf
\EndIf
\EndFor
\State Estimate the advantage using Equation \ref{eq:gae-2}.
\EndFor
\State Update Lagrangian multiplier $\lambda$ w.r.t. the objective function \ref{eq:ppo-2} with fixed $\pi_{\theta}$.
\State Update actor network $\pi_{\theta}$ w.r.t. the objective function \ref{eq:ppo-2} based on constant $\lambda$.
\State Update reward-related critic network $\boldsymbol{W}_{cri}^{r}$ w.r.t. the objective function \ref{eq:critic-1}.
\State Update cost-related critic network $\boldsymbol{W}_{cri}^{c}$ w.r.t. the objective function \ref{eq:critic-1}.
\EndFor
\State \Return the charging station recommendation policy $\pi_\theta$.
\end{algorithmic}
\end{algorithm}

\section{Experimental studies}
\label{sec:cases}

\subsection{Experimental settings}

We conduct two case studies with extensive experiments to validate the effectiveness of the proposed OP-SRL method for CS recommendation problem, and examine the additional value of incorporating the Lagrangian multiplier and state augmentation. In case A, we conduct numerical studies on a synthetic transportation network, i.e, the Nguyen-Dupuis network \citep{nguyen1984efficient}, which has been widely used in experiments concerning traffic \citep[e.g.,][]{bao2021optimal,chen2023network}. We first conduct a comparative analysis against baselines to demonstrate the effectiveness of the proposed method. Then sensitivity analysis for variations of different parameters is carried out to show the robustness and flexibility of the proposed method. In case B, we perform numerical studies on a large-scale real-world road network, i.e., an area of the Kowloon region in Hong Kong, to demonstrate the applicability in the practical context, where historical traffic demand profiles from actual observations are utilized as inputs. The road networks in case A and case B are integrated with the IEEE 33-bus distribution system \citep{baran1989network} and the IEEE 69-bus system \citep{savier2007impact}, respectively, for the PDN modeling. Both IEEE 33-bus and 69-bus systems are well-known standard benchmarks in the power system and have been widely used for various distribution network analyses. As their names suggest, the two power systems refer to distribution networks with 33 and 69 nodes, respectively. A simulation platform is developed for the coupled transportation-power systems by integrating SUMO \citep{behrisch2011sumo} and Pandapower \citep{thurner2018Pandapower}, where SUMO is an open-source microscopic simulation tool and is used for dynamic network loading to model traffic flow propagation at the individual level, and Pandapower, a Python-based package, is employed for power flow calculation using Newton-Raphson method.

\begin{table}[h!]
\centering
\caption{\label{tab:two-networks} Experimental settings in case A and case B.}
\begin{tabular}{ccc} 
\toprule 
&  Case A & Case B \\ 
\midrule 
Road network & Nguyen's network & Specific area in Kowloon \\
Power grid & IEEE 33-bus system & IEEE 69-bus system \\
Area size & $\sim$3 km${}^2$ & $\sim$12 km${}^2$ \\
No. of controlled EVs & 300 & 356\\
Proportion of controlled EVs & 50\% & 8\%\\
No. of CSs & 5 & 12\\
No. of charging piles per CS & 60 & 30\\
\bottomrule 
\end{tabular}
\end{table}

The details of the experimental setup in two cases are listed in Table \ref{tab:two-networks}. In both cases, the duration of warm-up and control is separately set as 20 minutes and 1 hour, respectively. The traffic demand in case A is set as totally 600 vehicles per hour. The penetration of EVs to be controlled is set to 50\%, except in the sensitivity analysis of EV penetration. {Both EVs and non-EVs are generated on the road network at uniform departure intervals, with random origins and destinations.} On the other hand, case B is established with a practical road network setup and real-world observed traffic demand in a specific area of Kowloon region, Hong Kong. Case B has a total traffic demand of 4,442 vehicles per hour, retrieved from the TomTom Move platform\footnote{https://move.tomtom.com} based on real-world observations during the morning peak hours, i.e., 7 a.m. to 9 a.m., in October 2022. {Specifically, the Kowloon area is divided into 12 OD zones and, similar to case A, individual trips for both EVs and non-EVs among each OD pair are generated at identical intervals, with randomly selected starting links within the origin zone and random destinations within the destination zone.} According to a report by the Hong Kong Environmental Protection Department, the percentage of EVs in Hong Kong is around 8\% at the end of November 2023 \citep{hkepd2023}. Then the EV penetration in case B is set accordingly. 

Additionally, the number of charging stations in case B is set to 12, with the average distance to CSs of around 500 meters in Hong Kong as suggested in \citet{he2022spatial}. Considering that the commonly recognized planning goal or assumption for charging infrastructure is to establish a 1:1 ratio between EVs and charging piles, such as in Hong Kong \citep{hk1v2019}, the total number of charging piles for all CSs is determined based on this assumption. Assuming that the number of charging piles at each CS is the same. {Additionally, in the sensitivity analysis on on different EV penetration rates in Case A, the ratio of EV-to-charging ports ranges from 1:3 to 2:1.} We consider each EV with the same battery capacity $E_{\text{bat}}$ of 24 kWh and the charging efficiency $\eta$ of 0.9 \citep{ireshika2021voltage}. The power consumption per kilometer $\rho$ is designated as about 0.15 kWh/km \citep{behrisch2011sumo}. {For EVs with charging request due to low SoC, assume that their initial SoC when initiating charging requests follows a uniform distribution between 30\% and 60\%.} And EVs terminate charging when SoC achieves 80\% for battery health and longevity. The time interval for the charge controller is set to 10min, except in the related sensitivity analysis experiment; we consider $V_{\text{ref,1}}=0.9$ and  $V_{\text{ref,2}}=0.95$ as voltage reference values \citep{najera2019strategies}; besides, the maximum charging power $P_\text{max}=50$ kW, and the minimum proportion of charging power $c=30\%$.

In terms of the training process, we conduct 1,000 episodes of training for the proposed method and baselines, each with five times of different random seeds. The Adam optimization method is applied in all models. The implementation details of the method concerning the hyper-parameters are provided in \ref{appendix} (see Tables \ref{tab:parameters-SRL} and \ref{tab:parameters-predictor}). {The experimental studies were conducted on an Ubuntu server AMD Ryzen Threadripper PRO with 3975WX 32-Cores at 2.2 GHz using Python with the Pytorch package.}

\subsection{Baseline methods}
We benchmark the performance of our OP-SRL method {against eight baseline methods, including a greedy rule-based method, a model predictive control method,} and three categories of RL-based methods, given that the crucial components of the proposed method are constraint consideration, Lagrangian method, and predictor.
\begin{itemize}
\item {Shortest-path-based greedy method (Greedy): For each EV with charging request, this method calculates the distance-shortest path to each CS based on the Dijkstra algorithm and then the closest CS is picked as the designated CS.}
\item {Model predictive control (MPC): The model of coupled systems developed in Section \ref{Problem statement and model formulation} is used as the system model in MPC. Assuming that the MPC, in addition to the state information obtained by the SRL agent of the proposed method (see Equation \ref{eq:st-1}), has knowledge of the travel pattern of both EVs and non-EVs as well as SoC distribution for EVs with charging request. Upon receiving a new charging request from an EV (denoted as time step $k_{p}$), the MPC method forecasts all other possible charging requests that may occur during the control period, which is set to 2 min in the following experiments with about 11 EVs for charging guidance. Then the optimal CS allocation scheme for these charging requests is determined via brute-force search, and only the CS for the current EV is applied. After that, MPC proceeds to the next step and repeats the above process until all EVs and non-EVs involved have finished their trips.}

{The MPC method aims to optimize system performance over a future horizon (set to 5 min in the following experiments) in terms of both the traffic efficiency-related objective (i.e., $r_{k_p}$) and the voltage violation-related constraint (i.e., $c_{k_p}$). Similar to definitions of reward function and cost function in Equation \ref{eq:reward-6} and \ref{eq:cost-3} of the SRL agent, the $r_{k_p}$ and $c_{k_p}$ can be calculated using Equation \ref{eq:mpc-1} and Equation \ref{eq:mpc-2} respectively.}
\begin{align}
{r_{k_p}} &\,{=}\, {{w_1} (R_{max} - {w_3} TTT_{k_p}),}
\label{eq:mpc-1}\\
{c_{k_p}} &\,{=}\, {\frac{1}{\lvert \mathcal{U}^\mathrm{PDN} \rvert} \sum_{b \in \mathcal{U}^\mathrm{PDN}} v_{b,p_k^*}^\text{dev},}
\label{eq:mpc-2}
\end{align}
{where $TTT_{k_p}$ denotes the travel time segment in the $k_p$-th time step of the MPC, obtained from Equation \ref{eq:reward-new-eq2-add1}; $w_1$ and $w_3$ are the normalization coefficients of $r_{k_p}$; $p_k^*$ represents the peak power load moment in the $k_p$-th time step; $v_{b,p_k^*}^\text{dev}$ is the voltage violation of bus $b$ at time point $p_k^*$, as depicted in Equation \ref{eq:CVV-2}.}

{Considering both $r_{k_p}$ and $c_{k_p}$, the objective of MPC, $r^{\prime}_{k_p}$, can be formulated as the weighted average of $r_{k_p}$ and $c_{k_p}$ with coefficients of equal absolute value, as shown in the following Equation.}
\begin{equation}
{r^{\prime}_{k_p} = r_{k_p} - c_{k_p}.}
\label{eq:mpc-3}
\end{equation}

\item PPO with the Lagrangian multiplier (PPOlag): We contrast our proposed method with the configuration that excludes the predictor, aiming to assess the extra value brought by state augmentation.
\item Four RL methods without constraint consideration, including DQN \citep{mnih2013playing}, REINFORCE \citep{williams1992simple}, Actor-Critic \citep{sutton2018reinforcement}, and PPO \citep{schulman2017proximal}: 

This set of comparative experiments aims to evaluate the impact of constraint consideration in the policy. Specifically, these RL methods used for comparison solely focus on the objective-related reward (as defined in Equation \ref{eq:reward-6}), without considering the constraint or cost. For the sake of comprehensive coverage, the baseline methods we have selected incorporate both value-based methods (DQN) and policy gradient methods (REINFORCE, Actor-Critic, and PPO).
\item PPO with fixed penalty factor (PPOpenalty): To demonstrate the effectiveness of introducing the Lagrangian method in the proposed method, the PPO method with a fixed penalty coefficient for constraint violation is employed as another baseline. As shown in Equation \ref{eq:baseline-penalty}, PPOpenalty accounts for the cost when designing the reward function $r^{\prime}(s_t,a_t)$, without additional consideration of the cost function. The integrated reward function $r^{\prime}(s_t,a_t)$ is calculated as the weighted sum of reward $r$ and cost $c$ with coefficients of equal absolute value, after normalizing $r$ and $c$. The negative sign is used to account for the adverse effect of the cost.
\begin{equation}
r^{\prime}(s_t,a_t) = r(s_t,a_t)-c(s_t,a_t),
\label{eq:baseline-penalty}
\end{equation}
where $r(s_t,a_t)$ and $c(s_t,a_t)$ are the reward function and cost function in the proposed OP-SRL, calculated via Equations \ref{eq:reward-6} and \ref{eq:cost-3} respectively.
\item PPO with the Lagrangian multiplier (PPOlag): We contrast our proposed method with the configuration that excludes the predictor, aiming to assess the extra value brought by state augmentation.
\end{itemize}

\subsection{Evaluation metrics}
We devise three performance metrics to evaluate these methods, considering road network efficiency, power grid safety, and EV user satisfaction, as outlined below:
\begin{itemize}
\item \textbf{Total travel time (TTT)}\\
The objective of the proposed problem is to minimize the total travel time, which is also regarded as a representation of traffic network efficiency. The calculation of $TTT$ is depicted in Equation \ref{eq:TTT-1}.
\item \textbf{Cumulative voltage violation (CVV)}\\
We use the widely adopted voltage violation as the assessment indicator for power grid safety \citep{marouani2009application}. Specifically, we consider the nodal voltage deviation accumulated over an episode, as calculated in Equation \ref{eq:CVV-final}.
\item \textbf{Waiting time and charging time per EV (WCT)}\\
In order to evaluate the satisfaction of EV users, the average time consumed by EVs at CSs (including waiting time and charging time) is adopted as the evaluation metric, as expressed in Equation \ref{eq:metric-WCT}.
\begin{equation}
WCT = \frac{1}{\lvert \mathcal{I}^\mathrm{EV} \rvert} \sum_{i \in \mathcal{I}^\mathrm{EV}} (tt_i^\text{w} + tt_i^\text{c}).
\label{eq:metric-WCT}
\end{equation}
\item {\textbf{Execution time (ET)}}\\
{For all RL-based methods, the total execution duration to train the agent is recorded for comparison.}
\item {\textbf{Decision-making time (DT)}}\\
{The decision-making time of each method is also logged for comparison.}
\end{itemize}

\subsection{Case A: Nguyen-Dupuis network}
\begin{figure}[h!]
\center
\includegraphics[width=0.6\textwidth]{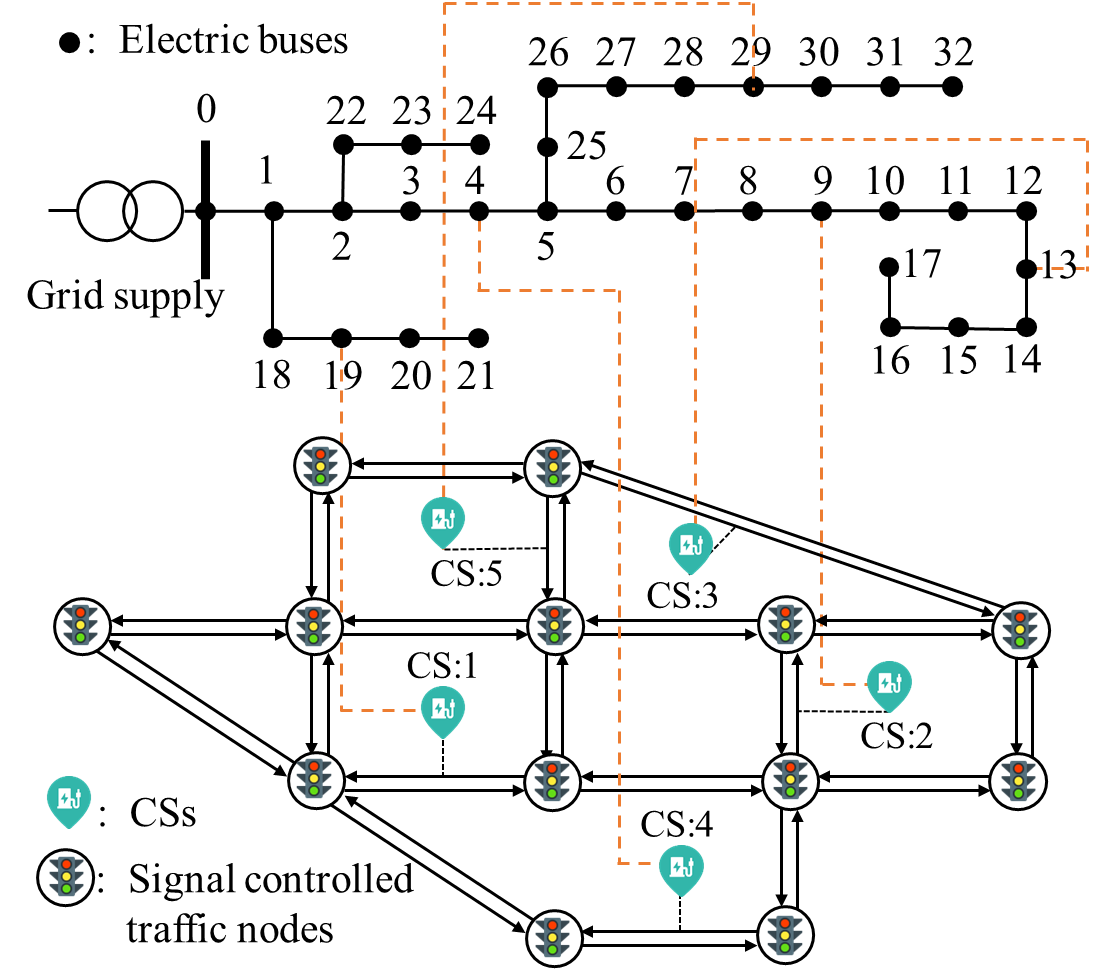}
\caption{Schematic diagram of the integration between the Nguyen-Dupuis transportation network and the IEEE 33-bus distribution system.}
\label{fig:caseA-coupled-sys}
\end{figure}

In this section, we carry out the experimental studies on the Nguyen-Dupuis transportation network integrated with the IEEE 33-bus distribution system, as illustrated in Figure \ref{fig:caseA-coupled-sys}. First, we benchmark the proposed method against baseline methods. Second, we implement sensitivity analysis for the variation of several parameters, including EV penetration, time interval of charge controller, and decoder length in the predictor.

\subsubsection{Comparative analysis}
To demonstrate the effectiveness of the proposed method, we compare the performance of the proposed OP-SRL method with eight baselines. {These baselines are either greedy-rule based method (Greedy), or model-based MPC (MPC),} or regular RL methods without considering constraint (DQN, REINFORCE, Actor-Critic, and PPO), or the constraint-aware RL-based method with fixed penalty coefficient (PPOpenalty), or the version resulting from ablating the predictor in the proposed method (PPOlag).

The detailed experimental results are shown in Table \ref{tab:1}. {The table indicates that, concerning the three system performance metrics (i.e., TTT, CVV, and WCT), the proposed method outperforms the baselines. The greedy algorithm performs worst in terms of CVV, although it outperforms REINFORCE and PPO in TTT and surpasses REINFORCE in WCT. This is because the greedy algorithm focuses solely on the shortest path distance when selecting target CSs, while neglecting the impact on the power grid, resulting in poor performance in CVV. In this case, the performance of MPC is similar to that of DQN, but inferior to PPOlag and the proposed method. Across the three metrics, the proposed method shows improvements of 7.1\%, 5.7\%, and 11.8\% over MPC. This could be attributed to MPC's lack of long-term consideration, despite its perfect model for system state prediction.}

Among RL-based methods, the proposed method performs the best and demonstrates the best stability with the smallest standard deviation, followed by PPOlag, with REINFORCE performing the worst. In PPO-based methods, the vanilla PPO performs the worst, while incorporating a weighted penalty term in the reward helps enhance performance (PPOpenalty), i.e., a reduction of 1.5\%, 0.8\%, 1.9\% in TTT, CVV, and WCT. Furthermore, replacing the fixed penalty factor with the Lagrangian method (PPOlag) leads to improvements of 10.1\%, 12.6\%, and 16.3\% in TTT, CVV, and WCT, respectively. By further incorporating state augmentation, the proposed OP-SRL shows the best performance, which decreases TTT, CVV, and WCT by around 3.5\%, 5.1\%, and 6.5\%, respectively, compared to PPOlag (also the best baseline). 

\begin{table}[h!]
  \centering
  \caption{Comparative analysis against baselines for Nguyen-Dupuis network}
  \label{tab:1}
\begin{threeparttable}
  \begin{tabular}{@{} cccccc @{}}
    \toprule
    Method& TTT ($\times$1e+4 sec) & CVV (pu/bus) & WCT (min/EV) & {ET (d)} & {DT (sec)}\\
    \midrule
    {Greedy} & {61.48} & {24.20} & {22.63} & {-} & {0.144}\\
    {MPC} & {57.39} & {19.59} & {18.71} & {-} & {1.198}\\
    
  DQN  & 57.89 $\pm$ 2.35 & 20.55 $\pm$ 0.75 & 19.06 $\pm$ 1.28 & {0.53} & {0.012} \\
   REINFORCE & 67.01 $\pm$ 1.03 & 24.06 $\pm$ 0.25 & 24.15 $\pm$ 0.62 & {0.51} & {0.010} \\
   Actor-Critic & 61.00 $\pm$ 4.04 & 21.42 $\pm$ 1.81 & 20.90 $\pm$ 2.10 & {0.51} & {0.010} \\
   PPO & 62.36 $\pm$ 4.37 & 22.47 $\pm$ 1.74 & 21.60 $\pm$ 2.38 & {0.53} & {0.011}\\
   PPOpenalty & 61.40 $\pm$ 2.32 & 22.29 $\pm$ 1.11 & 21.18 $\pm$ 1.15 & {0.52} & {0.011}\\
   PPOlag & 55.20 $\pm$ 0.84 & 19.48 $\pm$ 0.54 & 17.64 $\pm$ 0.49 & {0.52} & {0.011}\\
   \textbf{OP-SRL} & \textbf{53.29} $\pm$ 0.73 & \textbf{18.48} $\pm$ 0.26 & \textbf{16.50} $\pm$ 0.29 & {1.26} & {0.012}\\
    \bottomrule
  \end{tabular}
\begin{tablenotes}
\footnotesize
\item[]The results show mean $\pm$ std for RL-based methods of five runs with different random seeds. All the evaluation metrics favor lower values and and the best results are highlighted in bold.
\end{tablenotes}
\end{threeparttable}
\end{table}

{Besides, the rankings of all methods remain consistent across the three metrics, except for the locally optimal greedy algorithm that only considers a single objective.} This is because different stakeholders' benefits are interconnected in the context of the coupled system. Given the queuing and charging time of EVs constitute a significant portion of trip time (about 56\% to 65\% in Table \ref{tab:1}), the overall traffic efficiency demonstrates an upward trend when queuing and charging time decreases. Furthermore, EVs' queuing time and charging duration largely depend on charging power. When the generated strategies fail to sufficiently consider the effects on the power grid and mislead EVs, it can cause imbalanced power load and thus voltage deviations. In such cases, the voltage-responsive controller will reduce power to prevent further voltage drop and mitigate adverse impacts on the power grid (see the following Figure \ref{fig:baseline-power} on time-varying charging power curves for detailed illustration). Hence, the queuing and charging time of EVs are intricately linked to the extent of voltage deviation in this context. It should be highlighted that this is also why certain constraint-ignorant methods (i.e., DQN and Actor-Critic) still contribute to the improvement of the power grid safety index compared to PPOpenalty with a simple penalty term.

\begin{figure}[h!]
  \centering
  \begin{minipage}[t]{0.355\linewidth}
  \begin{tikzpicture}
    \node[anchor=south west, inner sep=0] (image) at (0,0) {\includegraphics[width=1\textwidth]{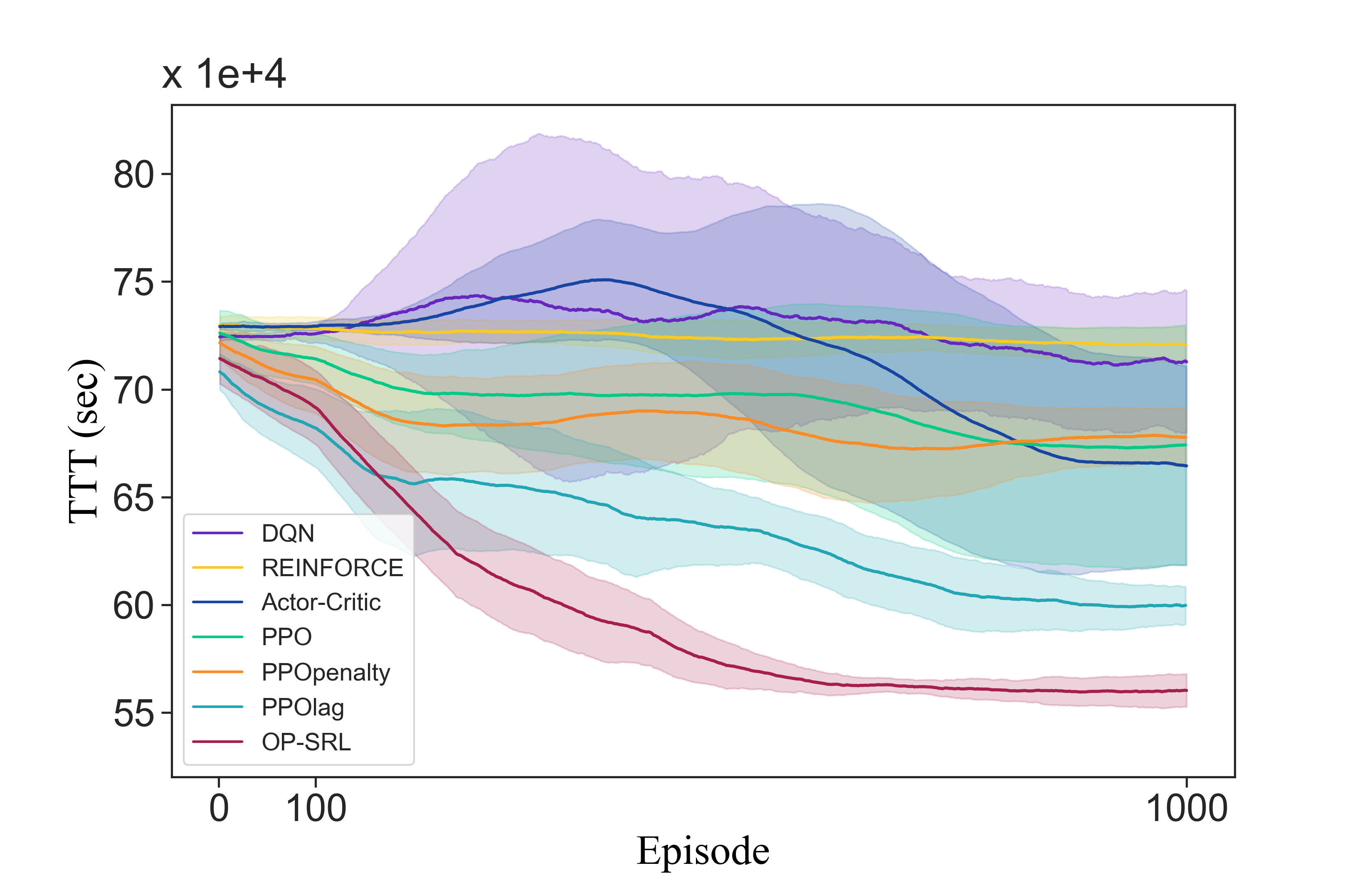}};  
    \begin{scope}[x={(image.south east)},y={(image.north west)}]
      \draw[dashed, black, line width=0.6pt] (0.2317, 0.528) -- (0.2317, 0.11);
    \end{scope}
  \end{tikzpicture}
  \caption*{(a) Traffic efficiency}
  \end{minipage}\hfill\hspace{-20pt}
  \begin{minipage}[t]{0.355\linewidth}
  \begin{tikzpicture}
    \node[anchor=south west, inner sep=0] (image) at (0,0) {\includegraphics[width=1\textwidth]{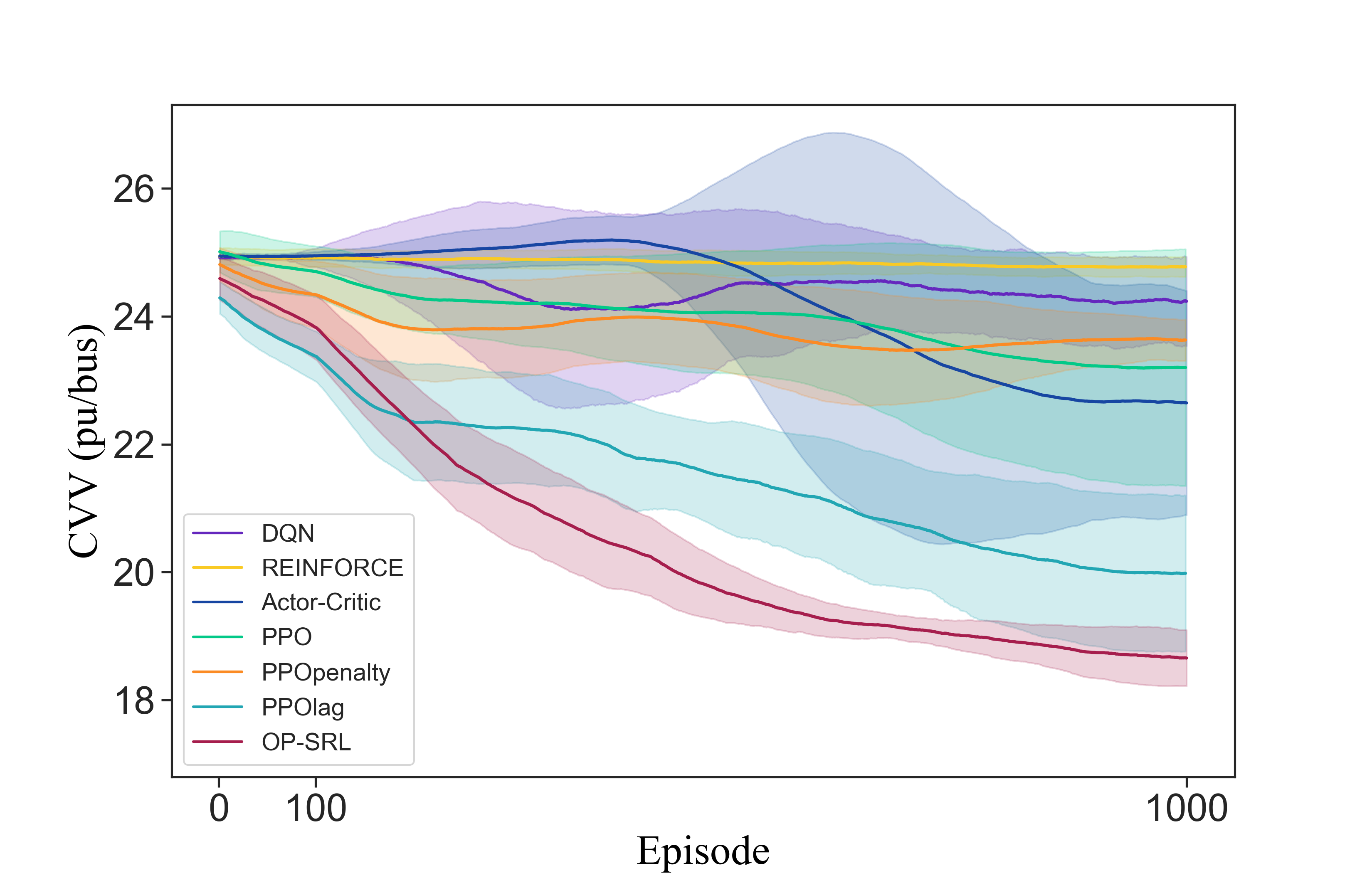}};  
    \begin{scope}[x={(image.south east)},y={(image.north west)}]
      \draw[dashed, black, line width=0.6pt] (0.2317, 0.626) -- (0.2317, 0.11);
    \end{scope}
  \end{tikzpicture}
  \caption*{(b) Grid safety}
  \end{minipage}\hfill\hspace{-20pt}
  \begin{minipage}[t]{0.355\linewidth}
  \includegraphics[width=1\textwidth]{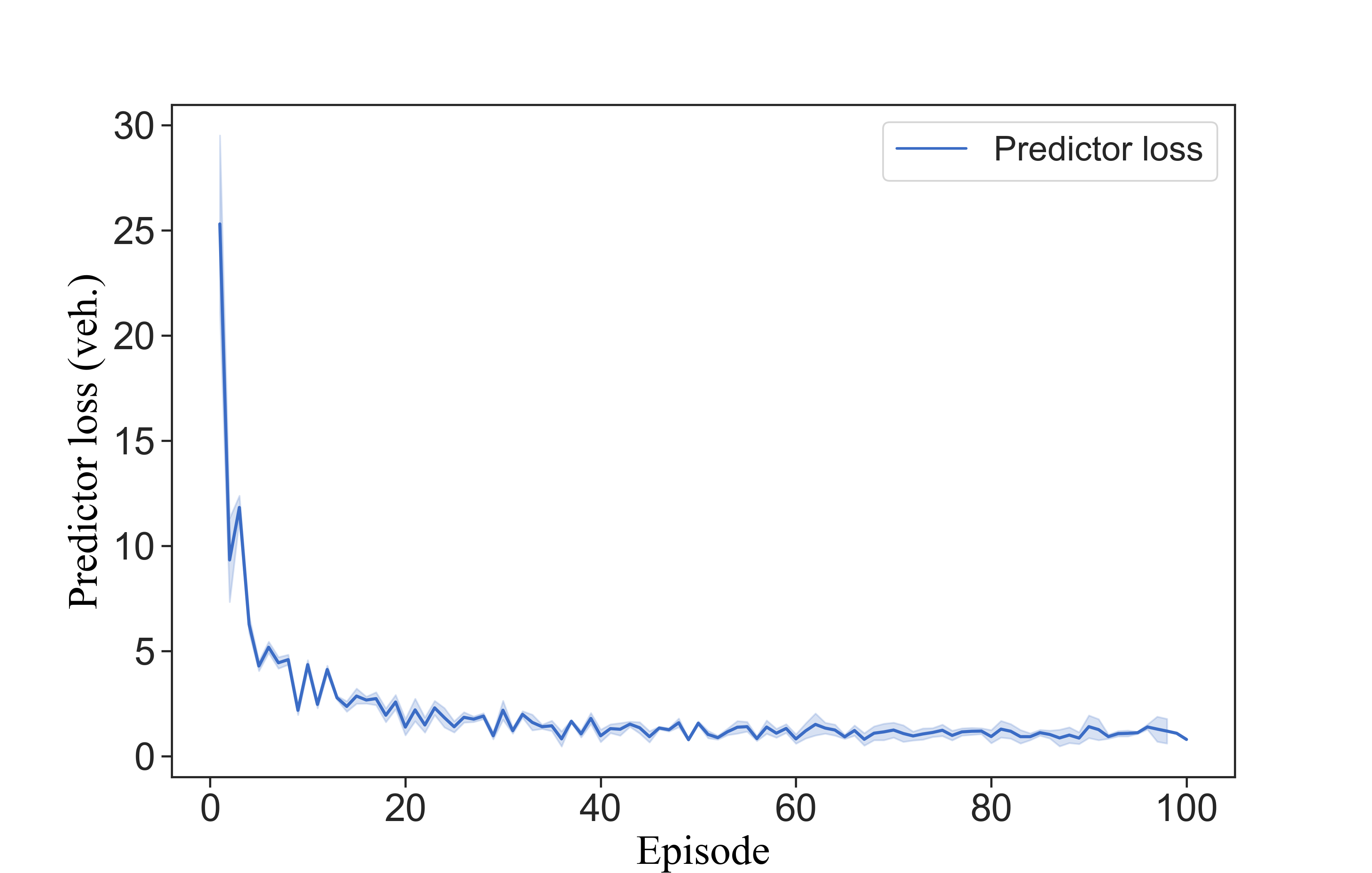}
  \caption*{(c) Seq2Seq predictor's loss}
  \end{minipage}
  \caption{Training curves of the proposed method against six RL-based baselines in terms of TTT (i.e., objective value) in (a) and CVV (i.e., constraint value) in (b), along with training curve of Seq2Seq predictor over RL episodes in (c). All the curves are plotted using five training runs with different random seeds. Solid lines are the mean and the shaded areas stand for the standard deviation region. Besides, Black dashed lines in (a) and (b) indicate the convergence boundary of the predictor.}
\label{fig:baseline-train-curve}
\end{figure}

The training process of the proposed OP-SRL method against six RL-based baselines is depicted in Figure \ref{fig:baseline-train-curve}. {The first and second sub-graphs represent the training curves for traffic efficiency and grid safety, respectively, which correspond to the objective and constraint guiding the agent's learning process. Within the given 1000 episodes, the training curves of both metrics gradually stabilize for all RL-related methods, leading to improved strategies and convergence of agents.} The figure indicates that the performances of TTT and CVV during training are generally consistent for each method. Just as mentioned earlier, in the coupled systems of UTN and PDN, the interests of different entities are closely intertwined via charging stations and EVs. Among all the methods, the training process of unconstrained ones (i.e., DQN, REINFORCE, Actor-Critic, and PPO) tends to exhibit more significant oscillations. This is because these methods solely focus on the influence of strategy on the road network and fail to take into account the other part of the coupled system, i.e., the power grid. Comparatively, methods that incorporate constraint (PPOpenalty, PPOlag, and OP-SRL) demonstrate a more stable training process in both metrics.

Furthermore, Figure \ref{fig:baseline-train-curve} demonstrates the enhancement of effectiveness and efficiency via the introduction of the Lagrangian method and predictor. From the figure, we can observe that the proposed OP-SRL shows notable improvements over PPO and PPOlag in terms of both objective and constraint. However, due to the long-term action delay, PPOlag exhibits slower convergence speed and demonstrates drastic oscillations in the performance of constraint. In contrast to PPOlag, the proposed OP-SRL achieves faster convergence to better values of both traffic efficiency and grid safety, with more stable training curves. Regarding the added value of the predictor, we can evaluate it by comparing the training curves of the proposed method before and after the predictor reaches convergence. Figure \ref{fig:baseline-train-curve}c indicates that the predictor converges around episode 100. Meanwhile, we can observe clear turning points of the proposed method in Figure \ref{fig:baseline-train-curve}a and \ref{fig:baseline-train-curve}b at about episode 100. Before reaching this turning point, the proposed method is inferior to PPOlag; but after that, the method exhibits faster convergence speed and rapidly surpasses PPOlag regarding both metrics. This is because, at the beginning of the training process, the predictor has significant errors and cannot provide accurate estimations of future states to assist the agent's training. Instead, it interferes with the agent's decision-making by providing incorrect information, resulting in OP-SRL initially performing worse than PPOlag. With an increasing number of training iterations, the predictor eventually converges around episode 100. After convergence, the predictor can offer more accurate predicted information to support the agent in making more insightful decisions, leading to the final superiority of OP-SRL over PPOlag.

{Regarding the execution time (ET) of RL-based baselines in Table \ref{tab:1}, methods that do not incorporate the Seq2Seq predictor have comparable execution times, approximately half a day, whether or not the Lagrangian method is utilized. With additional inclusion of the predictor, the proposed method's execution time is almost twice that of the other methods. Yet, when it comes to decision-making time (DT), all RL-based methods are approximately at 0.011 seconds, representing only 7.6\% and 0.9\% of that of greedy methods and MPC. This suggests that the inclusion of the predictor has not noticeably extend the decision time. Moreover, the computational expense required by RL-based methods after convergence is significantly lower than that of greedy method and MPC, enabling RL-based methods to make real-time decisions in complex environments with high-dimensional observation spaces.}

\begin{figure}[h!]
\center
\includegraphics[width=0.6\textwidth]{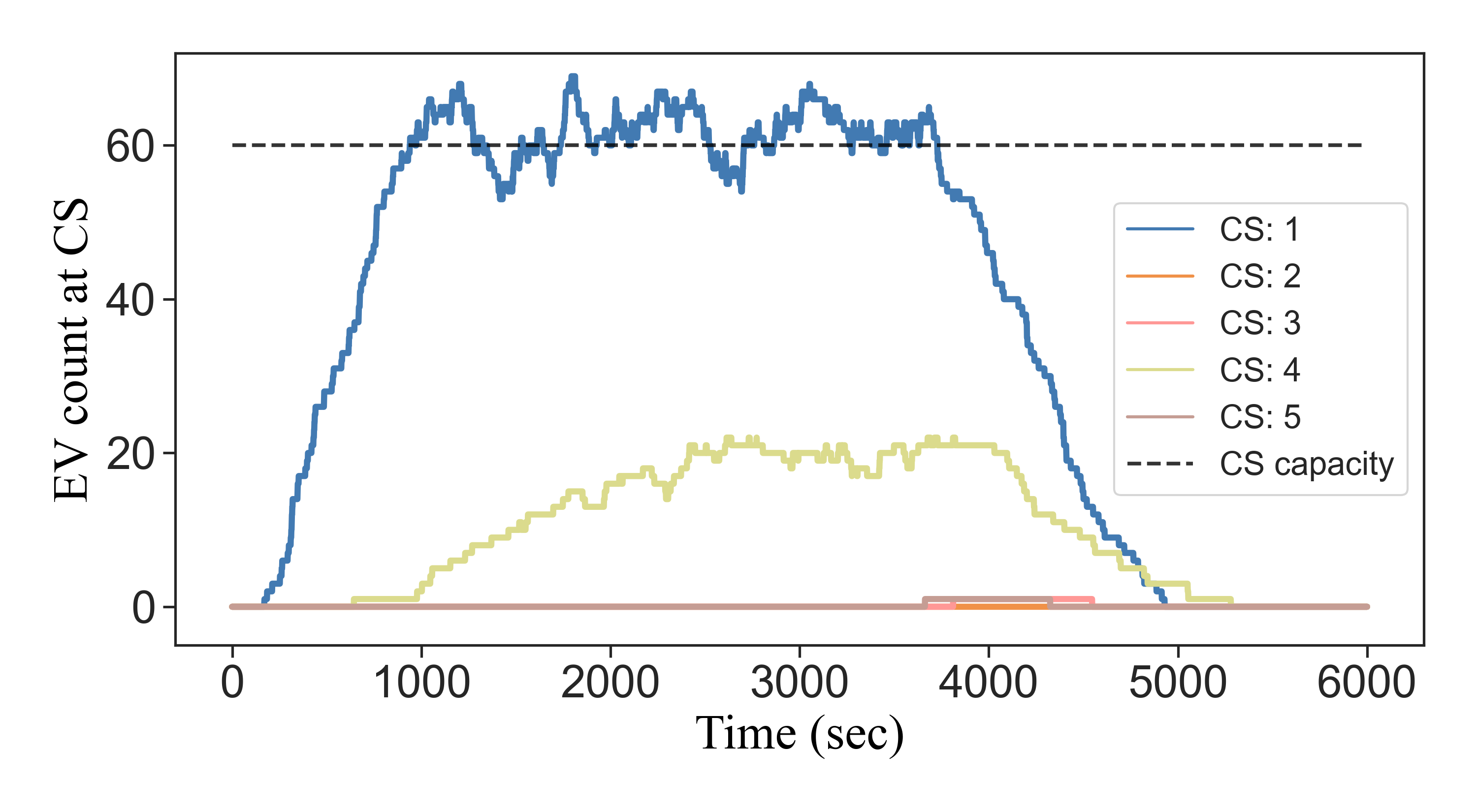}
\caption{Time-varying vehicle count at each charging station (CS) as a result of the proposed method. The vehicle count here comprises both queued vehicles and charging vehicles. The black dashed line represents the service capacity of each charging station, i.e., the number of charging piles.}
\label{fig:baseline-zzcls}
\end{figure}

The best strategy obtained in the proposed method is illustrated in Figure \ref{fig:baseline-zzcls}, where the time-varying vehicle counts at each charging station (marked as CS:1 to CS:5) represent the results of the charging station recommendation strategy. It can be seen from the figure that only CS:1 and CS:4 are effective charging stations, while the remaining ones basically do not provide any service and can be identified as redundant charging service facilities. Charging station 1 accounts for nearly 75\% of the charging demand, while the remaining 25\% is primarily handled by charging station 4. In the obtained strategy, despite the utilization of only two charging stations, and there are occasions when the charging demand of CS:1 exceeds its service capacity (crossing the dashed line), the performance in terms of coupled system performance and user satisfaction is superior to the results obtained from the baselines. The underlying reasons can be derived from Figure \ref{fig:baseline-power}, which displays time-varying curves of charging power using the proposed method against the baselines. The charging station recommendation strategy obtained from the proposed method allows for a reduction in voltage deviation, leading to a generally higher charging power output from the charging controller compared to other methods. This, in turn, contributes to a decrease in both vehicle waiting time and charging duration at charging stations, ultimately enhancing traffic efficiency. In this way, the proposed method can also identify and rank the effective positions for charging station deployment, to avoid extra construction costs and redundant service facilities.

\begin{figure}[h!]
\center
\includegraphics[width=0.6\textwidth]{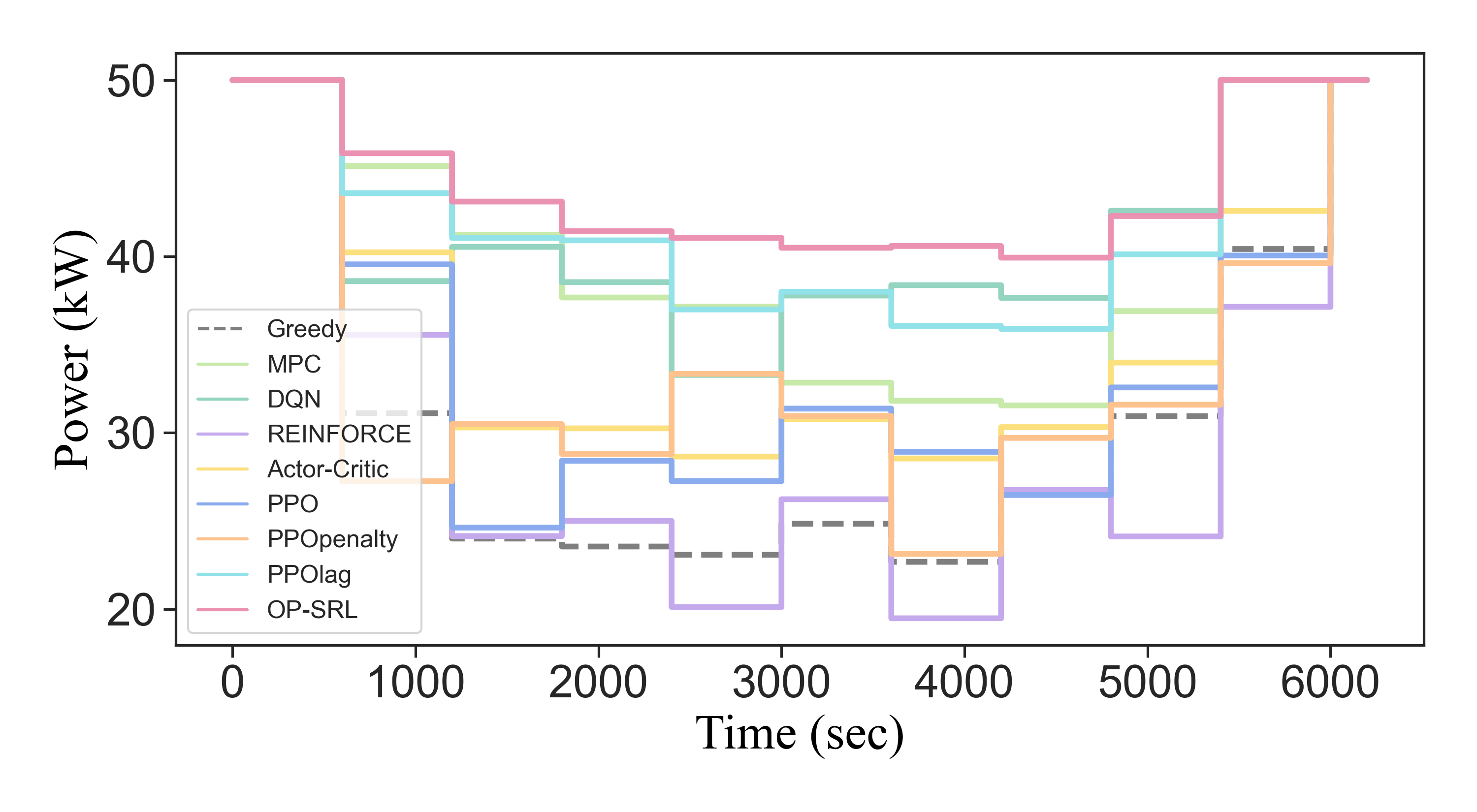}
\caption{{Time-varying charging power using different methods.}}
\label{fig:baseline-power}
\end{figure}

{\subsubsection{Discussion}}
\vspace{\baselineskip}
\subsubsubsection{{Discussion on different constraint-aware RL-based methods}}

{
In the comparative analysis, PPOpenalty is introduced as a constraint-aware RL-based method to investigate the effectiveness of the introduction of the Lagrangian method in the proposed method (i.e., PPOlag) to address the CMDP model. Besides PPOpenalty, we further discuss the performance of other RL methods (i.e., REINFORCE, Actor-Critic, and DQN) when considering the constraint violation using a fixed penalty coefficient, as shown in Equation \ref{eq:baseline-penalty}. These methods are referred to as REINpenalty, ACpenalty, and DQNpenalty, respectively.}

{
As illustrated in Table \ref{tab:add-pnlCompare}, compared to the other four RL methods that consider constraints, PPOlag achieves the best results across TTT (regarding the objective), CVV (regarding the constraint), and WCT (regarding user satisfaction). Among REINpenalty, ACpenalty, DQNpenalty, and PPOpenalty, DQNpenalty performs the best, followed by ACpenalty, while REINpenalty performs the worst, despite showing better stability. This result demonstrates consistency with the ranking of the unconstrained RL methods in Table \ref{tab:1}. In comparison to DQNpenalty, the best method with a fixed coefficient, PPOlag still improves the three metrics by approximately 4.2\%, 4.8\%, and 7.0\%, respectively. This is because the Lagrangian method can flexibly adjust the Lagrange multiplier according to the degree of constraint violation, while other methods fail to adaptively adjust the penalty coefficient. As a result, improvements in power grid constraints also facilitate the enhancement of the traffic efficiency objective and user satisfaction in the context of coupled systems. Thus, the additional value of introducing the Lagrangian method into the proposed OP-SRL method is demonstrated.
}

\begin{table}[h!]
  \centering
  \caption{{Evaluation of different constraint-aware RL-based methods.}}
  \label{tab:add-pnlCompare}
  \begin{threeparttable}
  \begin{tabular}{@{} cccc @{}}
    \toprule
    {Method} & {TTT ($\times$1e+4 sec)} & {CVV (pu/bus)} & {WCT (min/EV)} \\
    \midrule
    {REINpenalty} & {66.84 $\pm$ 0.33} & {24.07 $\pm$ 0.12} & {24.06 $\pm$ 0.21} \\
    {ACpenalty} & {58.71 $\pm$ 2.02} & {20.62 $\pm$ 1.12} & {19.50 $\pm$ 0.94} \\
    {DQNpenalty} & {57.59 $\pm$ 3.69} & {20.45 $\pm$ 1.67} & {18.96 $\pm$ 2.11} \\
    {PPOpenalty} & {61.40 $\pm$ 2.32} & {22.29 $\pm$ 1.11} & {21.18 $\pm$ 1.15} \\
    {PPOlag} & {55.20 $\pm$ 0.84} & {19.48 $\pm$ 0.54} & {17.64 $\pm$ 0.49} \\

    \bottomrule
  \end{tabular}
\begin{tablenotes}
\footnotesize
\item[]{The results show mean $\pm$ std for five runs with different random seeds.}
\end{tablenotes}
\end{threeparttable}
\end{table}

\begin{figure}[h!]
\center
\includegraphics[width=\textwidth]{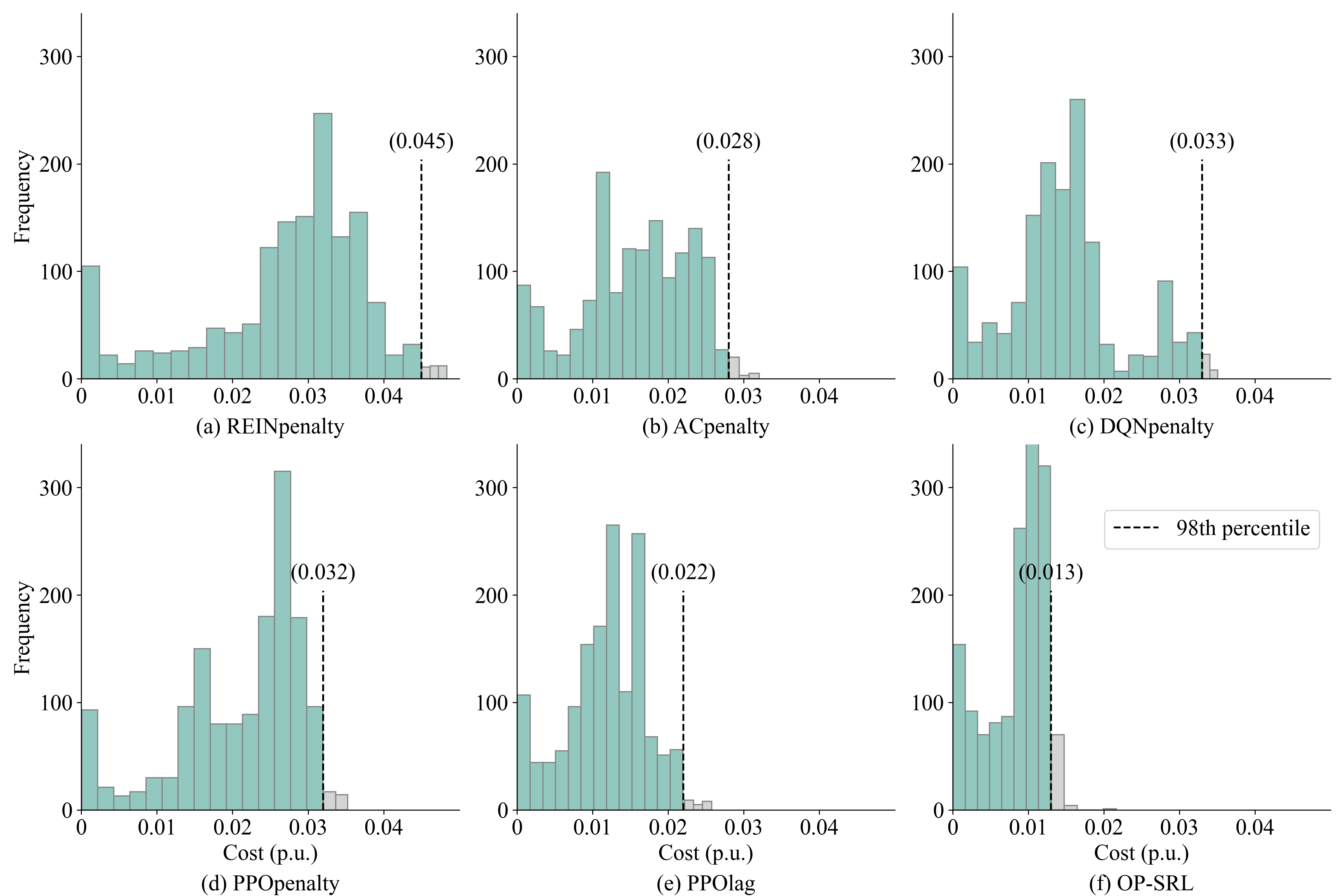}
\caption{{Distribution of cost values for solutions of different constraint-aware RL-based methods. Black dashed lines indicate the position of the 98th percentile which is noted in parentheses. Green bars show the region below the 98th percentile, while gray bars represent the portion above the 98th percentile.}}
\label{fig:discussion2-cost}
\end{figure}

To demonstrate the effectiveness of introducing the Lagrangian method to enhance action safety, we further explore and discuss the constraint satisfaction level of solutions from different methods. For each method, step-wise cost values from the best results of five runs are used to analyze the frequency distribution characteristic. As shown in Figure \ref{fig:discussion2-cost}, the cost value represents the voltage violation due to extra EV charging loads, where a lower value represents a safer strategy or a higher constraint satisfaction level. We can see that the Lagrangian-based methods, PPOlag and OP-SRL, have the smallest cost distribution range between approximately 0 and 0.025; ACpenalty, DQN penalty, and PPOpenalty extend to around 0.035; and REINpenalty performs the worst with a range extending to nearly 0.05. For the 98th percentile, the proposed OP-SRL achieves the smallest value, 0.013, followed by PPOlag with 0.022, and then ACpenalty with the 98th percentile of 0.028. Compared to ACpenalty, PPOlag and OP-SRL show an improvement of 21.4\% and 53.6\%, respectively.

\subsubsubsection{{Discussion on the effect of different RL methods with the predictor}}

{For a more in-depth explanation of the efficacy of integrating PPOlag with the Seq2Seq predictor in the proposed approach, DQN, the best-performing baseline excluding PPOlag in the comparative analysis, is integrated with the Seq2Seq predictor, which yields a method named DQN+Seq2Seq.}

{Table \ref{tab:add-dqn+predictor} showcases the comparison results between the proposed OP-SRL method and DQN+Seq2Seq, as well as their results without the predictor. The results suggest that the effectiveness of both PPOlag and DQN is boosted by the inclusion of the predictor, with enhancements of 2.8\% and 3.5\% respectively. In comparing PPOlag and DQN, the effectiveness of PPOlag remains superior to DQN even after their incorporation of the predictor. Compared to DQN+Seq2Seq, the proposed method shows an improvement of 5.3\%.}

\begin{table}[h!]
  \centering
  \caption{{Evaluation of different RL-based methods with/without the predictor.}}
  \label{tab:add-dqn+predictor}
  \begin{threeparttable}
  \begin{tabular}{@{} cccc @{}}
    \toprule
    {Method} & {TTT ($\times$1e+4 sec)} & {CVV (pu/bus)} & {WCT (min/EV)} \\
    \midrule
    {DQN}  & {57.89 $\pm$ 2.35} & {20.55 $\pm$ 0.75} & {19.06 $\pm$ 1.28} \\
    {DQN+Seq2Seq} & {56.27 $\pm$ 1.29} & {19.55 $\pm$ 0.73} & {18.38 $\pm$ 0.79} \\
    {PPOlag} & {55.20 $\pm$ 0.84} & {19.48 $\pm$ 0.54} & {17.64 $\pm$ 0.49} \\
    {OP-SRL} & {53.29 $\pm$ 0.73} & {18.48 $\pm$ 0.26} & {16.50 $\pm$ 0.29} \\

    \bottomrule
  \end{tabular}
\begin{tablenotes}
\footnotesize
\item[]{The results show mean $\pm$ std for five runs with different random seeds.}
\end{tablenotes}
\end{threeparttable}
\end{table}

\subsubsubsection{{Discussion on the performance of different Seq2Seq predictors}}

{In the proposed OP-SRL framework, the Seq2Seq predictor is a key component to provide insightful information for state augmentation. Apart from using LSTM-based Seq2Seq model as a predictor component, we can also use other predictor models as alternatives. To explore and discuss the performance of various types of predictors in the proposed OP-SRL framework, we introduce five Seq2Seq predictors as summarized below:}
{
\begin{itemize}
\item Three encoder-decoder models based on classic RNN \citep{rumelhart1986learning} and its extensions, i.e., LSTM \citep{hochreiter1997long} and Gated Recurrent Unit (GRU) \citep{Cho2014GRU}. The OP-SRL methods with these predictors are denoted as SRL+RNN, SRL+LSTM (used in the comparative analysis), and SRL+GRU, respectively. These predictors share the same architectural parameters as shown in Table \ref{tab:parameters-predictor}.
\item Temporal Convolutional Network (TCN) \citep{bai2018empirical}, which is based on the convolutional architecture. The associated OP-SRL method is referred to as SRL+TCN.
\item The Transformer \citep{Ashish2017transformer}, which is based on the attention mechanism. The corresponding OP-SRL method is denoted as SRL+Transformer.
\end{itemize}}

{The hyper-parameters of various Seq2Seq predictors are provided in Table \ref{tab:parameters-predictor}. Each associated method is performed five runs with different random seeds. As illustrated in Table \ref{tab:add-predictorCompare}, the inclusion of various Seq2Seq predictors results in better performance than PPOlag without any predictor, in terms of all three metrics. Despite this, the effects of different predictors on performance enhancement fluctuate to some extent. Regarding TTT, RNN and LSTM perform the best; For CVV, GRU and LSTM deliver the best results; and the top performers for WCT are LSTM and Transformer. Overall, each predictor model can help the agent make better decisions in this scenario.}

\begin{table}[h!]
  \centering
  \caption{{Evaluation of different Seq2Seq predictors.}}
  \label{tab:add-predictorCompare}
  \begin{threeparttable}
  \begin{tabular}{@{} cccc @{}}
    \toprule
    {Method} & {TTT ($\times$1e+4 sec)} & {CVV (pu/bus)} & {WCT (min/EV)} \\
    \midrule

    {SRL+RNN} & {53.79 $\pm$ 0.61} & {18.66 $\pm$ 0.25} & {16.92 $\pm$ 0.35} \\
    {SRL+GRU} & {54.41 $\pm$ 1.28} & {18.45 $\pm$ 0.14} & {17.24 $\pm$ 0.78} \\
    {SRL+LSTM} & {53.29 $\pm$ 0.73} & {18.48 $\pm$ 0.26} & {16.50 $\pm$ 0.29} \\
    {SRL+TCN} & {54.16 $\pm$ 0.38} & {18.77 $\pm$ 0.34} & {17.02 $\pm$ 0.19} \\
    {SRL+Transformer} & {54.15 $\pm$ 0.19} & {18.66 $\pm$ 0.17} & {16.92 $\pm$ 0.10} \\

    \bottomrule
  \end{tabular}
\begin{tablenotes}
\footnotesize
\item[]{The results show mean $\pm$ std for five runs with different random seeds.}
\end{tablenotes}
\end{threeparttable}
\end{table}

\subsubsubsection{{Discussion on EV compliance rate}}

{We extend our investigation into the effects of the CS recommendation strategy under varying compliance rates. Building upon the optimal strategy derived from the proposed method, we explore six scenarios with varying compliance rates, ranging from 0 to 100\% in increments of 20\%, where EVs not following charging guidance always choose the CS with the shortest path distance. The compliance rate of 100\% corresponds to the results obtained using one of the optimal strategies based on the proposed method. At a compliance rate of 0, all EVs select the nearest CS in terms of path distance. For compliance rates ranging from 20\% to 80\%, we calculate the average and standard deviation of results from five different random seeds.}

\begin{figure}[h!]
\center
\includegraphics[width=1\textwidth]{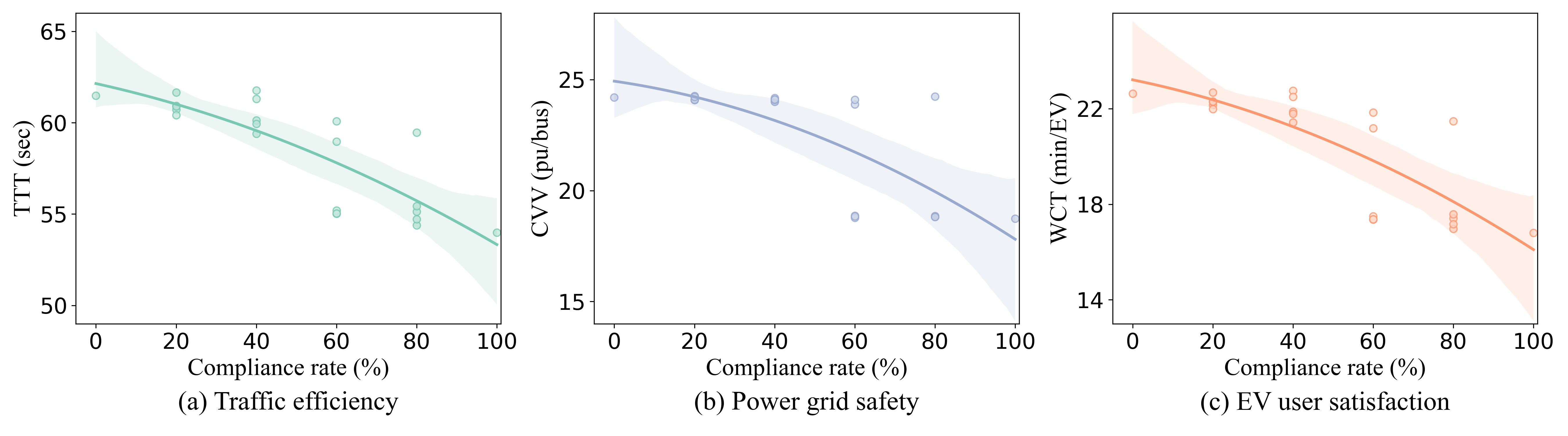}
\caption{{TTT, CVV, and WCT as a function of EV compliance rate.}}
\label{fig:add-complianceRate}
\end{figure}

\begin{table}[h!]
  \centering
  \caption{{Evaluation of different Seq2Seq predictors.}}
  \label{tab:add-compliance-rate}
  \begin{threeparttable}
  \begin{tabular}{@{} cccc @{}}
    \toprule
    {Method} & {TTT ($\times$1e+4 sec)} & {CVV (pu/bus)} & {WCT (min/EV)} \\
    \midrule
    {0\%} & {61.48} & {24.20} & {22.63} \\
    {20\%} & {60.92 $\pm$ 0.41} & {24.15 $\pm$ 0.08} & {22.28 $\pm$ 0.23} \\
    {40\%} & {60.51 $\pm$ 0.88} & {24.08 $\pm$ 0.06} & {22.07 $\pm$ 0.48} \\		
    {60\%} & {56.86 $\pm$ 2.2} & {20.89 $\pm$ 2.53} & {19.05 $\pm$ 2.02} \\		
    {80\%} & {55.83 $\pm$ 1.85} & {19.91 $\pm$ 2.16} & {18.12 $\pm$ 1.69} \\
    {100\%} & {53.98} & {18.73} & {16.8} \\
    \bottomrule
  \end{tabular}
\begin{tablenotes}
\footnotesize
\item[]{The results show mean $\pm$ std of five runs for compliance rates spanning from 20\% to 80\%.}
\end{tablenotes}
\end{threeparttable}
\end{table}

As shown in Figure \ref{fig:add-complianceRate}, as the compliance rate rises, an increasing number of EVs follow the optimal charging guidance strategy, resulting in improvements across all three metrics. Based on Table \ref{tab:add-compliance-rate}, the level of improvement in each metric is more significant at 60\% and 100\% compliance rates, whereas the improvements are comparatively smaller at 20\%, 40\%, and 80\%. In particular, the metrics increased by 6.0\%, 13.2\%, and 13.7\% at a compliance rate of 60\% compared to 40\%. In comparison to 80\% compliance rate, the metrics increased by 3.3\%, 5.9\%, and 7.3\% at 100\% compliance rate.

\subsubsection{Sensitivity analysis on EV penetration}

\begin{figure}[h!]
  \centering
  \begin{minipage}[t]{0.34\linewidth}
  \includegraphics[width=1\textwidth]{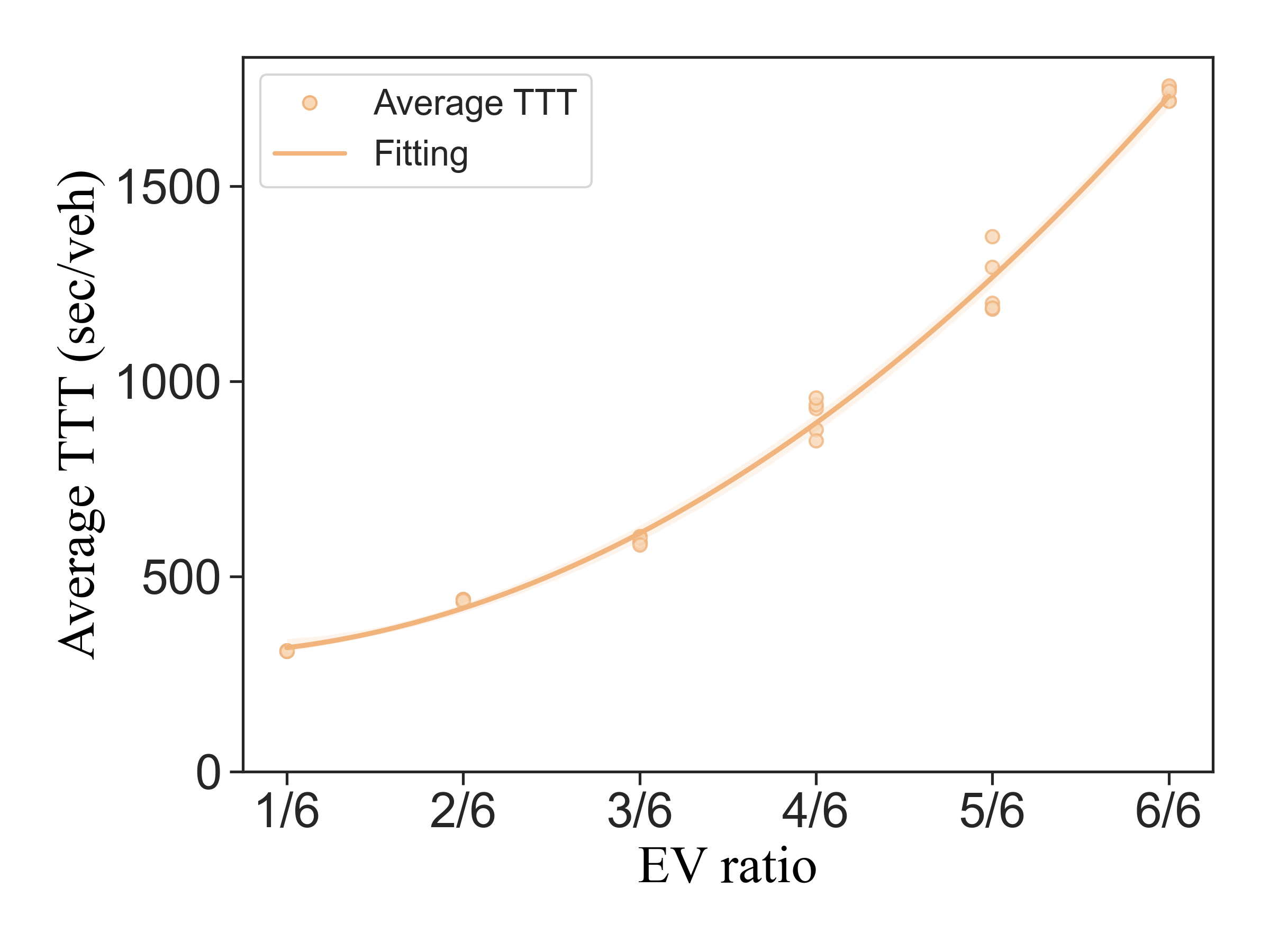}
  \caption*{(a) Traffic efficiency}
  \end{minipage}\hfill\hspace{-10pt}
  \begin{minipage}[t]{0.34\linewidth}
  \includegraphics[width=1\textwidth]{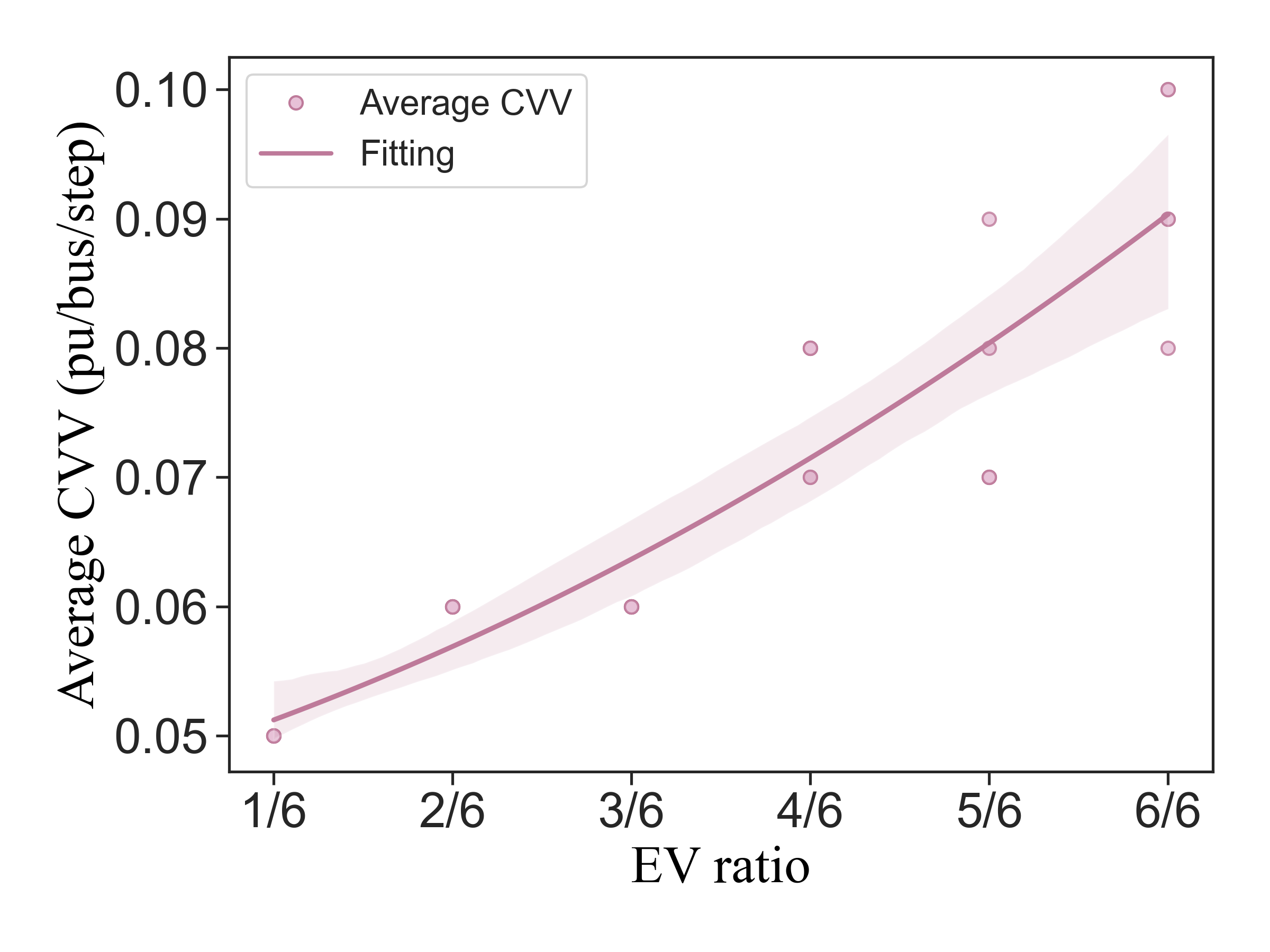}
  \caption*{(b) Power grid safety}
  \end{minipage}\hfill\hspace{-10pt}
  \begin{minipage}[t]{0.34\linewidth}
  \includegraphics[width=1\textwidth]{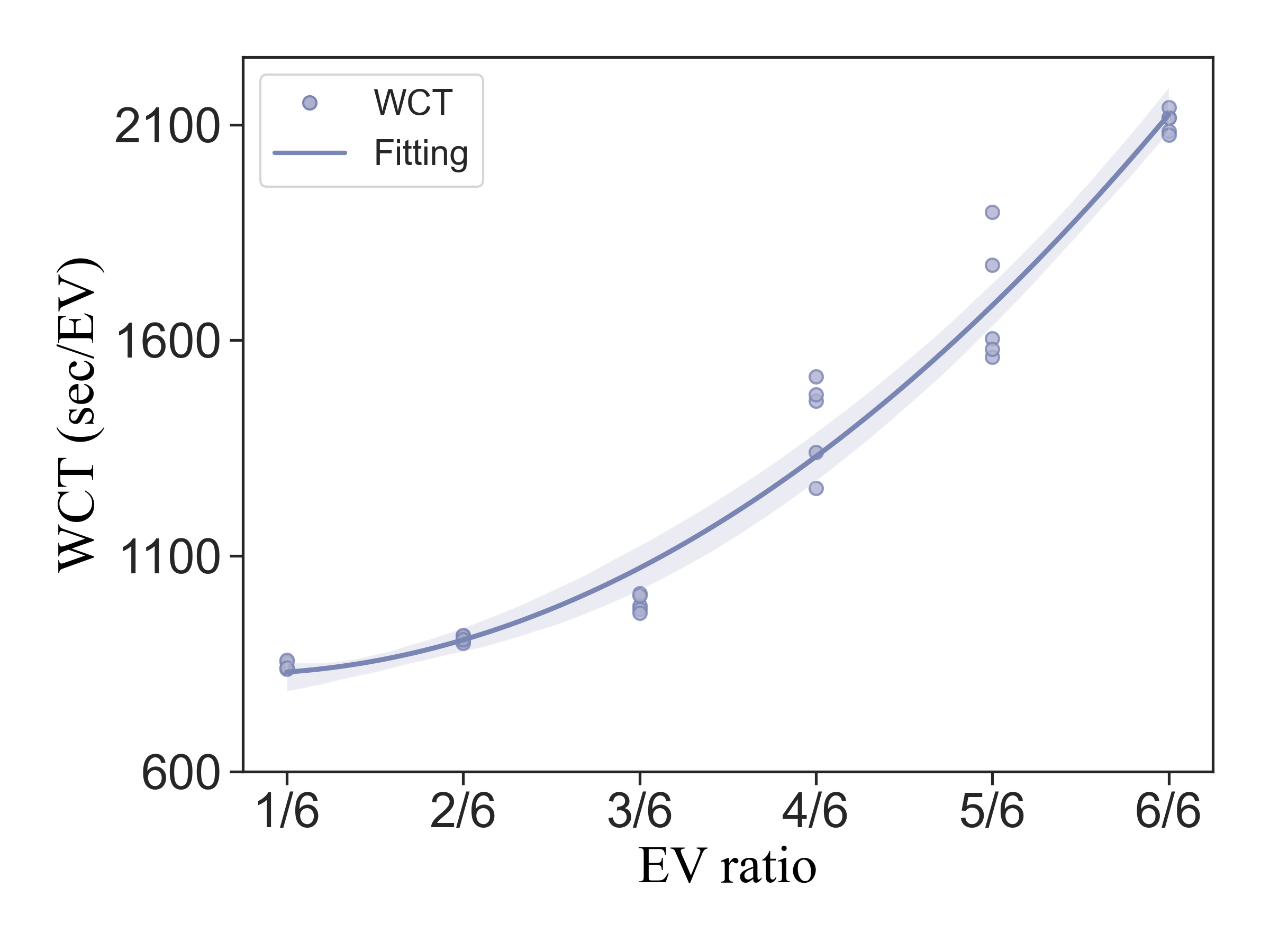}
  \caption*{(c) EV user satisfaction}
  \end{minipage}
  \caption{Average TTT, average CVV, and WCT as a function of EV ratio in (a), (b), (c) respectively. Solid lines represent the fitting results using a quadratic polynomial. The shaded areas depict the 95\% confidence intervals for five runs.}
\label{fig:sens1-fit}
\end{figure}

To investigate the effect of EV penetration on the performance metrics of various stakeholders, we examine six scenarios with different levels of EV penetration using the proposed OP-SRL method. The total travel demand for all six scenarios is 600 vehicles per hour, with the number of EVs requiring charging varying from 100 to 600 vehicles per hour, with increments of 100, corresponding to EV ratios from 1/6 to 6/6 (i.e., roughly from 16.7\% to 100\%). For comparability, we evaluate the TTT averaged per vehicle, the CVV averaged per action step, and the WCT averaged per EV. The proposed method can converge across different EV penetration rates, and Figure \ref{fig:sens1-fit} illustrates the trends of various indicators under the best policy as the EV ratio varies. The results show that with increasing EV penetration, all three metrics exhibit an overall upward trend. This occurs because, with a fixed number of charging piles, the increase in the number of EVs will lead to a longer waiting time. In addition, the rise in the number of EVs adds an extra load to the power grid, causing an increase in voltage violation (Figure \ref{fig:sens1-fit}b) and triggering the controller to output lower charging power, thereby lengthening the charging time for EVs (Figure \ref{fig:sens1-fit}c). As significant parts of total travel time, the increase in the waiting and charging time of EVs will result in a longer travel duration (Figure \ref{fig:sens1-fit}a).

\begin{figure}[h!]
\center
\includegraphics[width=0.6\textwidth]{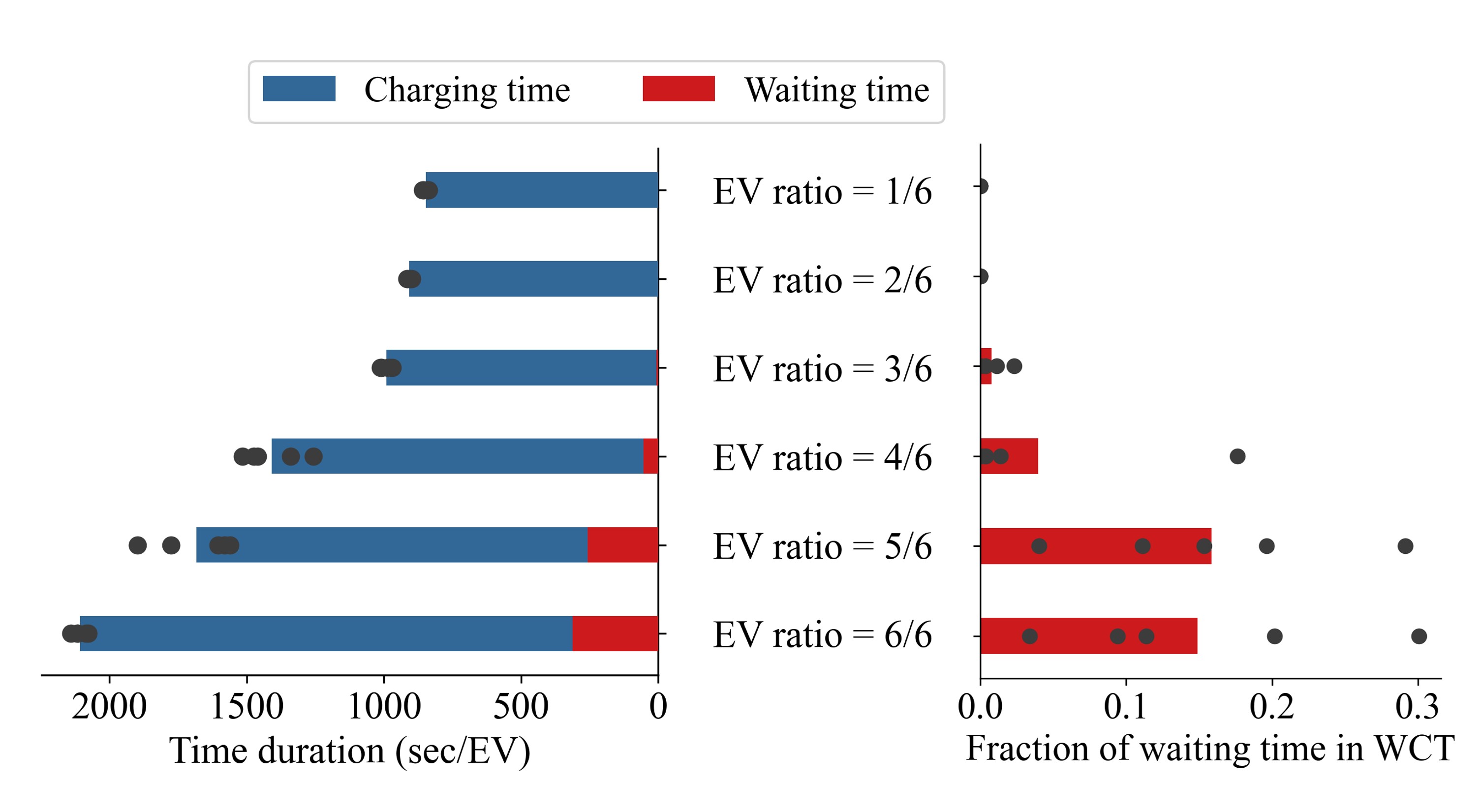}
\caption{The duration and fraction of waiting time and charging time across different EV ratios. On the left graph, the blue bars denote the average charging time, the red bars denote the average waiting time, and the black scattered points denote the WCT values for five runs. On the right graph, the red bars denote the percentage of waiting time in WCT, and the black scattered points denote the specific fractions in the five runs.}
\label{fig:sens1-wtct}
\end{figure}

Further details on the duration and fraction of waiting time and charging time are presented in Figure \ref{fig:sens1-wtct}. As shown in the left half of the graph, with the rise in EV ratio, the average waiting time for each EV grows from 0 to about 310 seconds, and the average charging time increases from around 850 seconds to nearly 1800 seconds. Also, the percentage of waiting time in WCT has increased from 0 to roughly 15\%, as illustrated in the right-hand graph.

\begin{figure}[h!]
  \centering
  \begin{minipage}[t]{0.34\linewidth}
  \includegraphics[width=1\textwidth]{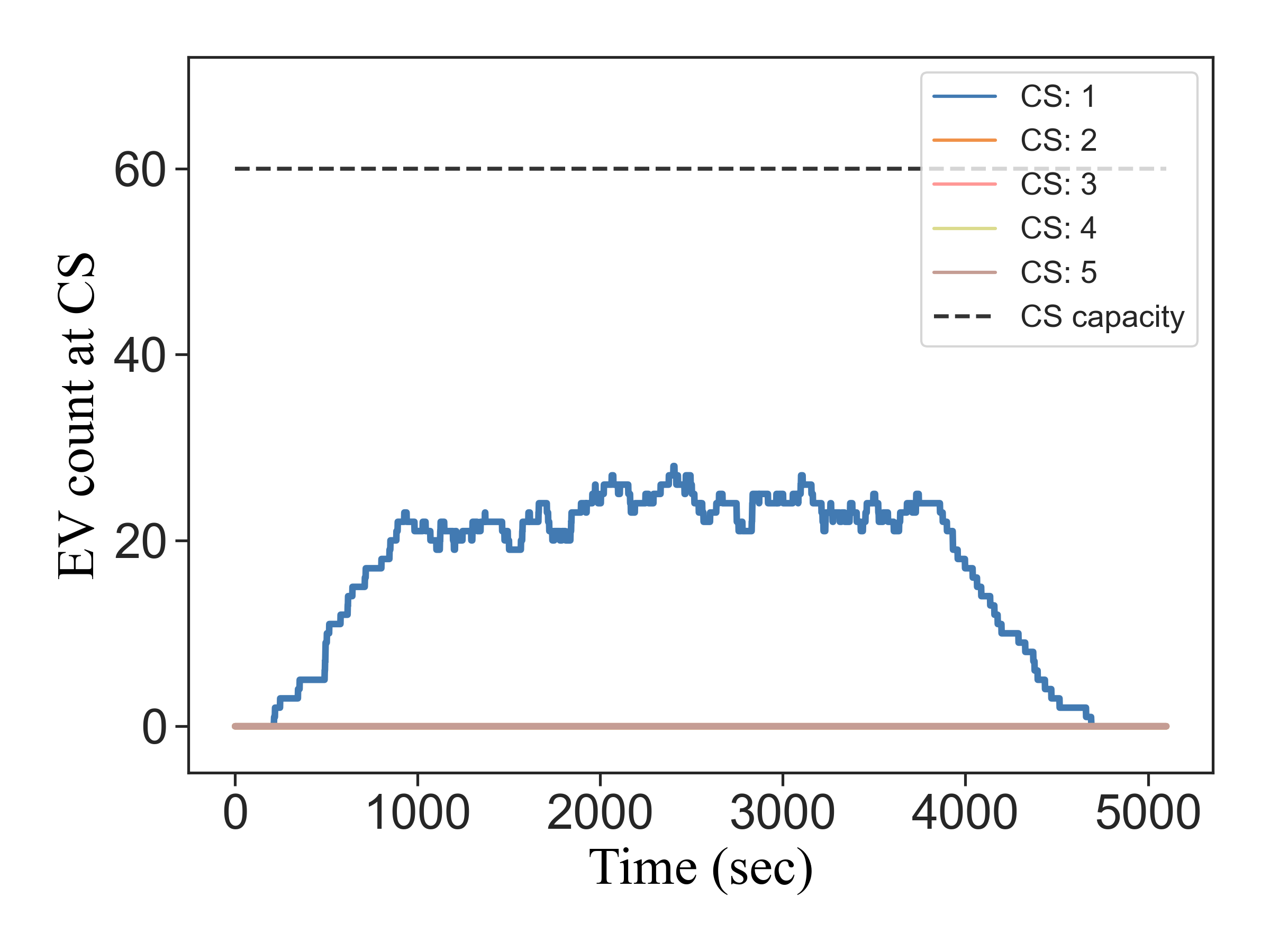}
  \caption*{(a) EV ratio = 1/6}
  \end{minipage}\hfill\hspace{-10pt}
  \begin{minipage}[t]{0.34\linewidth}
  \includegraphics[width=1\textwidth]{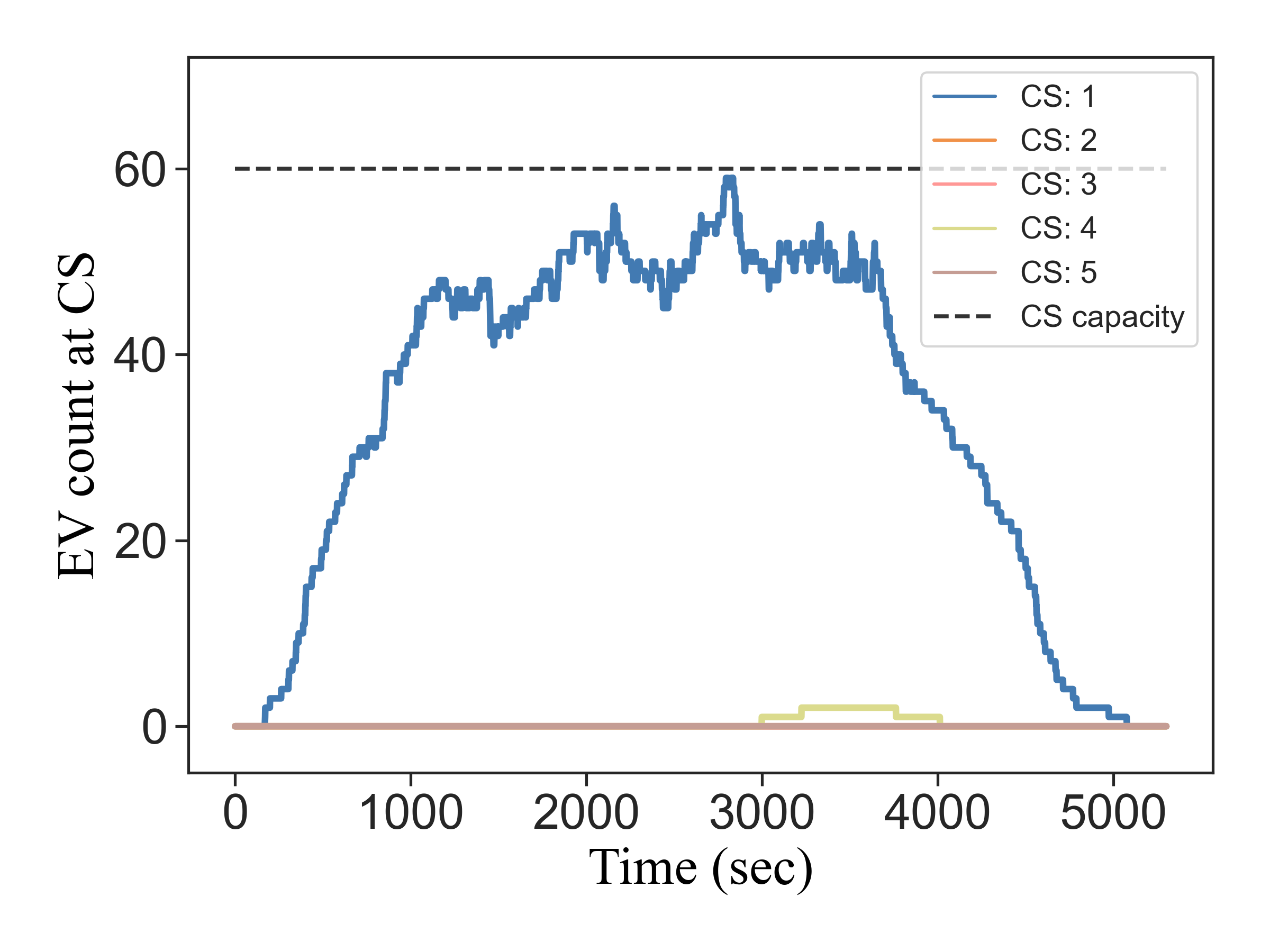}
  \caption*{(b) EV ratio = 2/6}
  \end{minipage}\hfill\hspace{-10pt}
  \begin{minipage}[t]{0.34\linewidth}
  \includegraphics[width=1\textwidth]{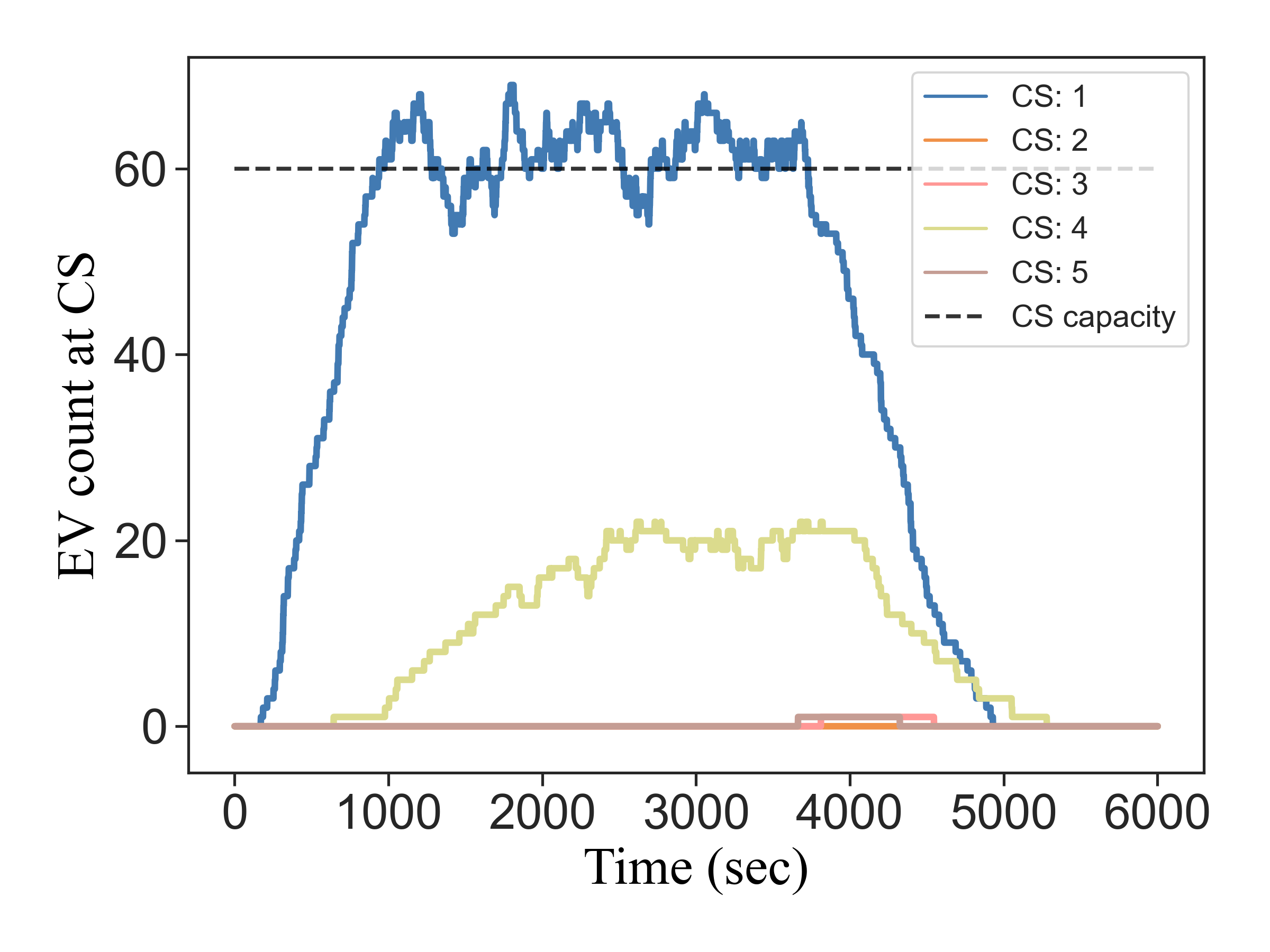}
  \caption*{(c) EV ratio = 3/6}
  \end{minipage}
  \begin{minipage}[t]{0.34\linewidth}
  \includegraphics[width=1\textwidth]{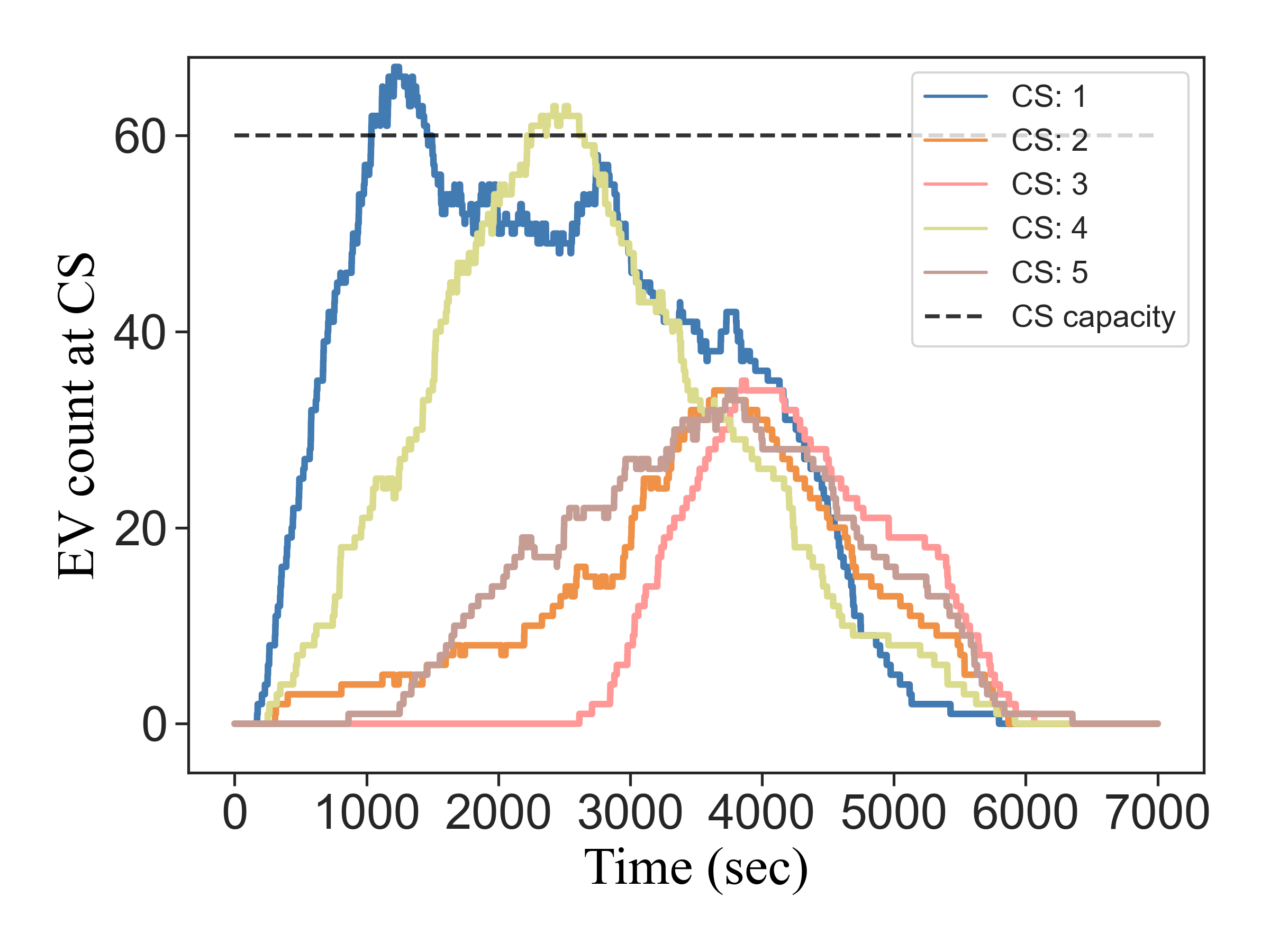}
  \caption*{(d) EV ratio = 4/6}
  \end{minipage}\hfill\hspace{-10pt}
  \begin{minipage}[t]{0.34\linewidth}
  \includegraphics[width=1\textwidth]{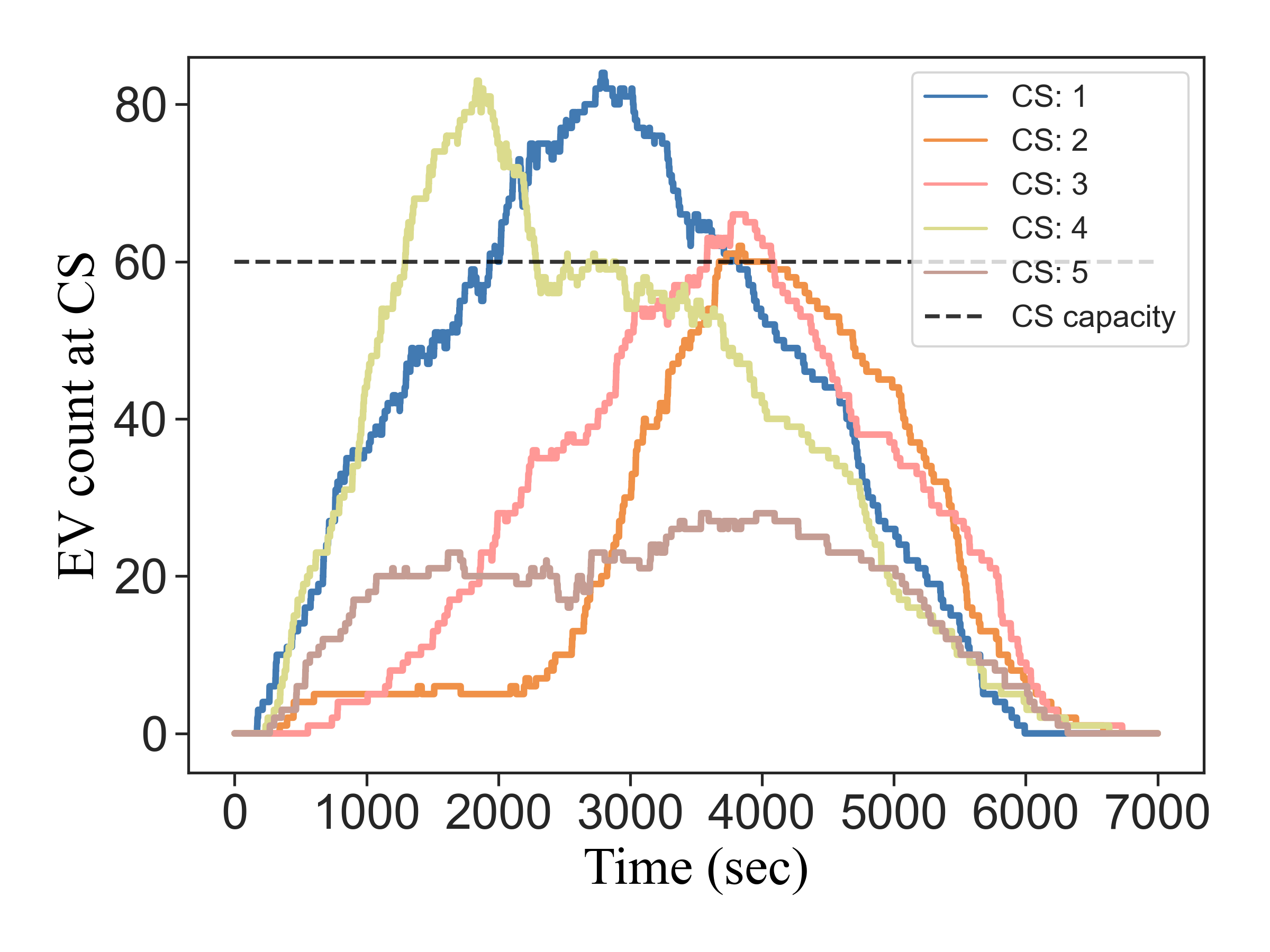}
  \caption*{(e) EV ratio = 5/6}
  \end{minipage}\hfill\hspace{-10pt}
  \begin{minipage}[t]{0.34\linewidth}
  \includegraphics[width=1\textwidth]{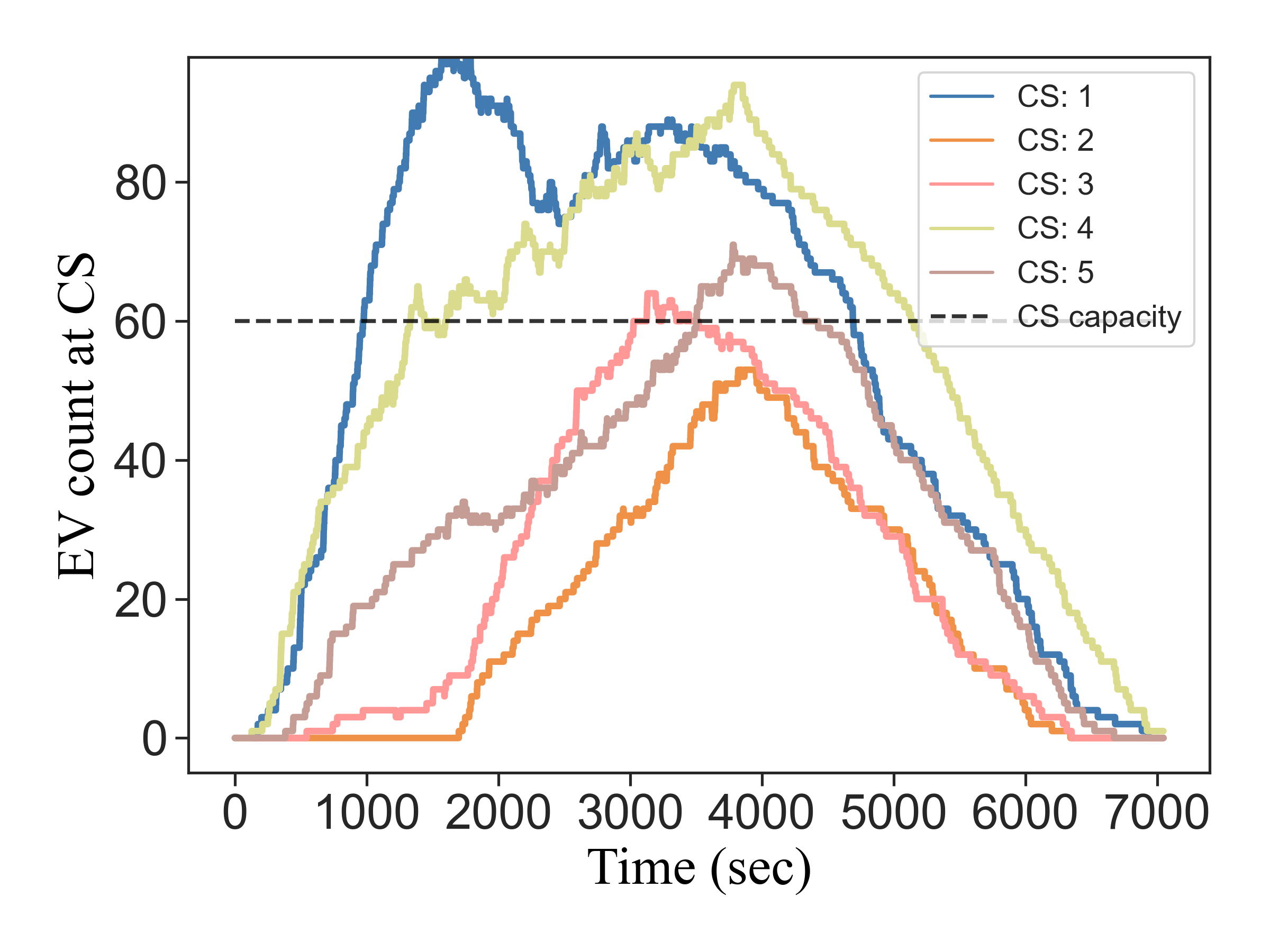}
  \caption*{(f) EV ratio = 6/6}
  \end{minipage}
  \caption{Time-varying vehicle count at each charging station (CS) under different EV penetration rates. The vehicle count here comprises both queued vehicles and charging vehicles. The black dashed line represents the service capacity of each charging station, i.e., the number of charging piles.}
\label{fig:sens1-dmd}
\end{figure}

Figure \ref{fig:sens1-dmd} demonstrates the time-dependent utilization of each charging station under varying EV penetration levels. As can be seen from the figure, when EV penetration is at low levels (i.e., 1/6 or 2/6), only the charging station labeled 1 is effective in providing charging services. When the EV penetration rate increases to 3/6, both charging stations labeled 1 and 4 serve as effective charging stations, with charging station 1 providing approximately 75\% of the charging services. Once the EV ratio is more than half, all charging stations undertake the charging service, despite CS:1 and CS:4 still providing the majority among them. As such, the proposed method allows for the importance evaluation for CSs regarding the extent of provided charging service in this setup and offers insights into CS site selection given a limited budget.

\begin{figure}[h!]
\center
\includegraphics[width=\textwidth]{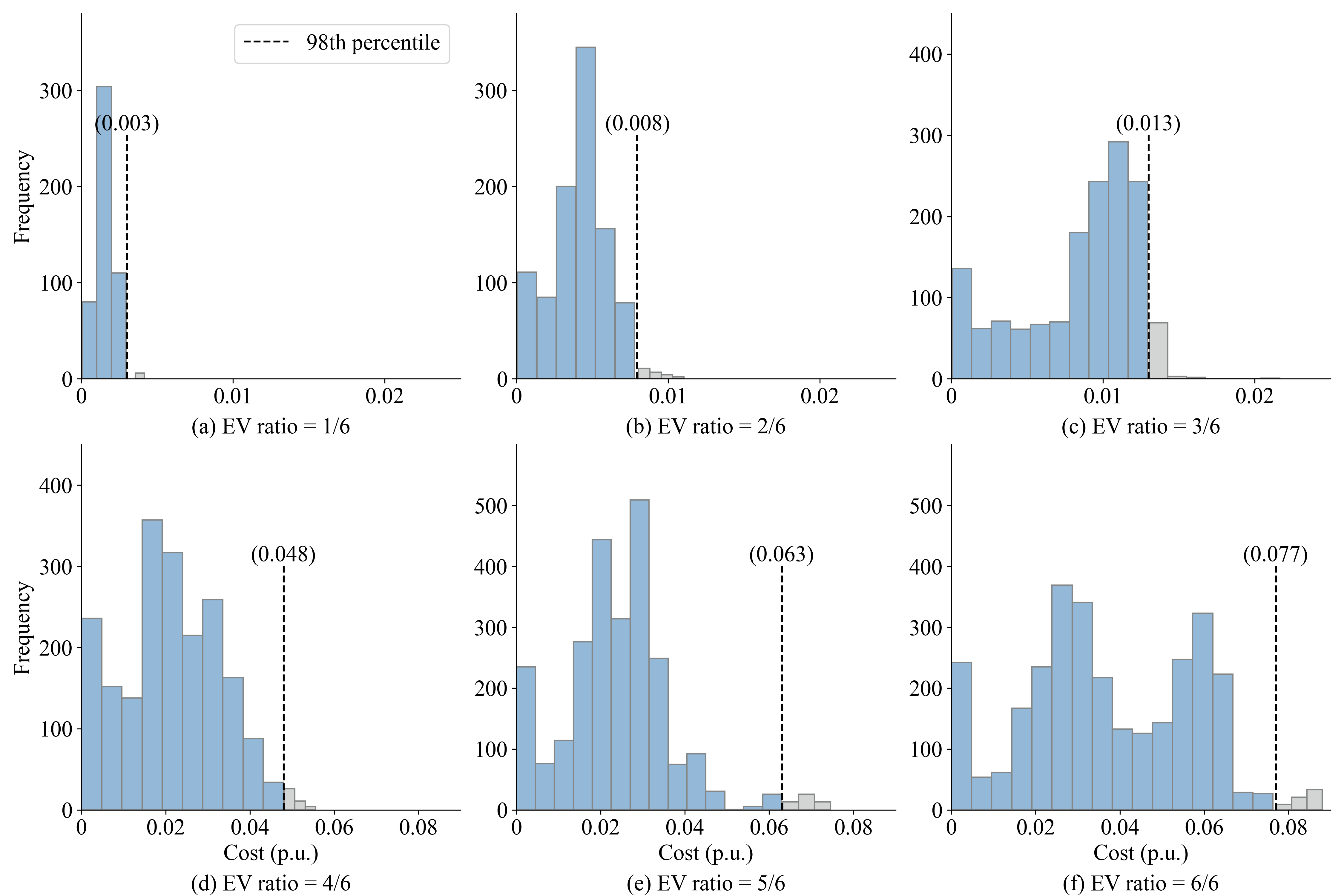}
\caption{{Distribution of cost values under EV penetrations from 1/6 to 6/6. Black dashed lines indicate the position of the 98th percentile which is noted in parentheses. Green bars show the region below the 98th percentile, while gray bars represent the portion above the 98th percentile.}}
\label{fig:EVsens-cost-distrib}
\end{figure}

Figure \ref{fig:EVsens-cost-distrib} illustrates the constraint satisfaction level of the proposed method under circumstances with different EV penetration ratios. For each scenario, step-wise cost values (i.e., the voltage violation due to extra EV charging loads) from the best results of five runs are used to analyze the frequency distribution characteristic. The results show that with increasing EV penetration and growing charging load, the distribution range of cost extends from less than 0.005 to more than 0.08, implying a decline in the safety of grid operation. Under low and moderate EV penetration rates, the 98th percentile of cost values remains below 0.015, suggesting a limited impact of EV load integration on grid security. However, with EV penetration rates exceeding 50\%, the 98th percentile of cost values notably increases to 0.048, 0.063, and 0.077, highlighting a more significant impact of EV charging load. This observed trend is in line with findings from various studies \citep[e.g.,][]{deb2018impact, tavakoli2020impacts}, underscoring the potential risk that the widespread adoption of EVs poses to the secure and stable operation of the grid.

\subsubsection{Sensitivity analysis on time interval of charge controller}
\begin{figure}[h!]
  \centering
  \begin{minipage}[t]{0.346\linewidth}
  \includegraphics[width=1\textwidth]{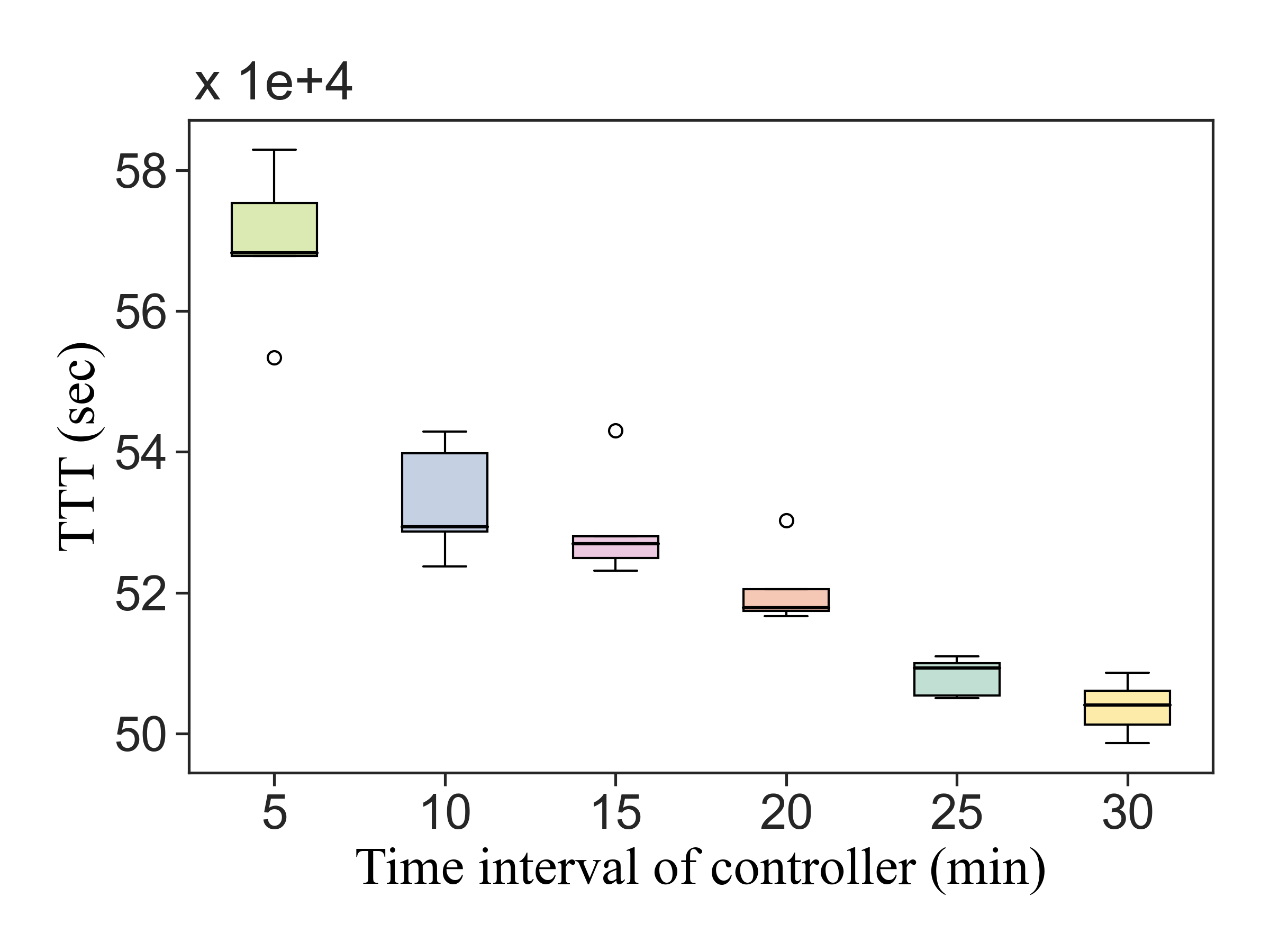}
  \caption*{(a) Traffic efficiency}
  \end{minipage}\hfill\hspace{-10pt}
  \begin{minipage}[t]{0.327\linewidth}
  \includegraphics[width=1\textwidth]{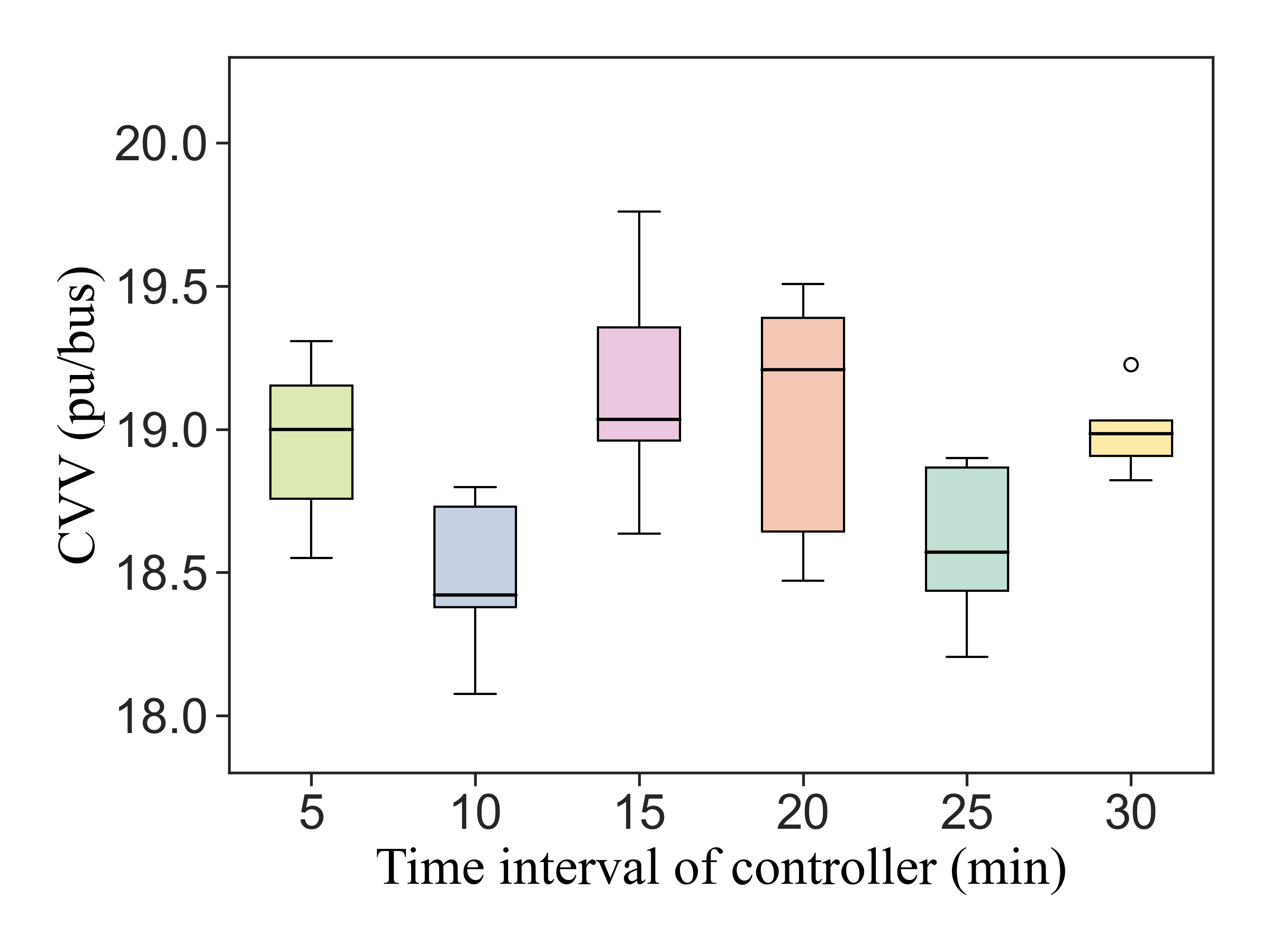}
  \caption*{(b) Power grid safety}
  \end{minipage}\hfill\hspace{-10pt}
  \begin{minipage}[t]{0.346\linewidth}
  \includegraphics[width=1\textwidth]{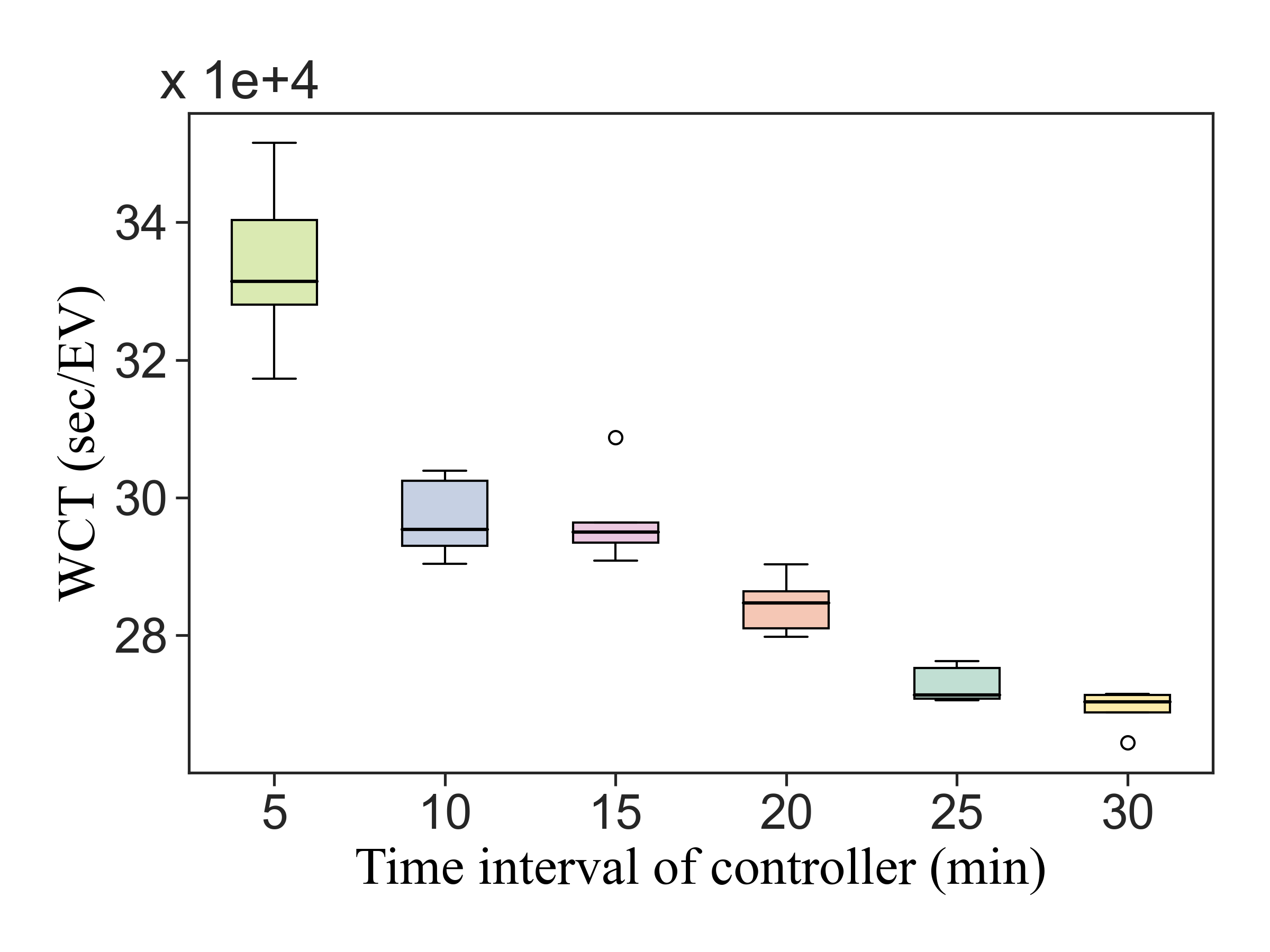}
  \caption*{(c) EV user satisfaction}
  \end{minipage}
  \caption{TTT, CVV, and WCT as a function of the time interval for charge controller in (a), (b), (c) respectively. Each box represents the distribution of performance metrics at a given time interval in five runs.}
\label{fig:sens2-fit}
\end{figure}

The time interval of the charge controller signifies the promptness of the controller's reaction to power load variations. The shorter the interval, the more immediate the controller's response. To estimate the impact of the controller's time interval on the coupled systems, we examine the proposed method with different control intervals. Figure \ref{fig:sens2-fit} shows the results of TTT, CVV, and WCT. Figure \ref{fig:sens2-fit}a and \ref{fig:sens2-fit}c illustrate that as the control interval grows from 5 minutes to 30 minutes, both TTT and WCT exhibit a decreasing trend. Figure \ref{fig:sens2-fit}b illustrates that CVV fluctuates up and down with the increase of the control interval. Hence, there is a trade-off between TTT (or WCT) and CVV when optimizing the control interval.

\subsubsection{Sensitivity analysis on decoder length in predictor}
The Seq2Seq predictor can manage situations where the input sequence (i.e., the encoder length) and the output sequence (i.e., the decoder length) are not equal. In this study, the output sequence of the decoder acts as an additional state to augment the state space of the SRL agent. To determine an appropriate decoder length value, we evaluate the performance of the proposed method under different decoder lengths, given a fixed encoder length of 5. The results suggest that this method delivers the best performance across all indicators when the decoder length is 5, just equal to the length of the encoder. When the decoder length is short (i.e., 1 or 3), the performance of each indicator is inferior and shows a larger fluctuation range. This is because a smaller output sequence length fails to provide enough future information for the agent to make decisions, causing the agent to possibly fall into local optimum and the decision-making effect exhibit greater randomness. On the contrary, when the decoder length is too long, it may bring redundant information, which could hinder the agent from making favorable decisions. 

\begin{figure}[h!]
  \centering
  \begin{minipage}[t]{0.346\linewidth}
  \includegraphics[width=1\textwidth]{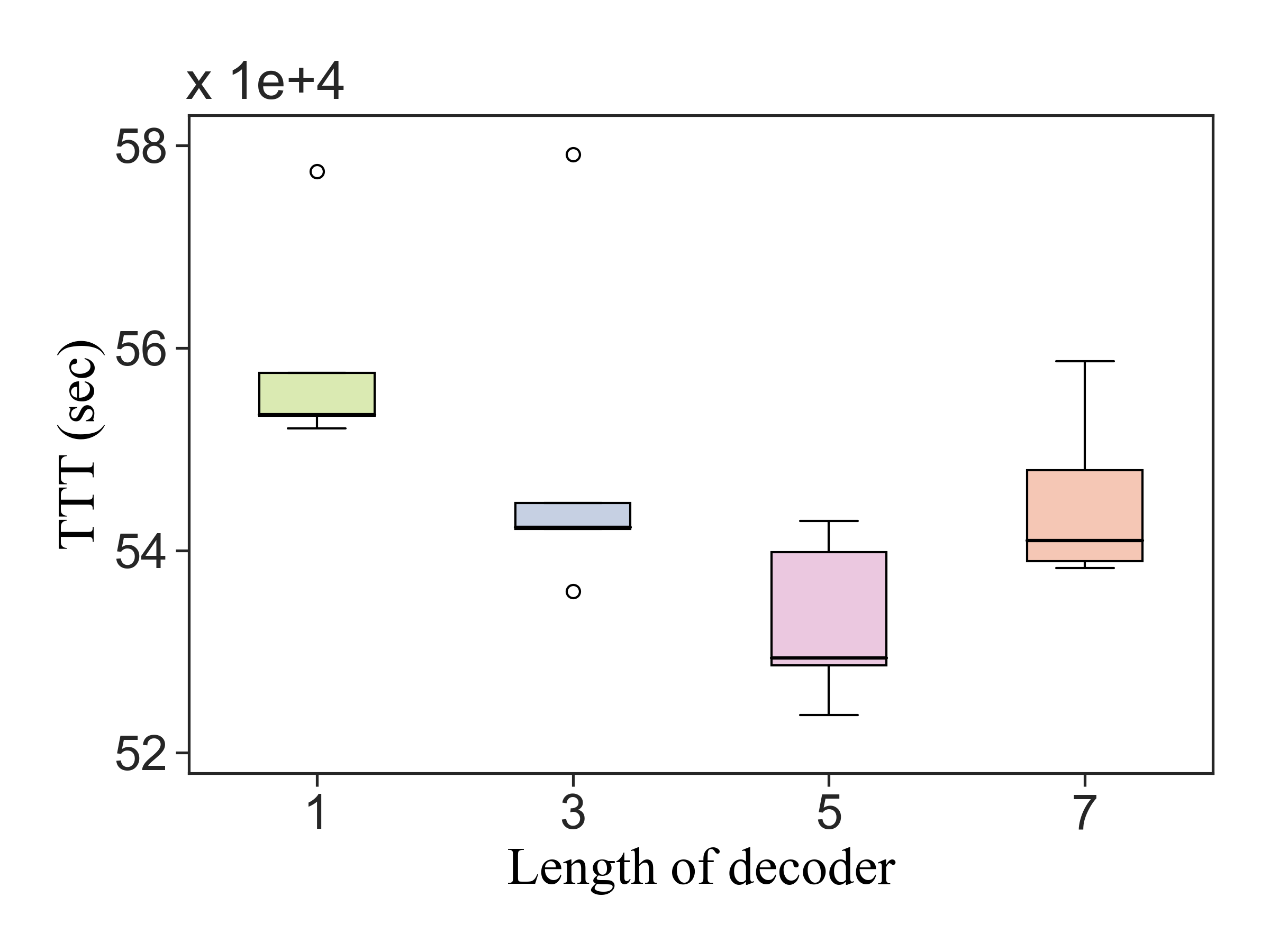}
  \caption*{(a) Traffic efficiency}
  \end{minipage}\hfill\hspace{-10pt}
  \begin{minipage}[t]{0.327\linewidth}
  \includegraphics[width=1\textwidth]{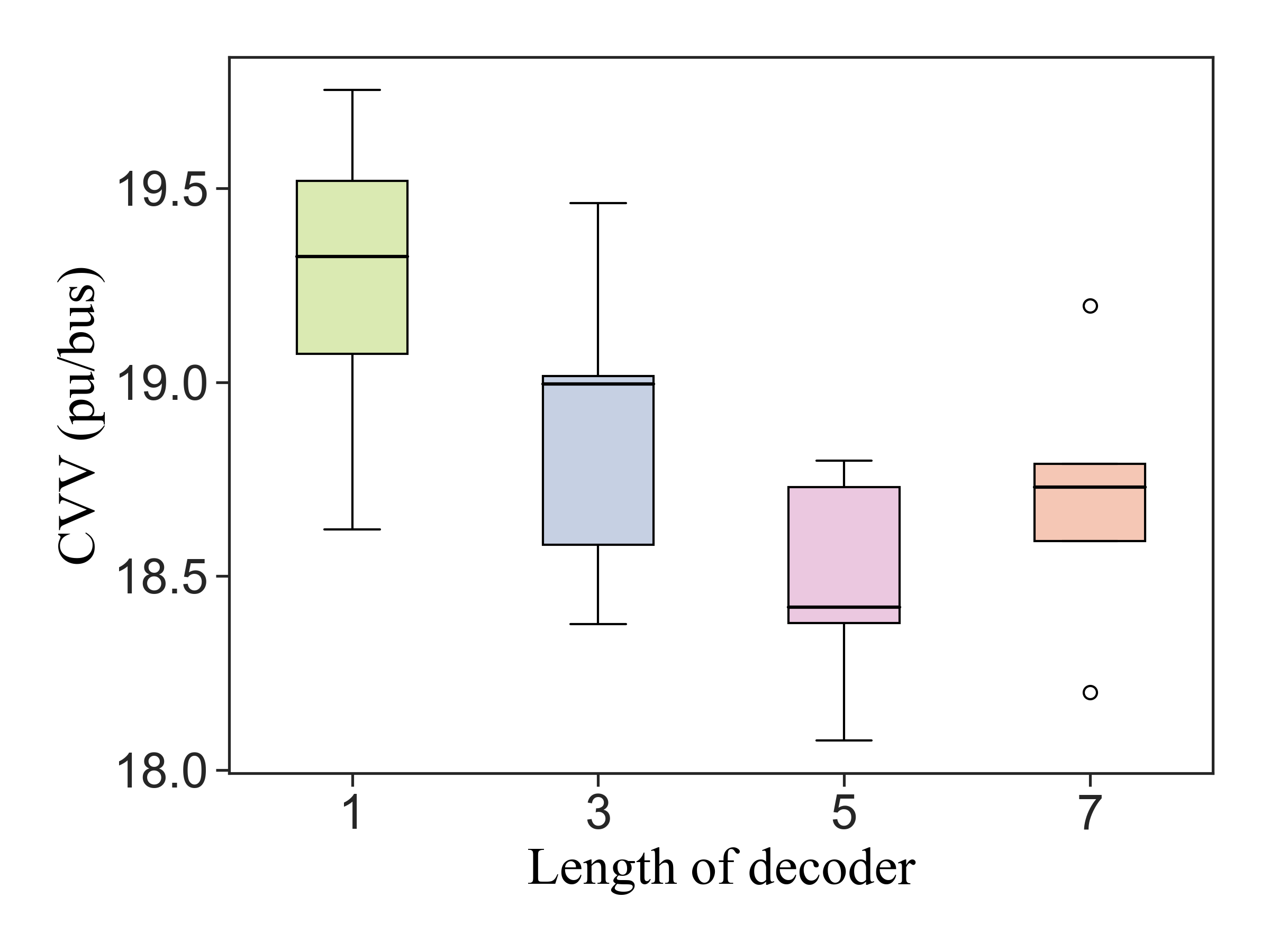}
  \caption*{(b) Power grid safety}
  \end{minipage}\hfill\hspace{-10pt}
  \begin{minipage}[t]{0.346\linewidth}
  \includegraphics[width=1\textwidth]{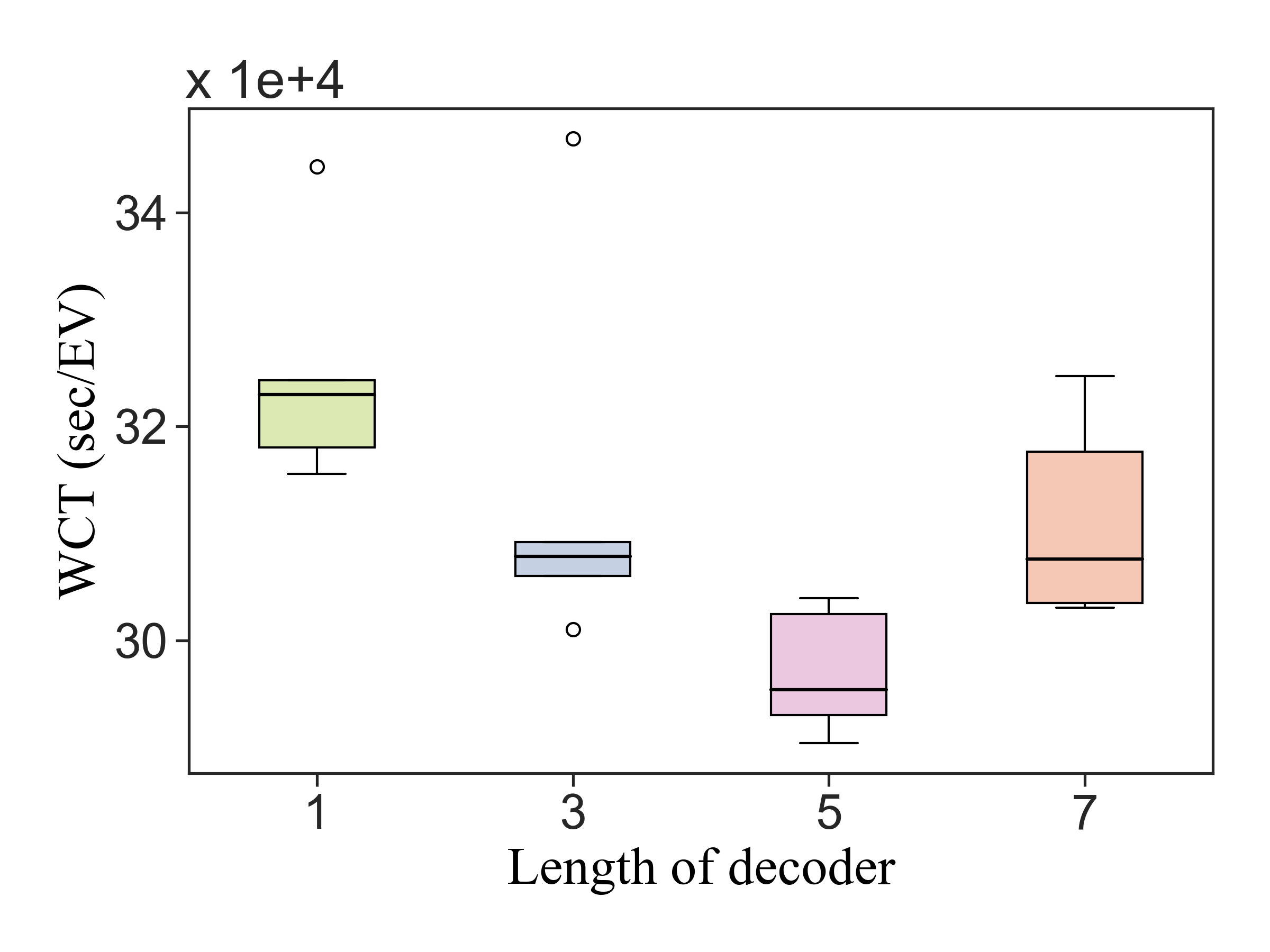}
  \caption*{(c) EV user satisfaction}
  \end{minipage}
  \caption{TTT, CVV, and WCT as a function of decoder length in (a), (b), (c) respectively. Each box depicts the distribution of performance metrics using a specific length of decoder.}
\label{fig:sens-dec-fit}
\end{figure}

\subsection{Case B: A large-scale real-world network}

\begin{figure}[h!]
  \centering
  \begin{minipage}[t]{0.32\linewidth}
    \includegraphics[width=1\textwidth]{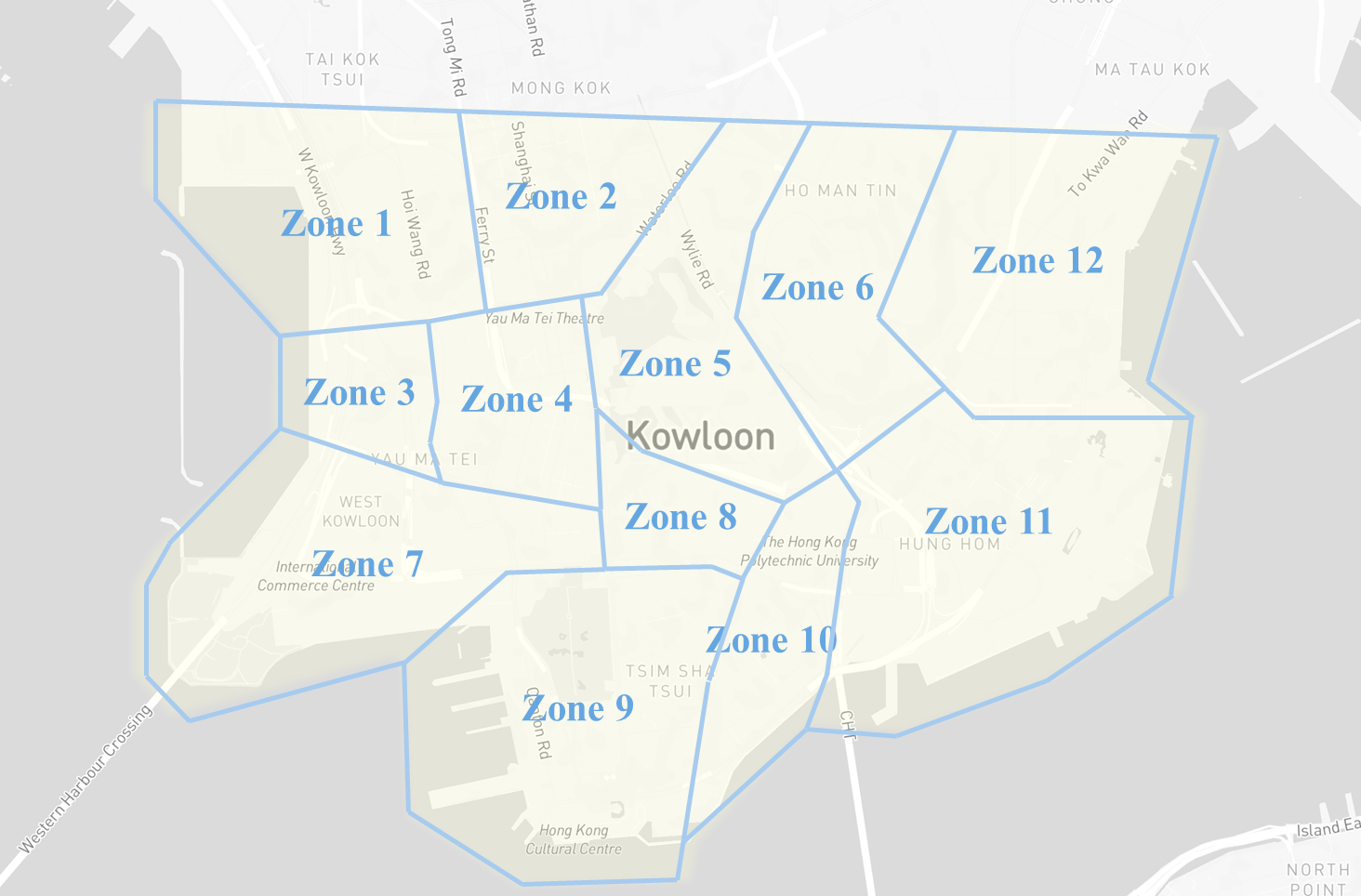}
  \caption*{(a) Kowloon area with 12 OD zones}
  \end{minipage}
  \begin{minipage}[t]{0.32\linewidth}
    \includegraphics[width=1\textwidth]{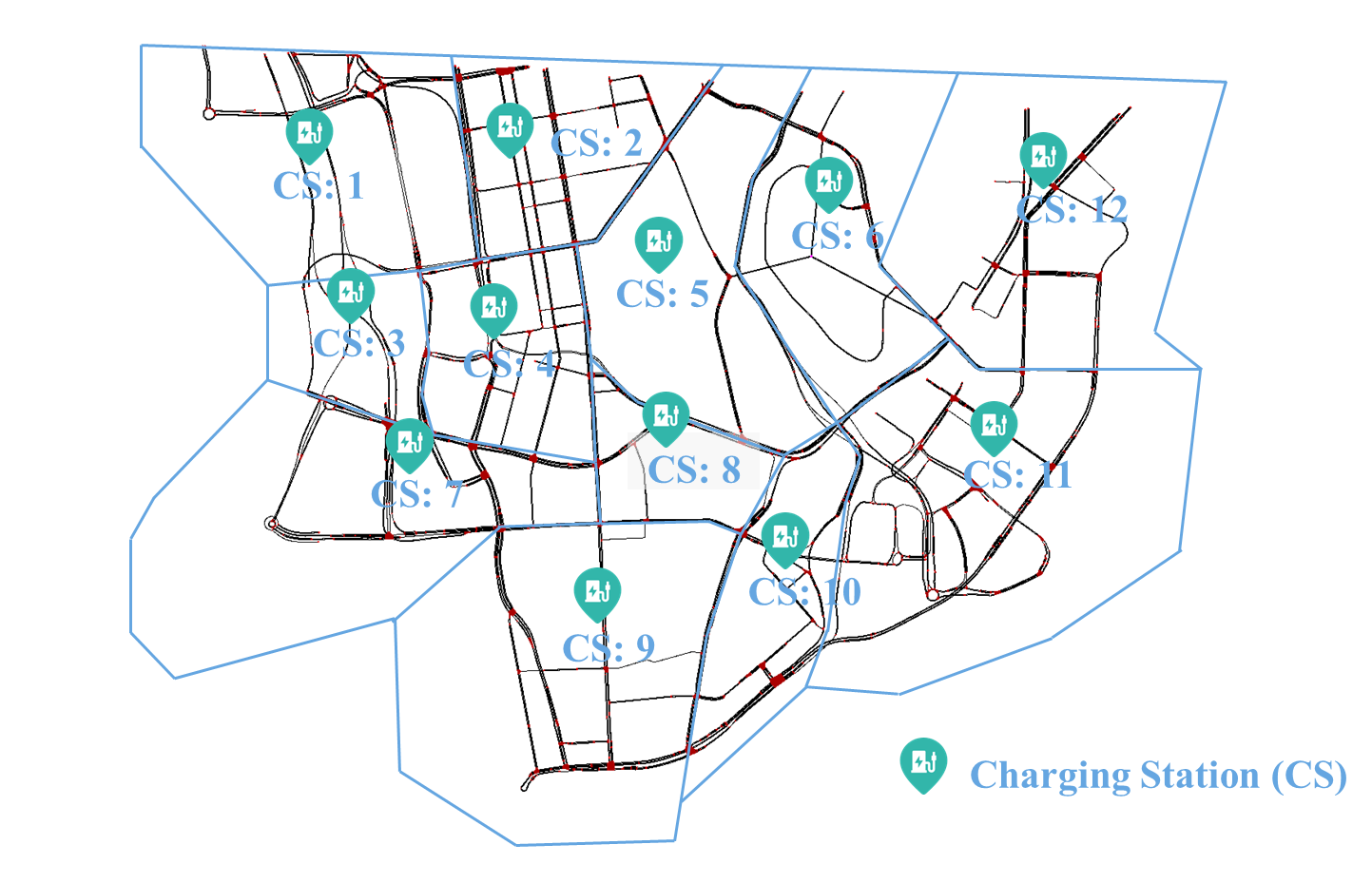}
  \caption*{(b) Road network in SUMO platform}
  \end{minipage}
 \begin{minipage}[t]{0.32\linewidth}
    \includegraphics[width=1\textwidth]{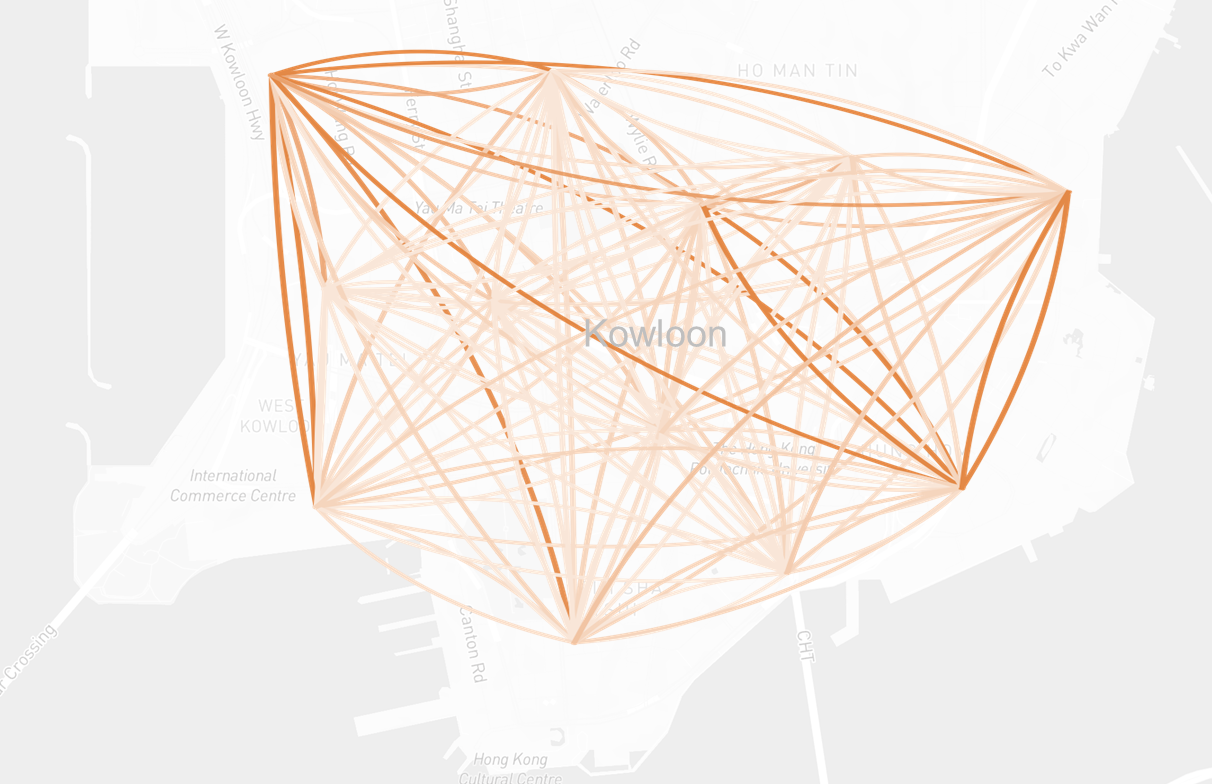}
  \caption*{{(c) OD demand distribution}}
  \end{minipage}
  \caption{{Study area in Kowloon region regarding (a) OD zone division, (b) road network in SUMO, and (c) demand distribution among 12 OD zones for totally 4,442 vehicles/hour, with darker connecting lines indicating higher traffic flow between two regions.}}
 \label{fig:caseB-networks}
 \end{figure}

To demonstrate the applicability of the proposed method in the practical context, we conduct another case study on a large-scale real-world traffic network, i.e., the specific area of the Kowloon region in Hong Kong, coupled with the IEEE 69-bus distribution system. Figure \ref{fig:caseB-networks}(a) illustrates the target area for study, which is divided into 12 OD zones (Zone 1 to Zone 12) for the allocation of individual trips from actual observations. Figure \ref{fig:caseB-networks}(b) presents the topological structure of this road network in the SUMO platform, comprising 817 nodes and 1,072 directed edges, as well as 12 CSs distributed across the centers of each OD zone. {Figure \ref{fig:caseB-networks}(c) shows the OD demand distribution across 12 OD zones}.

In this setup, we benchmark the proposed method against alternative methods that exclude key components (i.e., PPO and PPOlag) {as well as the greedy method and MPC}. The results are presented in Table \ref{tab:performance-kl}, including the impact on the coupled systems (i.e., TTT, CVV, and WCT) along with the time expenditure (i.e., ET and DT). {In this case, the proposed method also demonstrates the best performance across three system performance metrics, while the greedy method and the MPC performs the worst. In comparison to the greedy method, the proposed method has improved by 1.9\%, 9.4\%, and 7.0\% on the three metrics. When compared to MPC, the proposed method has exhibited improvements of 3.5\%, 4.9\%, and 11.0\% respectively.} Besides, the proposed method outperforms PPO by decreasing the TTT, CVV, and WCT by about 12,317 seconds (0.8\%), 0.94 pu per bus (5.5\%), and 26 seconds per EV (2.6\%). When compared with PPOlag, the TTT, CVV, and WCT of the proposed method shows a reduction of about 0.4\%, 1.5\%, and 1.1\%, respectively. Also, the rankings of these RL-based methods remain consistent across the three metrics, as derived in case A.

\begin{table}[h!]
  \centering
  \caption{Results of ablation study for the real-world network}
  \label{tab:performance-kl}
\begin{threeparttable}
  \begin{tabular}{@{} cccccc @{}}
    \toprule
    Method & TTT ($\times$1e+5 sec) & CVV (pu/bus) & WCT (min/EV) & {ET (d)} & {DT (sec)}\\
    \midrule
    {Greedy} & {16.452} & {17.86} & {17.20} & {-} & {0.471}\\
    {MPC} & {16.719} & {17.01} & {17.98} & {-} & {9.868}\\
    
    PPO & 16.257 $\pm$ 0.030 & 17.12 $\pm$ 0.45 & 16.4 $\pm$ 0.1 & {2.25} & {0.026} \\
    PPOlag & 16.205 $\pm$ 0.028 & 16.43 $\pm$ 0.42 & 16.2 $\pm$ 0.1 & {2.25} & {0.026}\\
    \textbf{OP-SRL} & \textbf{16.134} $\pm$ 0.028 & \textbf{16.18} $\pm$ 0.28 & \textbf{16.0} $\pm$ 0.1 & {3.85} & {0.027}\\
    \bottomrule
  \end{tabular}
\begin{tablenotes}
\footnotesize
\item[]The results show mean $\pm$ std for RL-based methods of five runs with different random seeds. All the evaluation metrics favor lower values and the best results are highlighted in bold.
\end{tablenotes}
\end{threeparttable}
\end{table}

\begin{figure}[h!]
  \centering
  \begin{minipage}[t]{0.355\linewidth}
  \begin{tikzpicture}
    \node[anchor=south west, inner sep=0] (image) at (0,0) {\includegraphics[width=1\textwidth]{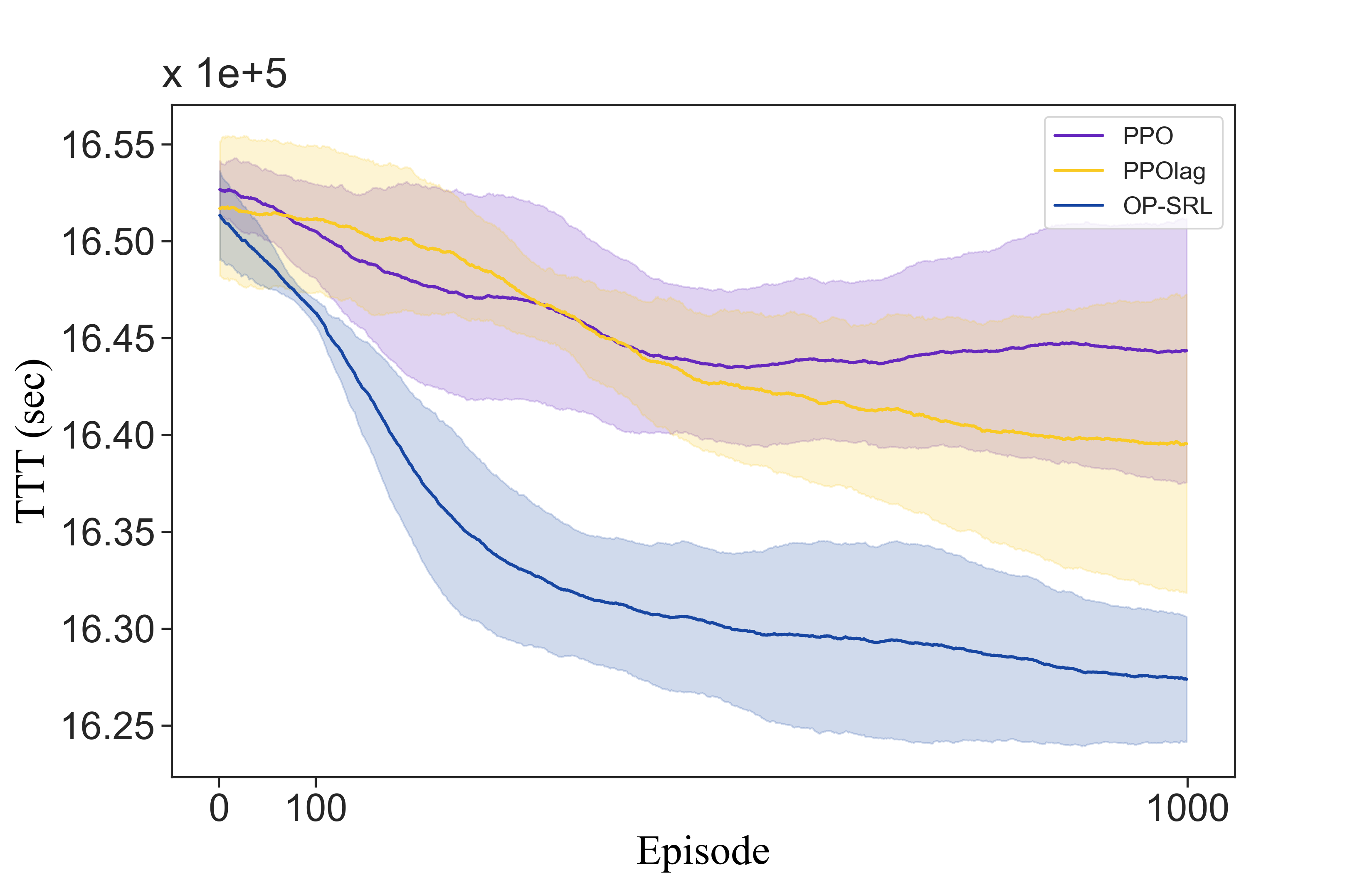}};
    \begin{scope}[x={(image.south east)},y={(image.north west)}]
      \draw[dashed, black, line width=0.6pt] (0.232, 0.641) -- (0.232, 0.11);
    \end{scope}
  \end{tikzpicture}
  \caption*{(a) Traffic efficiency}
  \end{minipage}\hfill\hspace{-20pt}
  \begin{minipage}[t]{0.355\linewidth}
  \begin{tikzpicture}
    \node[anchor=south west, inner sep=0] (image) at (0,0) {\includegraphics[width=1\textwidth]{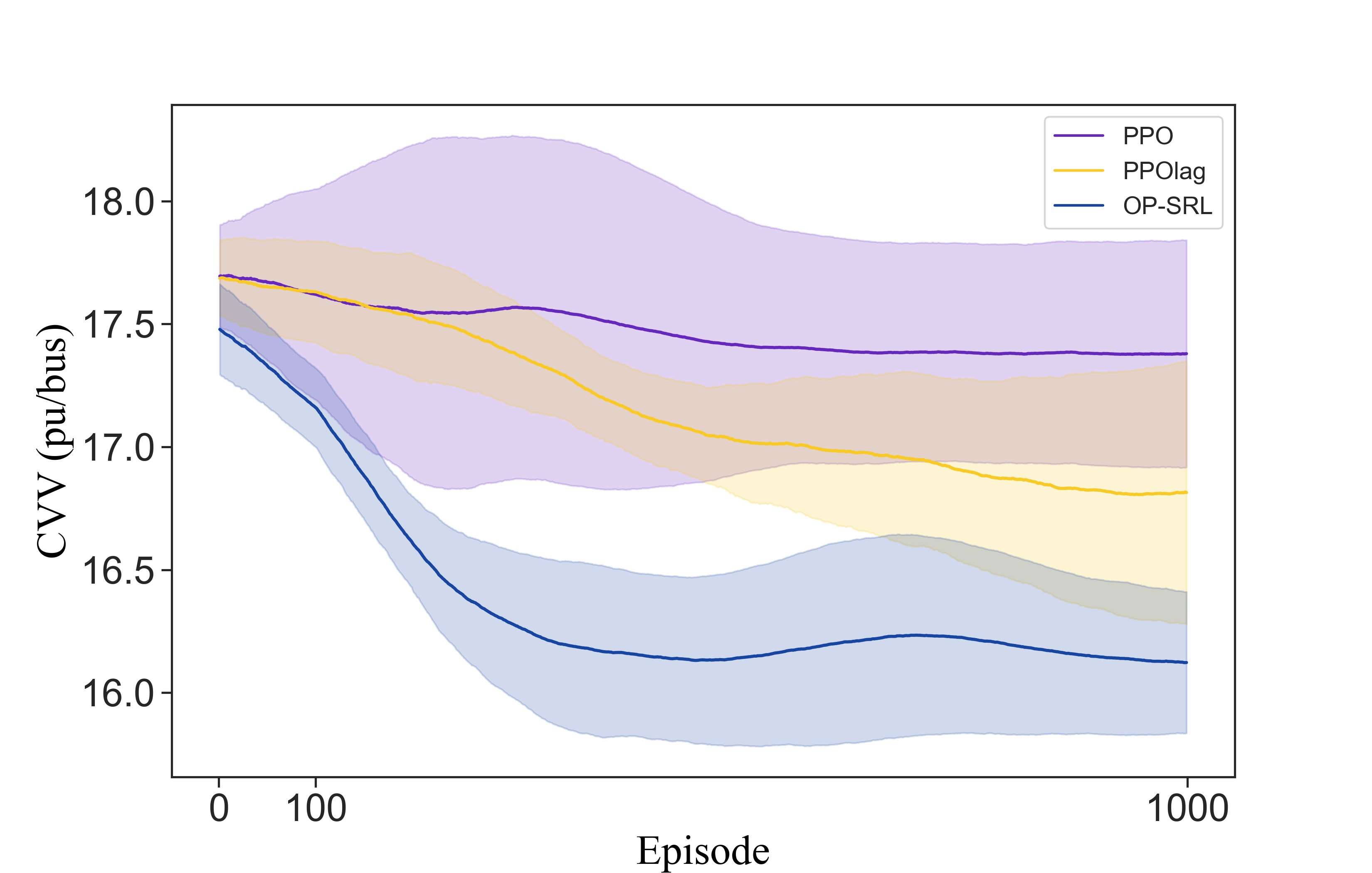}};
    \begin{scope}[x={(image.south east)},y={(image.north west)}]
      \draw[dashed, black, line width=0.6pt] (0.232, 0.5225) -- (0.232, 0.11);
    \end{scope}
  \end{tikzpicture}
  \caption*{(b) Grid safety}
  \end{minipage}\hfill\hspace{-20pt}
  \begin{minipage}[t]{0.355\linewidth}
  \includegraphics[width=1\textwidth]{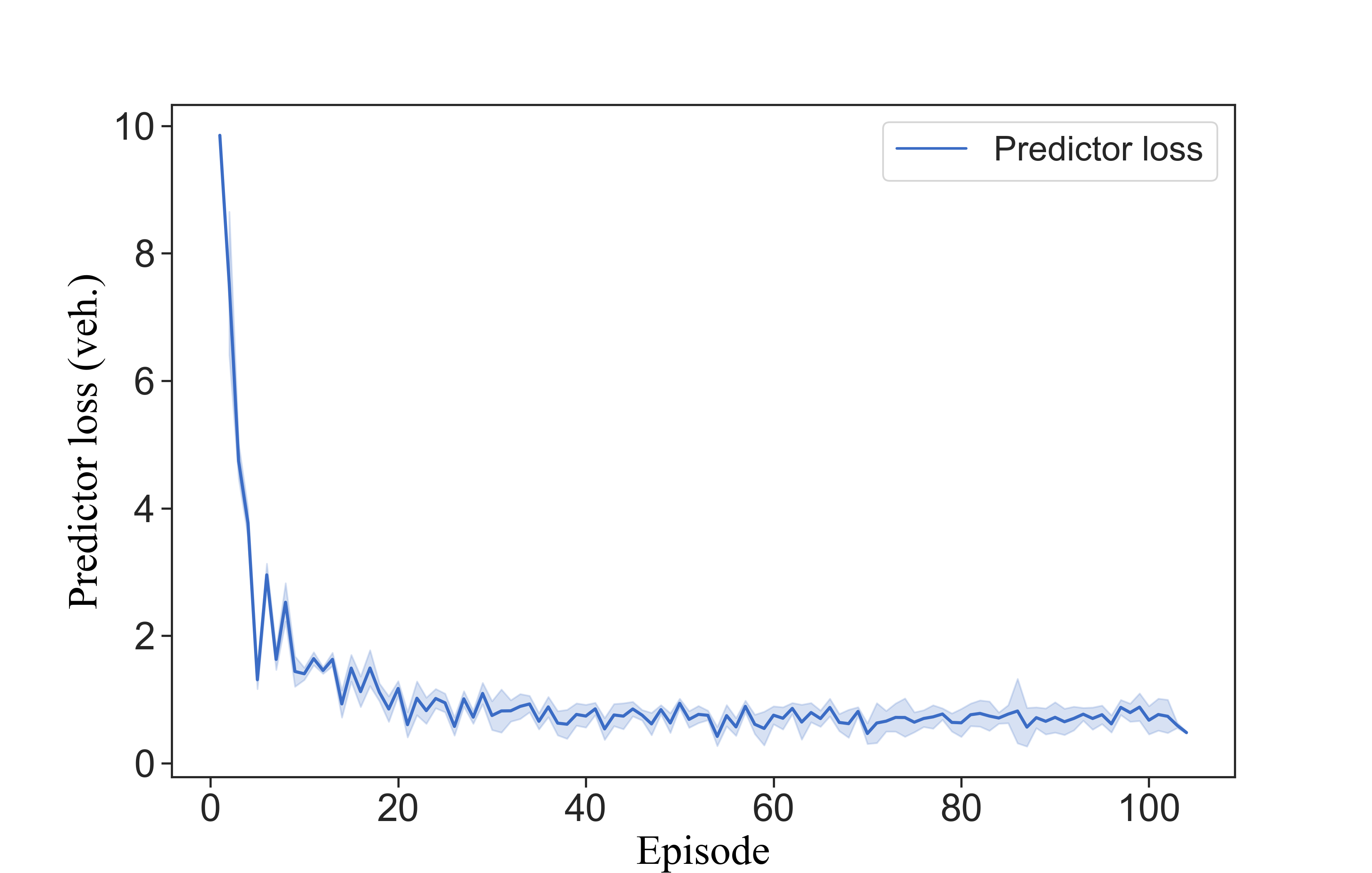}
  \caption*{(c) Seq2Seq predictor's loss}
  \end{minipage}
  \caption{Training curves of the proposed method, PPO, and PPOlag for the real-world network, in terms of TTT (i.e., objective value) in (a) and CVV (i.e., constraint value) in (b), along with training curve of Seq2Seq predictor over RL episodes in (c). All the curves are plotted using five training runs with different random seeds. Solid lines are the mean and the shaded areas stand for the standard deviation region. Besides, Black dashed lines in (a) and (b) indicate the convergence boundary of the predictor.}
 \label{fig:caseB-train-curve}
 \end{figure}
The training process of RL-based methods is shown in Figure \ref{fig:caseB-train-curve}. From this figure, similar results to case A can be obtained. Due to the lack of consideration for constraint, the PPO policy only demonstrates a tendency for improvement in traffic efficiency, whereas there is limited enhancement in grid safety. PPOlag improves upon PPO by incorporating the constraint via the Lagrangian method, resulting in enhanced effectiveness in both the objective and constraint. The proposed OP-SRL further integrates the prediction results from the predictor for state augmentation. Given the imperfect nature of the initial prediction, we can observe a distinct inflection point in the proposed method before and after the convergence of the predictor (around episode 100). After the convergence of the predictor, the proposed method exhibits faster convergence towards a superior strategy regarding both traffic efficiency and grid safety.

{In addition, concerning the execution time, the time expenditure for PPOlag and PPO without the Lagrangian method is similar, approximately around 2.25 days. On the other hand, the OP-SRL method, with the inclusion of the predictor, necessitates an extra 1.5 days for training. Nevertheless, the decision-making time for these RL-based methods is nearly the same, roughly 0.026 seconds, while the greedy method and MPC, respectively, require about 0.5 seconds and 9.9 seconds for making a decision. The decision time based on RL methods only accounts for 5.5\% and 0.26\% of the greedy method and MPC method. Moreover, in case B, with a significant increase in road network scale and decision space, the decision time of the proposed method is 2.4 times that of case A, whereas MPC is 8.2 times longer. This implies that the MPC method struggles to scale effectively as the problem size increases.}

\begin{figure}[h!]
\center
\includegraphics[width=0.6\textwidth]{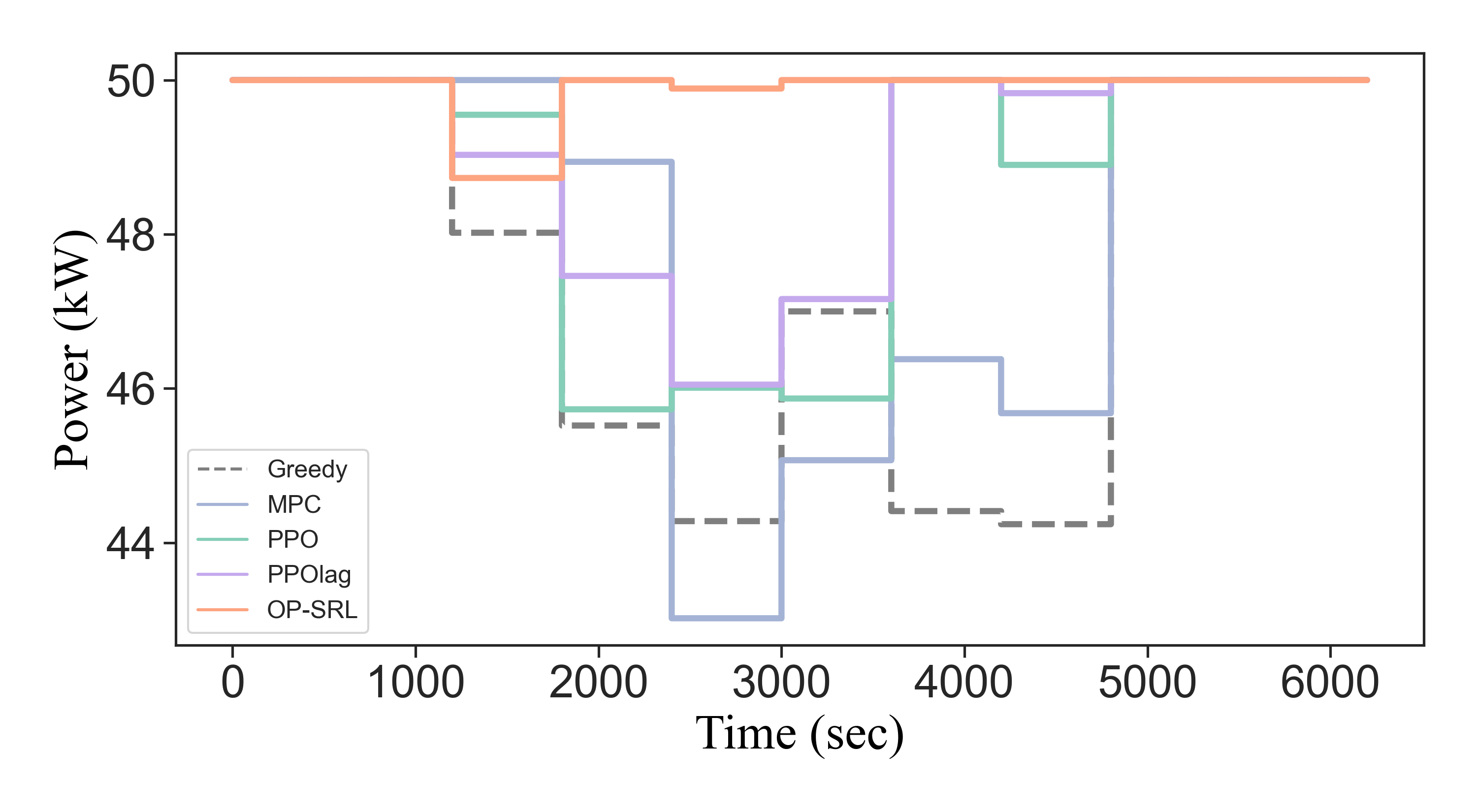}
\caption{{Time-varying charging power using different methods for the real-world network.}}
\label{fig:caseB-power}
\end{figure}

As a result of CS recommendation, Figure \ref{fig:caseB-power} shows the time-varying curves of charging power using different methods. The CS recommendation strategy obtained from the proposed method achieves a better spatial and temporal balance of charging demand, thereby minimizing its impact on the power grid, which enables the highest power output to be maintained for the majority of the time. To be specific, {the proposed method obtains a slightly lower charging power than other baselines excluding the greedy method around 1500 sec. However, this compromise leads to almost full charging power between 1800 sec and 3600 sec, while during this time range, the charging power of PPO and PPOlag has already decreased to below 47 kW. Besides, the power of MPC drops to almost below 46 kW between 2400 sec to 4800 sec and that of the greedy method remains relatively low during the whole control period.}

\section{Conclusion}
\label{sec:conc}

In this paper, we focus on the problem of public CS recommendation for en-route EVs in the context of dynamically coupled transportation-power systems, aiming at maximizing the system-level traffic efficiency while {enhancing} the safety of the power grid regarding voltage violation. Notably, the deterioration of grid safety also leads to a decrease in traffic efficiency, due to the potentially compulsory control for voltage stability in the power grid. The proposed problem is formulated as a CMDP model and then a novel OP-SRL method is developed to find the optimal and secure strategy.

{Extensive numerical studies were carried out on the Nguyen-Dupuis network (case A) and a large-scale real-world road network with observed traffic demand (case B), coupled with IEEE 33-bus and IEEE 69-bus distribution systems, respectively. Three performance metrics were devised to evaluate the proposed method, in terms of road network efficiency, power grid safety, and EV user satisfaction. The comparative analysis against baselines in case A showed that 1) the proposed method achieves superior performances with the best stability; 2) the rankings of different methods in terms of the three metrics are generally consistent due to closely intertwined interests among different stakeholders in the context of coupled systems; 3) after the convergence of the predictor, the proposed method exhibits faster convergence towards a superior strategy and rapidly surpasses the best baseline method (i.e., PPOlag); 4) the inclusions of incorporating Lagrangian multiplier and state augmentation bring added value for the method. Besides, sensitivity analysis for the variation of several parameters (i.e., EV penetration, control interval of the voltage-responsive charge controller, and decoder length in the predictor) showed the robustness and flexibility of the proposed method. In addition, the comprehensive discussion of different methods and settings provides opportunities to choose the most appropriate strategies and methods. Furthermore, results in case B showed the applicability of the proposed method in the practical context.}

The findings of this study also deliver several valuable managerial insights for the EV service provider. First, the service provider should implement charging guidance strategies in the context of the coupled transportation-power systems, since considering the coupling effects benefits the operation of both the traffic network and the power grid. This conclusion regarding the importance of integrating considerations from both systems is consistent with the insights from \citet{yang2024optimal}, despite their focus on managing shared autonomous EVs (SAEVs) mobility and charging dynamics. Second, the uneven distribution of charging demand across various CSs can be obtained from the optimal charging guidance strategy, offering the system planner the potential to optimize the layout and capacity configuration of CSs. Third, the service provider should operate with charging demand forecasts to generate forward-looking charging station recommendation strategies.

This research paves the way for several promising avenues toward more deliberate and effective methods to address challenges in the EV-related control strategy. First, investigating SRL methods with hard constraints would bring more opportunities for broader applications. Second, the integration of transfer learning methods would facilitate the development of a more generalized approach. Then, exploring the recommendation problem involving multiple types of refueling services (e.g., slow and fast charging modes, mobile charging, and battery swapping) would help to develop a more generalized control strategy. Besides, considering the charging urgency and expected charging time to further incorporate individual preferences in decision-making is an interesting research direction. Finally, developing the models and methods suited for applications in large-scale road networks would be a topic of practical significance.

\section*{Acknowledgments}
The work described in this paper is supported by National Natural Science Foundation of China (72471057, 52131203), Natural Science Foundation of Jiangsu Province (BK20232019), the Research Grants Council of the Hong Kong Special Administrative Region, China (Project No. PolyU/25209221 and PolyU/15206322), and the Otto Poon Charitable Foundation Smart Cities Research Institute (SCRI) at the Hong Kong Polytechnic University (Project No. P0043552). The contents of this article reflect the views of the authors, who are responsible for the facts and accuracy of the information presented herein.

\appendix

\section{Hyper-parameters used in the OP-SRL method}  

Table \ref{tab:parameters-SRL} and Table \ref{tab:parameters-predictor} provide the implementation details of the proposed OP-SRL method concerning the hyper-parameters.
\label{appendix}
\setcounter{table}{0}
\renewcommand{\thetable}{A\arabic{table}}
\begin{table}[h!]
\centering
\caption{\label{tab:parameters-SRL} SRL parameters in case A and case B.}
\begin{threeparttable}
\addtolength{\tabcolsep}{20pt}
\begin{tabular}{lc} 
\toprule 
\qquad\qquad\qquad Hyper-parameter & Value \\ 
\midrule 
No. of MLP hidden layers in SRL-related networks\tnote{a} & 2\\
Hidden dimension of MLP in SRL-related networks & 64\\
{No. of epochs to train the SRL agent} & {200}\\
{No. of episodes in each SRL epoch} & {5}\\
{Batch size} & {64}\\
{No. of training iterations in each epoch} & {40}\\
Learning rate of SRL in case A and case B & 3e-4, 1e-4\\
Learning rate of Lagrangian multipiler & 0.035\\
Trade-off coefficients of GAE ($\eta^r$ and $\eta^c$) & 0.95\\
Discount factors ($\gamma^r$ and $\gamma^c$) in case A and case B & 0.97, 0.99\\
\bottomrule 
\end{tabular} 
\begin{tablenotes}
\footnotesize
\item[a] Including actor network, reward critic network, and cost critic network.
\end{tablenotes}
\end{threeparttable}
\end{table}

\begin{table}[H]
\centering
\caption{\label{tab:parameters-predictor} Hyper-parameters of various Seq2Seq predictors.}
\begin{tabular}{ccc} 
\toprule 
Model & Hyper-parameter & Value \\ 
\midrule 
\multirow{9}{*}{Shared parameters} & No. of iterations in each training step & 20\\
& Sequence length of encoder (exclude TCN) & 5\\
& Sequence length of decoder (exclude TCN) & 5\\
& Time interval per step in Seq2Seq ($w$) & 4 min\\
& Sampling interval & 1 min\\
& Minimum training sample size ($n_1^{th}$) & 64\\
& Sample size in a training interval ($n_2^{th}$) & 50\\
& Batch size & 64\\
& Learning rate & 1e-3\\
\midrule
\multirow{3}{*}{RNN/GRU/LSTM} & No. of hidden layers & 2\\
& Hidden dimension & 256\\
& Dropout & 0.5\\
\midrule
\multirow{5}{*}{{TCN}} & {No. of hidden layers} & {2}\\
& {Hidden dimension} & {256}\\
& {Number of channels for each layer} & {64, 64}\\
& {Kernel size} & {3}\\
& {Dropout} & {0.5}\\
\midrule
\multirow{6}{*}{{Transformer}} & {Number of expected features in the encoder and decoder} & {64}\\
& {Number of heads} & {2}\\
& {Number of layers in the encoder} & {2}\\
& {Number of layers in the decoder} & {2}\\
& {Dimension of the feedforward network model} & {64}\\
& {Dropout} & {0.1}\\
\bottomrule 
\end{tabular} 
\end{table}

\bibliography{main_myref}

\end{document}